\documentclass[a4paper,11pt]{article}
\usepackage[latin1]{inputenc}
\usepackage{indentfirst}
\usepackage[left]{lineno}
\usepackage{amsfonts}
\usepackage{float}

\usepackage{latexsym,amssymb,amsmath}
\usepackage[dvips]{graphics,graphicx,epsfig,color}
\usepackage[lofdepth,lotdepth]{subfig}
\usepackage[lofdepth,lotdepth]{subfig}
\begin{document}
%\linenumbers
\title{\bf Resonant amplified seismic  response  within a hill or mountain}
\author{Armand Wirgin\thanks{LMA, CNRS, UPR 7051, Aix-Marseille Univ, Centrale Marseille, F-13453 Marseille Cedex 13, France, ({\tt wirgin@lma.cnrs-mrs.fr})} }
\date{\today}
\maketitle
\begin{abstract}
We show theoretically what is meant by the term '(surface shape) resonance' in connection with the seismic response of a protuberance (emerging from flat ground) such as a hill or mountain of arbitrary shape.
We address the specific problem of cylindrical protuberances of rectangular shape submitted to a SH plane wave. We find that the principal (i.e., qualitative) characteristics of the seismic response of a mountain are quite similar to those of a hill, and that the occurrence of significative amplification of the displacement field within these structures is due to the coupling of the incident wave to (surface shape) resonances.
\end{abstract}
\newline
\newline
Abbreviated title: Seismic motion within  a hill or mountain
\newline
\newline
Corresponding author: Armand Wirgin, \\ e-mail: wirgin@lma.cnrs-mrs.fr
%%%%%%%%%%%%%%%%%%%%%%%%%%%%%%%%%%%%%%%%%%%%%%%%%%%%%%%%%%%%%%%%%%%%%%%%%%%%%%%%%%%%
\newpage
%%%%%%%%%%%%%%%%%%%%%%%%%%%%%%%%%%%%%%%%%%%%%%%%%%%%%%%%%%%%%%%%%%%%%%%%%%%%%%%%%%%%
\tableofcontents
%%%%%%%%%%%%%%%%%%%%%%%%%%%%%%%%%%%%%%%%%%%%%%%%%%%%%%%%%%%%%%%%%%%%%%%%%%%%%%%%%%%%
\newpage
\newpage
%%%%%%%%%%%%%%%%%%%%%%%%%%%%%%%%%%%%%%%%%%%%%%%%%%%%%%%%%%%%%%%%%%%%%%%%%%%%%%%%%%%%%%%%%%%%%%%%%%%%%%
%%%%%%%%%%%%%%%%%%%%%%%%%%%%%%%%%%%%%%%%%%%%%%%%%%%%%%%%%%%%%%%%%%%%%%%%%%%%%%%%%%%%%%%%%%%%%%%%%%%%%%
\section{Introduction}\label{intro}
 This investigation is concerned with a problem of determining the response of an object to an incident wave \cite{ps01,ke72a},  the object being such that its interior is accessible to the incident wave.

 The wavefield in the interior of the object, as well in its (external) vicinity can be qualified as its 'near-field response' \cite{cb01,ha12,hs19,lh17,lw17,ll12,mp71,pe04,sb12}. The wavefield in the exterior of the object, and rather far from the latter can be qualified as  its 'far-field response' \cite{ha12,lb11,mb06,ma63,og91,pa95,rw82,ri51}.  The term 'scattering' is usually associated with the far-field response, but since the distinction between the near and far fields is not always clear, 'scattering' \cite{si78,ps01,tc09b,wi73} is often employed to qualify the global response of the object to a dynamical solicitation. Nevertheless,  there is an intimate connection between the the near- and far-field response, as well as between the interior and exterior response, as is  illustrated in two recent publications \cite{wi19,wi20}.

 Herein, we shall be particularly interested in the internal response of an isolated  feature (loosely-termed  'structure' \cite{ce08,fv99,go13,jb13,jl19,jo89,lw82,ro13,sa98,sd94}) located near or on the earth's nominally-flat boundary  to an elastic wave (more precisely, seismic wave) whose sources are within the earth. The social and economic implications of this problem are considerable since they have to do with the disasters \cite{he83} provoked by earthquakes: destruction of the natural \cite{fp18} and built environment \cite{as17,ba05,bg07,bi99,cg89,de05,fv99,gr05,gu15,mu08,md08,me06,os92,pt05,sh00} often resulting in considerable harm to the population.

 The feature of the earth's boundary that first comes to mind is the above-ground structure (AGS), often termed    topography \cite{be14,ag05,an97,as14,ac06,an10,as17,ap99,bh12,ba82,bc09,bg07,bo72,bo73a,bo96,bp05,bf14,ce87,ce88,ch19,ca95,dw73,gb88,gr09,gb79,hm14,
  kv19,kn18a,kn18c,le09,ma07,pr14,ri97,ri03,ro96,sa10,tk13,tc09b,wl12,wo82,zc06,ma10,mc12,mc15,ts99,rk74,rr12,ls95,da12,gb13,mb14,
  bm14,aj13,aw07,my12,bd84,kn18a,my12,mb97,nb95,ps94,pc10,rr15,sc91,sm05,sh96,ts99,wd18,hh73,wt77},
  convex surface irregularity \cite{bh12,lo17,lc09,pa02,si78,vc99,zc06,al70,at15,wi64a,da12,sd94,vc99,ka08}
  or slope \cite{pa11,sm05,tk13,an97,bp05,bg10,ke94,aw07,ke72a,ke72b,mk66,mg11,ra19},
  roughness \cite{hu98,ii09,ma82,ma88,og91,ri51,vw70,wi88,bs87,ab62,mi69,rw82,tc09a,wi68,kt03,lh85,ld67,ma63,pc88},
  cliff \cite{ak05,mk66,pa02},   ridge \cite{bt85,kn18c,pl94,zm93},   protuberance \cite{wi20},
  or bump \cite{mp18}.

   The  AGS's affected  by seismic waves can be divided into two classes: i) a 'natural' AGS such as a hill \cite{as14,bh12,bb96,bu08,cl11,ch19,gr09,hw11,kj08,kn18a,lh99,la13a,la13b,ll06,lc10,mn75,os92,pa11,pc10,ql05,pa02,rk08,so01,yx14a,
ym92,da12,dg99,jc00,bb96,lo17,ra15,gb13,ys20,tc10,yx14a,tc09b,ub18,pc10}, volcano \cite{sm18,ri03,so01,mb06,ca13,mb06,sm18},
mountain \cite{ks91,kw92,wi88,wi90a,wi90b,wk92,wk93,bg10,bg10,gb79,ma07,dw73,sm75,ra19,he83,la13b,la13c,ma10} and ii) a 'manmade' AGS such as a building (composed or not of a  limited number of smaller units) \cite{gr05,br06,ll12,lc09,wi18a,wg06a,wg06b,wt75,zs11,ue10,to01,ps15,tt92,jb73,jl19,mf14,gu15,my12,ga19,aa15,aa16,bf13,ck15,co14,
dg07,dm10,ho57,it14,it15,jb73,kk06,lp10,mu08,md08,rg12,tr72,vz14},
industrial facility such as a nuclear power plant \cite{fd16,nb73,yk14,ww18,zh18b,ho12,gl18,nk98,gw08},
dike \cite{ht01,tv18,dc19,dc19,ys20}, dam \cite{bo73a,kt91a,kt91b,gw05a,gw05b},
town or city \cite{gr05,pt05,ga19,br04,br06,cg89,cs00,cl15,ca01,dg07,db13,fb06,ga19,gi09,gt05,gw08,gb02,gc16,gs13,it14,it15,kj93,
kh04,kn18b,kn17,ll12,lc06,my12,pa09,ro06,sn15,sb15,sb16,sd00,sk08,sk09,ta10,tb11a,tb11b,tl12,ue10,vl16,wi18a,wb96,wg06a,wg06b,wt75}, or layer of some sort (such as landfill \cite{mk06,zt06,ld67,td93,vc99}).

Other structures affected by seismic waves are the below-ground structures (BGS's) which can also  be divided into two classes: 1) 'natural' such as a cavity, canyon \cite{hw11,vs17,tc10,ct15,hl10},
unfilled valley \cite{bb85,bo73b,kn18c,ep14,jc00,fa95,kn18c},
basin or sediment-filled valley \cite{bs90,wi95,bl71,kn19,pe15,ri97,sn15,sk08,sk05,to92,zt06,mk06,sd00,hy96,bb80,bb85,ka96,ka08,ad00,ar19,zh18a,zr18,zt16,zt15,
vs17,va19,fl07,rm16,bs89,wi19,sd94,fa95,fs97,ct14,ga19,fa95,ka03,kh04,ks06,ms16,sk09,sm75,vl16,wt74},
or other  underground layer of some sort \cite{aa11,wo79,ba95,br00,br11,ej57,jp53,km12,ke01,kn64,le05,ll14,lo11,my87,sk05,st28a,
st28b,td93,vi67,vs17,wt77,zt20}, and 2) 'manmade' such as a
pipeline \cite{da99}, tunnel \cite{kz08,hl10}, or mine \cite{gl19}.

From the practical point of view all these structures have in common that their seismic response is marked by augmented duration of the shaking \cite{tb75,wb96,wg06a,wg06b,gr05,gt05,gw08} as well as  amplification of the wavefield \cite{an10,gr05,dw73,gb79,wk93,zr18} in their interiors  and in their neighborhood  relative to this response on flat ground overlying a homogeneous underground. This is known, or thought, to be the principal causal agent of: damage or destruction of buildings in cities (often located on flat ground underlain by a soft soil layer) \cite{de05,fn87,ka96,ri97,tl12},  landslides on hills and mountains \cite{ke94,pj00,zs13,sm05,mg11,hj02,hs03,fp18,fh18,ra19}, damage or destruction of dikes \cite{ht01,tv18,dc19} and dams \cite{kt91a,kt91b,sm05} resulting in massive flooding, damage of towns built near or on hill and mountain  summits, the liberation of poisonous gases in landfills \cite{mk06,zt06} and even the triggering of volcanic eruptions \cite{sm18,mb06}.

From the theoretical point of view, there are  obvious connections between the seismic response of 'natural' AGS's and that of both 'natural' and 'manmade' BGS's, but since this general subject is vast, we shall be more-specifically interested  in the seismic response of an isolated 'natural' AGS, which we assume to be entirely located above, and attached to, the (flat) ground. Nevertheless, we  offer some bibliographic references and evoke the issue of resonances in connection with the seismic response of 'manmade' AGS's and 'natural' as well as 'manmade' BGS's.

It so happens that similar problems are  of great interest in other fields of physics such as electrodynamics (e.g, optics and microwaves)  \cite{ba70,lw84,mo85,lm98,ma82,dg82,mv85,wl84,gs99,ts12,si19,as87,wm85,wm86,lh17,wk01,sb12,gd18,ha12,lw17,cb01,dn19,lp19a,
lp19b,mo98,gi07,kz07,ra07,bb95,ez06,ma12,og91,pa95,pe04,pe80,rw82,ri51,sb12,tc09a,vw70,wi64a,wi64b,wi68,wi73,wi02},
acoustics \cite{vw70,ub92,mf14,ve93,gd11a,gd11b,wi18b,hu98,ma88,ma63,mi69}, hydrodynamics \cite{zz08} and atomic physics \cite{hr85,fr98,gg79,hu10,og91,ra07,ra45}, even though what was traditionally-important in these contexts was the far-field response of the scattering structures usually termed periodically-rough surfaces,
corrugations and gratings \cite{fa36,fa41,kt03,ld67,pc88,as66,lp19b,lm98,pe80,ra07,ra45,wi64b}, or irregular surfaces \cite{hu98,ii09,ma88,og91,ri51,vw70,bs87,ab62,ma82,mi69,rw82,tc09a,wi68,ii09,wi64a,wi82,wi02}), rather than their near-field response.

In recent years, research in these fields has largely switched to the near-field, stimulated by the discovery of such spectacular effects as SERS (Surface-Enhanced Raman Scattering)\cite{lw84,mo85,tl12,wk01,wl84},
enhanced frequency-selective (total) absorption (as applied notably to energy harvesting \cite{kw15,lp19a} or noise reduction \cite{gd11a,gd11b,wi18b,lp19a}), and negative refraction \cite{kz07,lh11} in media (recently dubbed metamaterials \cite{ww18,zx18}) bounded by, or including, periodic structures  \cite{ez06,lw84,lm98,ma82,dg82,mr88,wl84,gs99,ts12,si19,as87,wm85,wm86,lh17,wk01,sb12,gd18,ha12,lw17,cb01,dn19,lp19a,lp19b,be19,
dg16,fe17,ha12,lh11,pe04}.

The core of the present contribution is the notion of 'resonance' and its connection with wavefield amplification
 \cite{ra85,ge19,vi67,jp53,wi88,wi90a,wi90b,ha64,lo11,mr88,ma88,mv85,fn87,ub92,ud96,mm80,bt17,zr18,lw15,wo79,bb85,ep14,hl84,lh85,
lc06,lc09,lm98,lw84,mf14,ve93,wi95,wm85,wm86,zz08,dg16,dn19,ej57,gr05,gw05a,gw05b,gw08,jo89,kj08,kv19,ks91,kn18b,kn17,la85,
lp19a,lp19b,lh85,lc09,ma12,mo98,mp71,pa02,ra85,ra89,sb12,si19,se11,st28a,st28b,vz14,wi88,wk93,wl84}. This notion has often been invoked, but, in our opinion not well-understood,  in connection with scattering of a seismic wave by one or several protuberances. The reason for this may be linked to the fact that the majority of previous studies on the seismic response of structures on or below the earth's surface were purely-numerical (e.g., \cite{gb89,ve12, aa12,aa15,aa13,ar19,bp05,ck15,dm10, go13,gc16,ks06,ro96,sd00,va19,wl12,zt06,ht01,bl71,wt77,si78,bo73b,da12,dg99,hw11,vc99,jc00,kk06,le09,ls95,lc09,ms16,ts99,mb97,
pa09,pa02,pa11,ps94,rk08,rm16,ri03,ro96,sc91,sc89,se11,sd00,sk08,sk09,sk05,ta10,tb11a,tb11b,tr72,ts99,vc99,vl16,wl12,wi73,wo82,
wt74,ys20,yx14a,ym92,zm93}).

In optics, the  topic has been dubbed 'surface shape resonances' \cite{ej57,pe80,ma82,dg82,ma88,mr88,mv85,wi88,wi90b,wi95,lw84,lm98,lm98,mo85}, and appears to be well-understood at present, notably from the theoretical viewpoint, so that it is opportune to make use of this knowledge in the present investigation.
%%%%%%%%%%%%%%%%%%%%%%%%%%%%%%%%%%%%%%%%%%%%%%%%%%%%%%%%%%%%%%%%%%%%%%%%%%%%%%%%%%%%%%%%%%%%%%%%%%%%%%
\section{Description of what is  meant here by the seismic response of a protuberance}\label{desc}
The general problem is identical to the one studied in our recent contribution \cite{wi20}. In the first approximation, the earth's surface is considered to be (horizontally-) flat (termed "ground" for short) and to separate the vacuum (above) from a linear, isotropic, homogeneous (LIH) solid (below), so as to be stress-free. In the second approximation the flat ground is locally deformed so as to penetrate into what was formerly the vacuum half space. We now define the protuberance as the region between the locally-deformed stress-free surface and what was formerly a portion of the flat ground. This protuberance is underlain by the same LIH solid as previously, but the solid material within the protuberance is now assumed to be  only linear and isotropic (i.e., not homogeneous). In fact, we consider the specific  case in which the material within the protuberance is in the form of a horizontal bilayer so as to be able to account for various empirically-observed effects that are thought to be due to inhomogeneity of the protuberance material. Furthermore, we assume that: the protuberance is of infinite extent along one ($z$) of the cartesian ($xyz$) coordinates,  and its stress-free boundary to be of arbitrary shape (in its $xy$ cross-section plane). The underlying problem  of much of what follows  is the prediction of the seismic wave response of this earth model.

The earthquake sources are assumed to be located in the lower half-space and to be infinitely-distant from the ground so that the seismic (pulse-like) solicitation takes the form of a body (plane) wave in the neighborhood of the protuberance. This plane wavefield is assumed to be of the shear-horizontal ($SH$) variety, which means that: only one (i.e., the cartesian coordinate $z$) component of the incident displacement field is non-nil and this field does not depend on $z$.

We  assume, not only that the protuberance boundary does not depend on $z$ but also, that the (often relatively-soft) medium filling the protuberance as well as the (usually relatively-hard) medium below the protuberance are both linear and isotropic. Furthermore the medium of the below-ground half space is assumed to be homogeneous, whereas that of the protuberance to be piecewise homogeneous (however, this heterogeneity is such as to not depend on $z$). It ensues that the scattered and total displacement fields within and outside the protuberance do not depend on $z$. Thus, the problem we are faced with is 2D ($z$ being the ignorable coordinate), and it is sufficient to search for the $z$-component of the scattered displacement field, designated by $u_{z}^{s}(\mathbf{x};\omega)$ in the sagittal (i.e., $x-y$) plane, when $u_{z}^{i}(\mathbf{x};\omega)$ designates the incident displacement field, with $\mathbf{x}=(x,y)$ and $\omega=2\pi f$  the angular frequency, $f$ the frequency. Since we now know that only the $z$ component of the field is non-nil, we drop the index $z$ in the incident, scattered, and total displacement fields. The temporal version of the displacement field is $u_{z}(\mathbf{x};t)=2\Re\int_{0}^{\infty}u_{z}^{i}(\mathbf{x};\omega)\exp(-i\omega t)d\omega$ wherein $t$ is the temporal variable. Since we now know that only the $z$ component of the field is non-nil, we drop the index $z$ in the incident, scattered, and total displacement fields in all that follows.
\begin{figure}[ht]
\begin{center}
\includegraphics[width=0.75\textwidth]{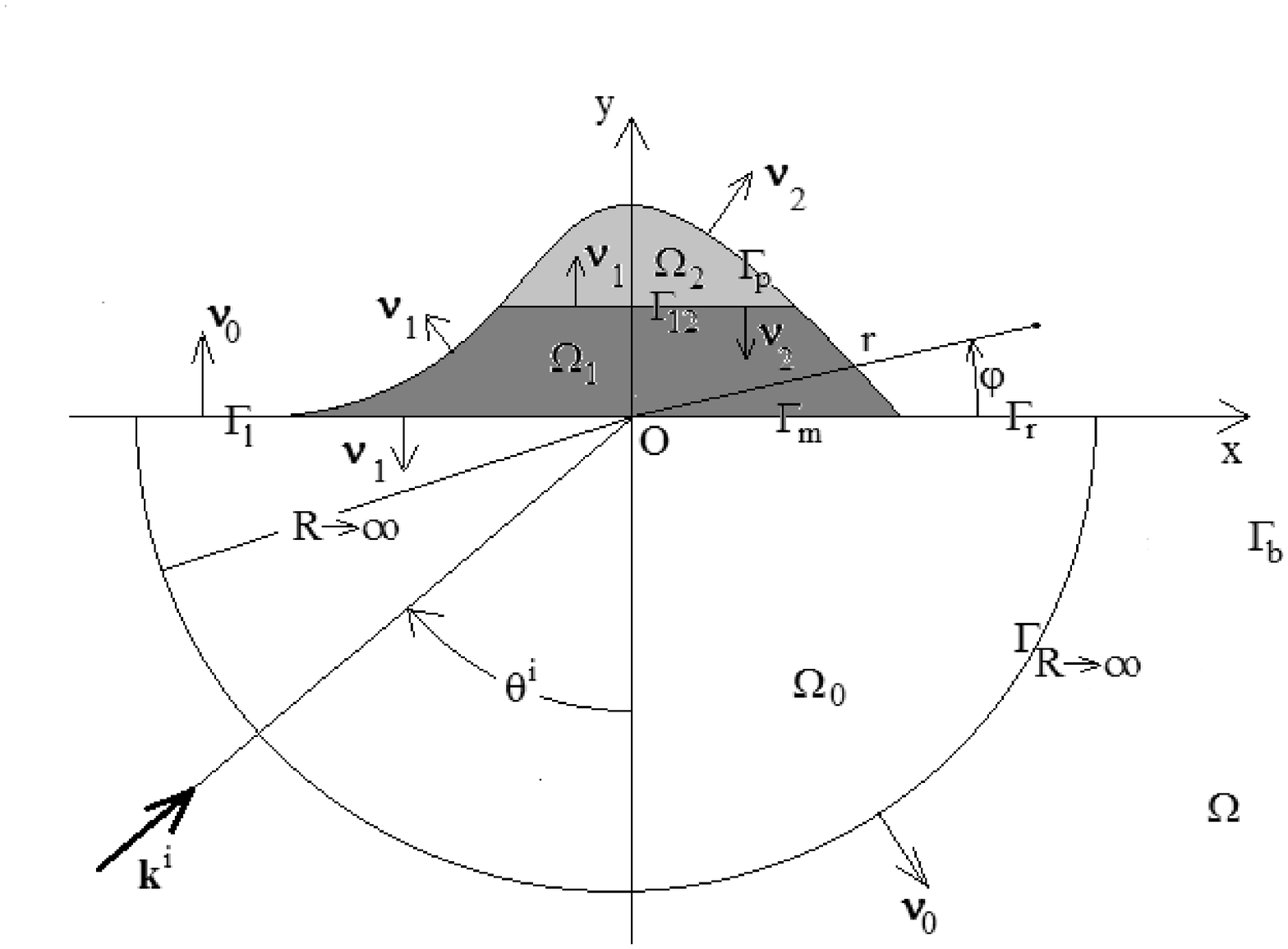}
\caption{Sagittal plane view of the 2D scattering configuration. The protuberance occupies the shaded areas and the medium within it is a horizontal bilayer.}
\label{protuberance}
\end{center}
\end{figure}

Fig. \ref{protuberance} describes the scattering configuration in the sagittal ($xy$) plane. In this figure, $\mathbf{k}^{i}=\mathbf{k}^{i}(\theta^{i},\omega)$ is the incident wavevector oriented so that its $z$  component is nil, and $\theta^{i}$ is the angle of incidence.

The portion of the ground outside the protuberance is stress-free but since the protuberance is assumed to be in welded contact with the surrounding below-ground medium, its lower, flat, boundary is the locus of continuous displacement and stress. Thus,  the incident field is able to  penetrate into  the protuberance and then  be scattered outside the protuberance in the remaining lower half space.

The three media (other than the one of the portion of the  space above the protuberance, being occupied by the vacuum, is of no interest since the field cannot penetrate therein)  are $M^{[l]}~;~l=0,1,2$  within which the real shear modulii $\mu^{[l]}~;~l=0,1,2$  and the generally-complex shear body wave velocities are $\beta^{[l]}~;~l=0,1,2$  i.e., $\beta^{[l]}=\beta^{'[l]}+i\beta^{''[l]}$, with $\beta^{'[l]}\ge 0$, $\beta^{''[l]}\le 0$, $\beta^{[l]}=\sqrt{\frac{\mu^{[l]}}{\rho^{[l]}}}$, and  $\rho^{[l]}$ the (generally-complex) mass density. The shear-wave velocity $\beta^{[0]}$ is assumed to be real, i.e., $\beta^{''[0]}=0$.

The numerical results in this first  contribution  on the general problem of scattering of seismic waves by a protuberane will be devoted to two configurations in which; i) the single solid within the protuberance is the same as that in the lower half space, and ii) the single solid within the protuberance is different from the one in the lower half space. Thus, in both cases, these numerical results will apply to a homogeneous protuberance.

As will become apparent further on, as well as in our subsequent contributions to this general subject, the essential notion of 'resonance' does not imply any restrictions either on the shape of the stress-free portion of the boundary of the protuberance or on the (lossy or non-lossy, homogeneous or inhomogeneous) nature  of the medium filling the protuberance.
%%%%%%%%%%%%%%%%%%%%%%%%%%%%%%%%%%%%%%%%%%%%%%%%%%%%%%%%%%%%%%%%%%%%%%%%%%%%%%%%%%%%%%%%%%%%%%%%%%%%%%
\section{Boundary-value problem for the bilayer protuberance whose boundary is of arbitrary shape}\label{bvp}
The protuberance occupies (in the sagittal plane (SP)) the finite-sized region $\Omega_{1}\bigcup\Omega_{2}$ (see fig. \ref{protuberance}).  The  below-ground half-space occupies the region $\Omega_{0}$. $\Omega_{0}$ is entirely filled with $M^{[0]}$ whereas $\Omega_{1}$ is  filled with $M^{[1]}$ and $\Omega_{2}$ with  $M^{[2]}$.

Always in the sagittal plane, the flat ground is described by $\Gamma_{G}$, with $x,y$ the cartesian coordinates in the SP) and is composed  of three segments; $\Gamma_{l}$, $\Gamma_{m}$, and $\Gamma_{r}$, which designate the left-hand, middle, and right-hand portions respectively of $\Gamma_{G}$. The protuberance is an above-ground structure whose upper and lower boundaries (in the SP) are $\Gamma_{p}$ and $\Gamma_{m}$, the latter being a plane segment of width $w$.

The analysis takes place in the space-frequency framework, so that all constitutive and field variables depend on the frequency $f$. This dependence will henceforth be implicit (e.g., $u(\mathbf{x};f)$, with $\mathbf{x}=(x,y)$, will be denoted by $u(\mathbf{x})$).

The seismic solicitation is a shear-horizontal (SH)  plane wave field of the form
\begin{equation}\label{1-000}
u^{i}(\mathbf{x})=a^{i}\exp(i\mathbf{k}^{i}\cdot\mathbf{x})=a^{i}\exp[i(k_{x}^{i}x+k_{z}^{i}y)]~,
\end{equation}
wherein $a^{i}=a^{i}(\omega)$ is the spectral amplitude of the seismic pulse, $\mathbf{k}^{i}=(k_{x}^{i},k_{y}^{i})$, $k_{x}^{i}=k^{[0]}\sin\theta^{i}$, $k_{y}^{i}=k^{[0]}\cos\theta^{i}$, $k^{[l]}=\omega/\beta^{[l]}~;~l=0,1,2$.

Owing to the fact that the configuration comprises three distinct regions, each in which the elastic parameters are constants as a function of the space variables, it is opportune to employ domain decomposition and (later on separation of variables). Thus, we decompose the total  field $u$ as:
\begin{equation}\label{1-010}
u(\mathbf{x})=u^{[l]}(\mathbf{x})~;~\forall\mathbf{x}\in\Omega_{l},~l=0,1,2~,
\end{equation}
with the understanding that these fields satisfy the 2D SH frequency domain elastic wave equation (i.e., Helmholtz equation)
\begin{equation}\label{1-020}
\Big(\triangle+\big(k^{[l]}\big)^{2}\Big)u^{[l]}(\mathbf{x})=0~;~\forall\mathbf{x}\in\Omega_{l},~l=0,1,2~,
\end{equation}
 with the notations $\triangle=\frac{\partial^{2}}{\partial x^{2}}+\frac{\partial^{2}}{\partial y^{2}}$ in the cartesian coordinate system of the sagittal plane.

 In addition, the field $u^{[0]}$ satisfies the radiation condition
\begin{equation}\label{1-030}
u^{[0]}(\mathbf{x})-u^{i}(\mathbf{x})\sim \text {outgoing~wave} ~;~\|\mathbf{x}\|\rightarrow \infty~.
\end{equation}
due to the fact that $\Omega_{0}$ is unbounded (i.e., a semi-infinite domain).

The stress-free nature of the boundaries $\Gamma_{l}$, $\Gamma_{p}$, $\Gamma_{r}$, entail the boundary conditions:
\begin{equation}\label{bc-010}
\mu^{[0]}u_{,y}^{[0]}(\mathbf{x})=0~;~\forall\mathbf{x}\in\Gamma_{l}+\Gamma_{r}~,
\end{equation}
\begin{equation}\label{bc-020}
\mu^{[2]}u_{,y}^{[2]}(\mathbf{x})=0~;~\forall\mathbf{x}\in\Gamma_{p}~,
\end{equation}
wherein   $u_{,\zeta}$ denotes the first  partial derivative of $u$ with respect to $\zeta$.

The fact, that the horizontal segment $\Gamma_{12}$ between the two media filling the protuberance is assumed to be an interface across which two media are in welded contact, entails the continuity conditions:
\begin{equation}\label{bc-040}
u^{[2]}(\mathbf{x})-u^{[1]}(\mathbf{x})=0~;~\forall\mathbf{x}\in\Gamma_{12}~,
\end{equation}
\begin{equation}\label{bc-050}
\mu^{[2]}u_{,y}^{[2]}(\mathbf{x})-\mu^{[1]}u_{,y}^{[1]}(\mathbf{x})=0~;~\forall\mathbf{x}\in\Gamma_{12}~.
\end{equation}

Finally, the fact, that $\Gamma_{m}$ was assumed to be an interface across which two media are in welded contact, entails the continuity conditions:
\begin{equation}\label{bc-040}
u^{[0]}(\mathbf{x})-u^{[1]}(\mathbf{x})=0~;~\forall\mathbf{x}\in\Gamma_{m}~,
\end{equation}
\begin{equation}\label{bc-050}
\mu^{[0]}u_{,y}^{[0]}(\mathbf{x})-\mu^{[1]}u_{,y}^{[1]}(\mathbf{x})=0~;~\forall\mathbf{x}\in\Gamma_{m}~,
\end{equation}

The purpose of addressing such a boundary-value (direct) problem is to determine $u^{[l]}(\mathbf{x});~l=0,1,2$ for various solicitations  and parameters  relative to the  various geometries of, and media filling, $\Omega_{l}~;~l=0,1,2$.  Rather than carry out this too-ambitious  program, we shall treat only one specific configuration: a protuberance of rectangular shape.

Before doing this, it is opportune to recall some consequences, brought to fore in \cite{wi20}, of the above-evoked  boundary-value problem for bilayer protuberances of general shape. In this publication, we show that the wavefield is constrained by the following conservation law:
\begin{equation}\label{2-140}
-\frac{\mu^{[1]}}{\mu^{[0]}}\Im\int_{\Gamma_{m}}u^{[1]*}(\mathbf{x})\boldsymbol{\nu}_{1}\cdot\nabla u^{[1]}(\mathbf{x})+\\
\Im\int_{\Gamma_{\infty}}u^{[0]*}(\mathbf{x})\boldsymbol{\nu}_{0}\cdot\nabla u^{[0]}(\mathbf{x})d\Gamma=0~,
\end{equation}
or
\begin{equation}\label{2-150}
\Im\int_{\Gamma_{\infty}}u^{[0]*}(\mathbf{x})\boldsymbol{\nu}_{0}\cdot\nabla u^{[0]}(\mathbf{x})d\Gamma+
\frac{\mu^{[2]}}{\mu^{[0]}}\Im\big[\big(k^{[2]}\big)^{2}\big]\int_{\Omega_{2}}\|u^{[2]}(\mathbf{x})\|^{2}d\Omega+
\frac{\mu^{[1]}}{\mu^{[0]}}\Im\big[\big(k^{[1]}\big)^{2}\big]\int_{\Omega_{1}}\|u^{[1]}(\mathbf{x})\|^{2}d\Omega=0~.
\end{equation}
The term 'constrained' means that the solution (for the wavefield) of the equations in the boundary-value problem must be such as to satisfy either (\ref{2-140}) or  (\ref{2-150}), this being a necessary but not sufficient condition for this solution to be qualified as 'correct'. Thus, either of these two equations furnish a useful, although not foolproof, means of testing a method for solving a scattering problem such as the one we are about to evoke.
%%%%%%%%%%%%%%%%%%%%%%%%%%%%%%%%%%%%%%%%%%%%%%%%%%%%%%%%%%%%%%%%%%%%%%%%%%%%%%%%%%%%%%%%%%%%%%%%%%%%%%%%%%%%%%%%%%%
\section{Case of a bilayer protuberance  of rectangular shape}
%
%%%%%%%%%%%%%%%%%%%%%%%%%%%%%%%%%%%%%%%%%%%
\subsection{Description of the configuration}
\begin{figure}[ht]
\begin{center}
\includegraphics[width=0.85\textwidth]{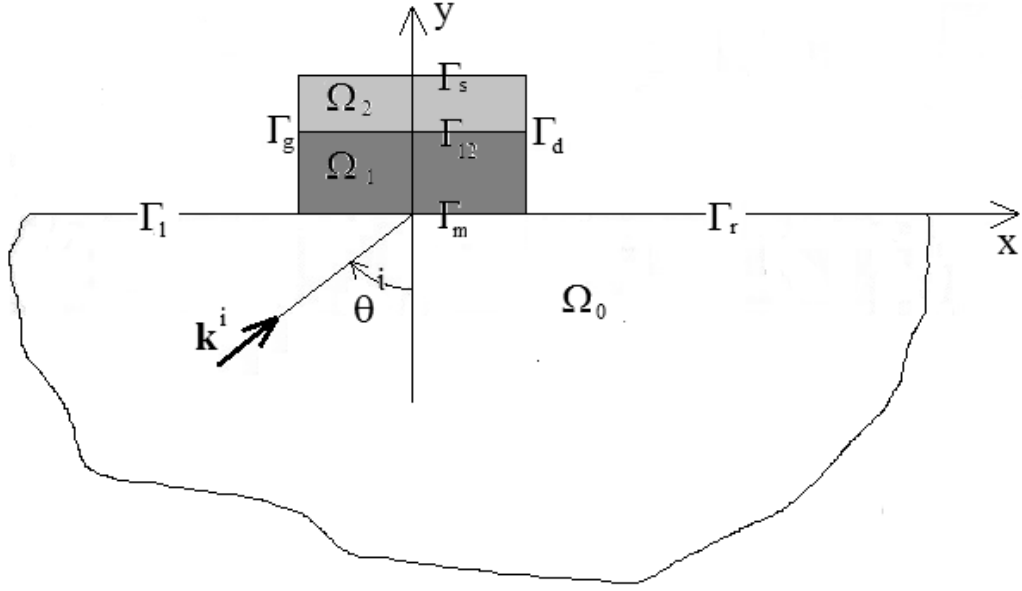}
\caption{Sagittal plane view of the 2D rectangular protuberance scattering configuration. Note that now the boundary $\Gamma_{p}$ of the above-ground feature is composed of three connected portions, $\Gamma_{g}$, $\Gamma_{s}$ and $\Gamma_{d}$.}
\label{hill}
\end{center}
\end{figure}

From now on, the option is to completely solve the forward scattering problem for the configuration depicted in
fig.\ref{hill}. The important feature thereof is the rectangular shape (in the sagittal plane) of the protuberance.
Note that we could just as well have chosen a feature of triangular shape as is done in e.g., \cite{kw92} and thus obtain essentially the same results (notably as concerns the resonant nature of the response) as in the present study.

The choice of protuberances with such simple shapes is dictated by the fact that key aspects of their seismic response can be unveiled in a relatively-simple manner, both from the theoretical and numerical angles, the latter  (numerical) feature being  very useful in a parametric study such as the one undertaken in the last part (i.e., starting with sect. \ref{Sills}) of the present investigation.

As previously, the width of the protuberance is $w$, and its other characteristic dimensions are the bottom ($h_{1}$) and top ($h_{2}$) layer thicknesses, with $h=h_{1}+h_{2}$ being the height of the protuberance. What was formerly $\Gamma_{p}$ is now $\Gamma_{g}\cup\Gamma_{s}\cup\Gamma_{d}$, wherein $\Gamma_{g}$ is the leftmost vertical segment of height $h$, $\Gamma_{s}$ is the top segment of width $w$ (located between $x=-w/2$ and $x=w/2$) and $\Gamma_{d}$ is the rightmost vertical segment of height $h$. Everything else is as in fig.\ref{protuberance}.
%%%%%%%%%%%%%%%%%%%%%%%%%%%%%%%%%%%%%%%%%%%%%%%%%%%%%%%
\subsection{Boundary-value problem}\label{bvp}
Owing to the fact that the configuration comprises three distinct regions, each in which the elastic parameters are constants as a function of the space variables, it is opportune to employ domain decomposition and separation of variables (DD-SOV). Thus, as previously, we decompose the total  field $u$ as:
\begin{equation}\label{5-005}
u(\mathbf{x})=u^{[l]}(\mathbf{x})~;~\forall\mathbf{x}\in\Omega_{l},~l=0,1,2~,
\end{equation}
with the understanding that these fields satisfy the 2D SH frequency domain elastic wave equation (i.e., Helmholtz equation),
$u^{[0]}$ the  radiation condition $u^{[0]}$, the stress-free boundary conditions on $\Gamma_{g}$, $\Gamma_{d}$, and $\Gamma_{p}$,  as well as the continuity conditions across  $\Gamma_{m}$ and $\Gamma_{12}$.
%%%%%%%%%%%%%%%%%%%%%%%%%%%%%%%%%%%%%%%%%%%%%%%%%%%%%%%%%%%%%%%%%%%%%%%%%%%%%%%%%%%%%%%%%%%%%%%%%%%
\subsection{Field representations via separation of variables (SOV)}\label{sov}
 As in the case of arbitrarily-shaped protuberances,  the SOV technique gives rise to the field representation \cite{wi20}
\begin{equation}\label{6-010}
u^{[0]}(\mathbf{x})=u^{i}(\mathbf{x})+u^{r}(\mathbf{x})+u^{s}(\mathbf{x})
\end{equation}
wherein $u^{r}(x,y)=u^{i}(x,-y)$ and
\begin{equation}\label{6-014}
u^{s}(\mathbf{x})=\int_{-\infty}^{\infty}\mathcal{B}(k_{x})f(k_{x},x)\exp(-ik_{y}y)\frac{dk_{x}}{k^{[0]}_{y}}~,
\end{equation}
with:
\begin{equation}\label{6-013}
f(k_{x},x)=\exp(ik_{x}x)~~,~~k_{y}^{[0]}=\sqrt{\big(k^{[0]}\big)^{2}-\big(k_{x}\big)^{2}}~~;~\Re k_{y}^{[0]}\ge 0~,~\Im k_{y}^{[0]}\ge 0~;~\text {for}~ \omega\ge 0~.
\end{equation}

Note that the scattered field $u^{s}$ satisfies the radiation condition and is expressed as a sum of plane waves, some of which are propagative (for real $k^{[0]}_{y}$) and the others evanescent (for imaginary $k^{[0]}_{y}$).

Within the  rectangular protuberance, the same SOV technique, together with the stress-free boundary conditions on $\Gamma_{p}$   give rise to the field representations:
\begin{equation}\label{6-030}
u^{[1]}(\mathbf{x})=\sum_{m=0}^{\infty}\left[a_{m}\exp\big(ik_{ym}^{[1]}y\big)+b_{m}\exp\big(-ik_{ym}^{[1]}y\big)\right]
f_{m}(x) ~,
\end{equation}
\begin{equation}\label{6-040}
u^{[2]}(\mathbf{x})=\sum_{m=0}^{\infty}d_{m}\cos\big[k_{ym}^{[2]}(y-h)\big]g_{m}(x)~,
\end{equation}
wherein:
\begin{equation}\label{6-045}
f_{m}(x)=\exp(ik_{xm}x)~~,~~g_{m}(x)=\cos[k_{xm}(x+w/2)]~,
\end{equation}
\begin{equation}\label{6-050}
k_{xm}=\frac{m\pi}{w}~~,~~k_{ym}^{[l]}=\sqrt{\big(k^{[l]}\big)^{2}-\big(k_{xm}\big)^{2}}~~,~~\Re k_{ym}^{[l]}\ge 0~~,~~\Im k_{ym}^{[l]}\ge 0~;~\omega\ge 0 ~,~l=1,2~.
\end{equation}
%
%%%%%%%%%%%%%%%%%%%%%%%%%%%%%%%%%%%%%%%%%%%%%%%%%%%%%%%%%%%%%%%%%%%%%%%%%%%%%%%%%%%%%%%%%%%%%%%%%%%%%%%%%%%%%%%
\section{Employment of the SOV field representations in the remaining boundary and continuity conditions}
As shown in \cite{wi20}, the stress-free boundary condition on the flanks of the protuberance , the  continuity condition across $\Gamma_{m}$, as well as the orthogonality conditions  of  $g_{m}(x)$ and $f(k_{x},x)$ give rise to the following two coupled sets of equations:
\begin{equation}\label{7-050}
d_{l}\frac{\kappa_{l}}{\epsilon_{l}}=2a^{i}I_{l}^{+}(k_{x}^{i})+
\int_{-\infty}^{\infty}\mathcal{B}(k_{x})I_{l}^{+}(k_{x})\frac{dk_{x}}{k_{y}^{[0]}}~;~l=0,1,2,....~,
\end{equation}
\begin{equation}\label{8-070}
\mathcal{B}(k_{x})=\frac{iw}{2\pi}\frac{\mu^{[1]}}{\mu^{[0]}}
\sum_{m=0}^{\infty}d_{m}^{(M)}\sigma_{m}k_{ym}^{[1]}I_{m}^{-}(k_{x})~;~\forall k_{x}\in\mathbb{R}~,
\end{equation}
as well as to the relations connecting $a_{l}$, $b_{l}$ to $d_{l}$:
\begin{equation}\label{8-073}
a_{l}=d_{l}\left(\frac{\exp(-ik_{yl}^{[1]}h_{1})}{2i\mu^{[1]}k_{yl}^{[1]}}\right)
\left[i\mu^{[1]}k_{yl}^{[1]}\cos(k_{yl}^{[2}h_{2})+\mu^{[2]}k_{yl}^{[2]}\sin(k_{yl}^{[2}h_{2})\right]~,
\end{equation}
\begin{equation}\label{8-075}
b_{l}=d_{l}\left(\frac{\exp(ik_{yl}^{[1]}h_{1})}{2i\mu^{[1]}k_{yl}^{[1]}}\right)
\left[i\mu^{[1]}k_{yl}^{[1]}\cos(k_{yl}^{[2}h_{2})-\mu^{[2]}k_{yl}^{[2]}\sin(k_{yl}^{[2}h_{2})\right]~,
\end{equation}
in which:
\begin{multline}\label{7-090}
I_{m}^{\pm}(k_{x})=\int_{-w/2}^{w/2}\exp(\pm ik_{x}x)\cos[k_{xm}(x+w/2)]\frac{dx}{w}=\\
\frac{i^{m}}{2}\text{sinc}\big[(\pm k_{x}+k_{xm})\frac{w}{2}\big]+\frac{(-i)^{m}}{2}\text{sinc}\big[(\pm k_{x}-k_{xm})\frac{w}{2}\big]
\end{multline}
(with sinc$(\zeta)=\sin(\zeta)/\zeta$ and sinc$(0)=1$)~,
\begin{equation}\label{8-040}
\kappa_{l}=\cos(k_{yl}^{[1]}h_{1})\cos(k_{yl}^{[2]}h_{2})-
\frac{\mu^{[2]}k_{yl}^{[2]}}{\mu^{[1]}k_{yl}^{[1]}}\sin(k_{yl}^{[1]}h_{1})\sin(k_{yl}^{[2]}h_{2})~,
\end{equation}
\begin{equation}\label{8-050}
\sigma_{l}=\sin(k_{yl}^{[1]}h_{1})\cos(k_{yl}^{[2]}h_{2})+
\frac{\mu^{[2]}k_{yl}^{[2]}}{\mu^{[1]}k_{yl}^{[1]}}\cos(k_{yl}^{[1]}h_{1})\sin(k_{yl}^{[2]}h_{2})~.
\end{equation}
Plugging (\ref{8-070}) into (\ref{8-050}) finally yields the single system of linear equations
\begin{equation}\label{10-030}
\sum_{m=0}^{\infty}E_{lm}^{(M)}d_{m}^{(M)}=c_{l}~;~l=0,1,2,...~,
\end{equation}
in which:
\begin{equation}\label{10-040}
E_{lm}^{(M)}=\delta_{lm}\frac{\kappa_{l}}{\epsilon_{l}}-
\frac{iw}{2\pi}\frac{\mu^{[1]}}{\mu^{[0]}}k_{ym}^{[1]}\sigma_{m}J_{lm}~~,~~c_{l}=2a^{i}I_{l}^{+}(k_{x}^{i})
~~,~~J_{lm}=\int_{-\infty}^{\infty}I_{l}^{+}(k_{x})I_{m}^{-}(k_{x})\frac{dk_{x}}{k_{y}^{[0]}}~,
\end{equation}
with  $\{d_{m}\}$ the to-be-determined  set of wavefield coefficients.

Until now everything has been rigorous provided the equations in the statement of the boundary-value problem are accepted as the true expression of what is involved in the seismic response of the  rectangular protuberance and certain summation interchanges (involved in the the obtention of the linear system) are valid. In order to actually solve for the sets $\{d_{m}\}$, and then for  $\{a_{m}\}$, $\{b_{m}\}$, $\{\mathcal{B}(k_{x})\}$ (each of whose populations is  considered to be infinite at this stage) we should usually resort  to numerics.  This is explained further on.

But before doing this we must evoke an encouraging result obtained in $\cite{wi20}$: the formal  solution to the scattering problem for  $\{d_{m}\}$, and then for  $\{a_{m}\}$, $\{b_{m}\}$, $\{\mathcal{B}(k_{x})\}$ (the term formal meaning implicit, because at this stage we have not solved explicitly for these quantities) turns out to be such as to wholly satisfy the conservation of flux law.
%%%%%%%%%%%%%%%%%%%%%%%%%%%%%%%%%%%%%%%%%%%%%%%%%%%%%%%%%%%%%%%%%%%%%%%%%%%%%%%%%%%%%%%%%%%%%%%%%%%%%%%%%%%%%%
\section{Some mathematical properties of the solution for $d_{m}$}
Before going into the details of numerics it is of considerable interest to evoke some mathematical properties  of the solution of the scattering problem.
%%%%%%%%%%%%%%%%%%%%%%%%%%%%%%%%
\subsection{General properties of $I^{\pm}_{lm}$ and their incidence on $J_{lm}$ and $E_{lm}$}
Eq. (\ref{7-090}) tells us that
\begin{equation}\label{4-120}
I_{2p+1}^{\pm}(k_{x})=i\frac{(-1)^{p}}{2}\left[\text{sinc}\big[(\pm k_{x}+k_{x2p+1})w/2\big]-
\text{sinc}\big[(\pm k_{x}-k_{x2p+1})w/2\big]\right]
~;~m=0,1,2,...~,
\end{equation}
from which we find
\begin{equation}\label{4-130}
I_{2p+1}^{\pm}(-k_{x})=-I_{2p+1}^{\pm}(k_{x})
~;~m=0,1,2,...~.
\end{equation}
Eq. (\ref{7-090}) also tells us that
\begin{equation}\label{4-140}
I_{2p}^{\pm}(k_{x})=\frac{(-1)^{p}}{2}\left[\text{sinc}\big[(\pm k_{x}+k_{x2p})w/2\big]+
\text{sinc}\big[(\pm k_{x}-k_{x2p})w/2\big]\right]
~;~m=0,1,2,...~,
\end{equation}
from which we find
\begin{equation}\label{4-150}
I_{2p}^{\pm}(-k_{x})=I_{2p}^{\pm}(k_{x})
~;~m=0,1,2,...~.
\end{equation}
It follows that
\begin{equation}\label{4-160}
J_{2p,2q+1}=J_{2p+1,2q}=0
~;~p,q=0,1,2,...~,
\end{equation}
whence
\begin{equation}\label{4-170}
E_{2p,2q+1}=E_{2p+1,2q}=0
~;~p,q=0,1,2,...~.
\end{equation}
%
%%%%%%%%%%%%%%%%%%%%%%%%%%%%%%%%
\subsection{Linear systems of equations for even  and odd orders of $\mathcal{F}_{m}$}
We had:
\begin{equation}\label{4-180}
\begin{array}{c}
\sum_{q=0}^{\infty}[E_{2p,2q}d_{2q}+E_{2p,2q+1}d_{2q+1}]=c_{2p}\\
\sum_{q=0}^{\infty}[E_{2p+1,2q}d_{2q}+E_{2p+1,2q+1}d_{2q+1}]=c_{2p+1}
\end{array}
~;~p=0,1,2,...~,
\end{equation}
so that
\begin{equation}\label{4-190}
\begin{array}{c}
\sum_{q=0}^{\infty}E_{2p,2q}d_{2q}=c_{2p}\\
\sum_{q=0}^{\infty}E_{2p+1,2q+1}d_{2q+1}=c_{2p+1}
\end{array}
~;~p=0,1,2,...~,
\end{equation}
which shows that the equations for the odd order $d_{m}$ are decoupled from those for  the even order $d_{m}$.
%%%%%%%%%%%%%%%%%%%%%%%%%%%%%%%%
\subsection{Normal incidence}
Normal incidence means that $\theta^{i}=0^{\circ}$.
%%%%%%%%%%%%%%%%%%%%%%%%%%%%%%%%%%%%%%%%%%%%%%%%%%%%%%%
\subsubsection{The odd-order diffraction coefficients}
Eq. (\ref{4-120}) tells us that, for $k_{x}=k^{i}_{x}$, and normal incidence, $k^{i}_{x}=0$,
\begin{equation}\label{4-200}
I_{2p+1}^{\pm}(k_{x}^{i})=I_{2p+1}^{\pm}(0)=0
~;~p=0,1,2,...~.
\end{equation}
whereas (\ref{4-140}) indicates
\begin{equation}\label{4-210}
I_{2p}^{\pm}(k_{x}^{i})=I_{2p}^{\pm}(0)=(-1)^{p}\text{sinc}(p\pi)=\delta_{p0}
~;~p=0,1,2,...~.
\end{equation}
It follows that:
\begin{equation}\label{4-220}
c_{2p}=2a^{i}\delta_{p0}~~,~~c_{2p+1}=0~;~p=0,1,2,...~,
\end{equation}
whence the   linear system of equations for the odd $d_{m}$
becomes
\begin{equation}\label{4-230}
\sum_{q=0}^{\infty}E_{2p+1,2q+1}d_{2q+1}=0
~;~p=0,1,2,...~,
\end{equation}
whose solution is necessarily
\begin{equation}\label{4-190}
d_{2q+1}=0
~;~q=0,1,2,...~,
\end{equation}
which means that, for normal incidence plane-wave solicitation, the odd-order diffraction coefficients are nil.
%%%%%%%%%%%%%%%%%%%%%%%%%%%%%%%%%%%%%%%%%%%%%%%%
\subsection{Iterative approach for the obtention of the even-order diffraction coefficients}
We can write $E_{lm}$ as $E_{lm}=\delta_{lm}\chi_{l}-F_{lm}$ so that the first equation of (\ref{4-190} becomes
\begin{equation}\label{4-200}
d_{2p}\chi_{2p}=c_{2p}+\sum_{q=0}^{\infty}F_{2p,2q}d_{2q}
~;~p=0,1,2,...~,
\end{equation}
which suggests solving for $d_{2p}$ by the iterative scheme:
\begin{equation}\label{4-210}
d_{2p}^{(0)}=\frac{c_{2p}}{\chi_{2p}}~~,~~d_{2p}^{(j)}=d_{2p}^{(0)}+\sum_{q=0}^{\infty}\frac{F_{2p,2q}}{\chi_{2p}}d_{2q}^{(j-1)}
~;~p=0,1,2,...~,
\end{equation}
%
%%%%%%%%%%%%%%%%%%%%%%%%%%%%%%%%%%%%%%%%%%%%%%%%%%%%%%%%%%%%%%%%%%%%%%%%%%%%%%%%%%%%%%%%%%%%%%%%%%%%%%
\subsubsection{Closed-form solution for $d_{2p}$ by the iterative scheme in a very special (VS) case}\label{VS}
We first note that $c_{2p}=2a^{i}I_{2p}^{+}(0)=2a^{i}\delta_{p0}$ so that $d_{2p}^{(0)}=\frac{2a^{i}\delta_{p0}}{\chi_{0}}$,
\begin{equation}\label{4-220}
d_{2p}^{(1)}=d_{2p}^{(0)}+\frac{F_{2p,0}}{\chi_{2p}}\frac{2a^{i}}{\chi_{0}}
~;~p=0,1,2,...~,
\end{equation}
and so on. Before going to higher-order iterates, first consider the first one, in which
\begin{equation}\label{4-230}
F_{2p,0}=\frac{iw}{2\pi}\frac{\mu^{[1]}}{\mu^{[0]}}k_{y0}^{[1]}\sigma_{0}J_{2p,0}
~.
\end{equation}
Secondly, we note that $\chi_{0}=\kappa_{0}$,  and by means of (\ref{8-040})-(\ref{8-050})
\begin{equation}\label{4-240}
\kappa_{0}=\cos(k^{[1]}h_{1}+k^{[2]}h_{2})+\left(\frac{\mu^{[2]}k^{[2]}}{\mu^{[1]}k^{[1]}}-1\right)
\sin(k^{[1]}h_{1})\sin(k^{[2]}h_{2})
~,
\end{equation}
\begin{equation}\label{4-250}
\sigma_{0}=\sin(k^{[1]}h_{1}+k^{[2]}h_{2})+\left(\frac{\mu^{[2]}k^{[2]}}{\mu^{[1]}k^{[1]}}-1\right)
\cos(k^{[1]}h_{1})\sin(k^{[2]}h_{2})
~,
\end{equation}
from which we find
\begin{equation}\label{4-260}
\sigma_{0}=0~~,~~\kappa_{0}=\cos N\pi=(-1)^{N}
~,
\end{equation}
in the very special (VS) case (recall that we are stll assuming that $\theta^{i}=0^{\circ}$) in which:
\begin{equation}\label{4-270}
k^{[1]}h_{1}+k^{[2]}h_{2}=N\pi~~,~~\mu^{[2]}k^{[2]}=\mu^{[1]}k^{[1]}~;~N=0,1,2,....
~.
\end{equation}
It follows (in this VS case) that $F_{2p,0}=0$ so that
\begin{equation}\label{4-280}
d_{2p}^{(1)}=d_{2p}^{(0)}
~;~p=0,1,2,...~,
\end{equation}
and, in the same manner, we find that the higher-order iterates are equal to the zeroth-order iterate, so that, in the very special case (recall that $\theta^{i}=0^{\circ}$ was also assumed) defined by (\ref{4-270}), the closed-form solution for the (even-order) diffraction coefficients is:
\begin{equation}\label{4-290}
d_{2p}^{(0)}=a^{i}(-1)^{N}\delta_{p0}~;~p=0,1,2,...., N=0,1,2,...~,
\end{equation}
it being recalled (always in the case $\theta^{i}=0^{\circ}$) that $d_{2p+1}=0~;~p=0,1,2,...$.
%%%%%%%%%%%%%%%%%%%%%%%%%%%%%%%%%%%%%%%%%%%%%%%%
\subsubsection{VS case: expressions of the fields and demonstration that the continuity conditions are satisfied}\label{VSC}
Inserting (\ref{4-290}) into (\ref{6-040}) gives
\begin{equation}\label{4-300}
u^{[2]}(\mathbf{x})=2a^{i}(-1)^{N}\cos\big(k^{[2]}(y-h)\big)~;~N=0,1,2,...~,
\end{equation}
whence
\begin{equation}\label{4-310}
u^{[2]}(x,h_{1})=2a^{i}(-1)^{N}\cos\big(k^{[2]}h_{2}\big)~~,
~~u_{,y}^{[2]}(x,h_{1})=2a^{i}(-1)^{N}k^{[2]}\sin\big(k^{[2]}h_{2}\big)~;~N=0,1,2,...~.
\end{equation}
Eqs. (\ref{6-030}) and (\ref{8-040})-(\ref{8-050}) lead to
\begin{equation}\label{4-320}
u^{[1]}(\mathbf{x})=\sum_{m=0}^{\infty}d_{m}\left[\kappa_{m}\cos\big(k_{ym}^{[1]}y\big)+\sigma_{m}\sin\big(k_{ym}^{[1]}y\big)\right]
\exp(ik_{xm}x)~,
\end{equation}
which, by means of (\ref{4-290}) (and the fact that the odd-order $d_{m}$ are nil), gives rise to
\begin{equation}\label{4-330}
u^{[1]}(\mathbf{x})=2a^{i}\cos(k^{[1]}y)~;~N=0,1,2,...~,
\end{equation}
whence
\begin{equation}\label{4-335}
u^{[1]}_{,y}(\mathbf{x})=-2a^{i}k^{[1]}\sin(k^{[1]}y)~;~N=0,1,2,...~,
\end{equation}
\begin{equation}\label{4-340}
u^{[1]}(x,h_{1})=2a^{i}(-1)^{N}\cos\big(k^{[2]}h_{2}\big)~~,
~~u_{,y}^{[1]}(x,h_{1})=2a^{i}(-1)^{N}k^{[1]}\sin\big(k^{[2]}h_{2}\big)~;~N=0,1,2,...~.
\end{equation}
Consequently
\begin{equation}\label{4-350}
u^{[1]}(x,h_{1})-u^{[2]}(x,h_{1})=0~;~\forall N=0,1,2,...~,
\end{equation}
\begin{equation}\label{4-360}
\mu^{[1]}u_{,y}^{[1]}(x,h_{1})-\mu^{[2]}u_{,y}^{[2]}(x,h_{1})=
2a^{i}(-1)^{N}\big[\mu^{[1]}k^{[1]}-\mu^{[2]}k^{[2]}\big]\sin\big(k^{[2]}h_{2}\big)~,
\end{equation}
of, on account of the second VS condition
\begin{equation}\label{4-370}
\mu^{[1]}u_{,y}^{[1]}(x,h_{1})-\mu^{[2]}u_{,y}^{[2]}(x,h_{1})=0~;~\forall N=0,1,2,...~,
\end{equation}
which shows that the  continuity conditions on the segment of separation at $y=h_{1}$ are satisfied by the VS case solution for the diffraction coefficients.

Introducing the fact that the odd-order $d _{m}$ are nil into (\ref{8-070}) gives
\begin{equation}\label{4-380}
\mathcal{B}(k_{x})=\frac{iw}{2\pi}\frac{\mu^{[1]}}{\mu^{[0]}}
\sum_{p=0}^{\infty}d_{2p}\sigma_{2p}k_{y2p}^{[1]}I_{2p}^{-}(K_{x})~;~\forall k_{x}\in\mathbb{R}~,
\end{equation}
so that, due to (\ref{4-290})
\begin{equation}\label{4-390}
\mathcal{B}(k_{x})=a^{i}(-1)^{N}\frac{iw}{\pi}\frac{\mu^{[1]}}{\mu^{[0]}}
\sigma_{0}k_{y0}^{[1]}I_{0}^{-}(K_{x})~;~\forall k_{x}\in\mathbb{R}~,~N=0,1,2,...~,
\end{equation}
or, on account of (\ref{4-260}),
\begin{equation}\label{4-400}
\mathcal{B}(k_{x})=0~.
\end{equation}
Consequently,
\begin{equation}\label{4-410}
u^{d}(\mathbf{x})=0~\Rightarrow~u^{[0]}(\mathbf{x})=u^{i}(x,y)+u^{i}(x,-y)=2a^{i}\cos(k^{[0]}_y)~,
\end{equation}
so that
\begin{equation}\label{4-420}
u^{[0]}_{,y}(\mathbf{x})=-2a^{i}k^{[0]}\sin(k^{[0]}_y)~.
\end{equation}
From (\ref{4-410}) and (\ref{4-330})we find $u^{[0]}(x,0)=2a^{i}$ and $u^{[1]}(x,0)=2a^{i}$ respectively, so that the  continuity condition across the base segment of the convex rectangular feature
\begin{equation}\label{4-430}
u^{[0]}(x,0)-u^{[1]}(x,0)=0~,
\end{equation}
is satisfied. Moreover,  (\ref{4-420}) and (\ref{4-335}) indicate that $u^{[1]}_{,y}(x,0)=0$ and $u^{[0]}_{,y}(x,0)=0$ respectively, so that the  continuity condition across the base segment of the convex rectangular feature
\begin{equation}\label{4-420}
\mu^{[0]}u^{[0]}_{,y}(x,0)-\mu^{[1]}u^{[1]}_{,y}(x,0)=0~,
\end{equation}
is satisfied, whatever be the relation of $\mu^{[0]}$ to $\mu^{[1]}$.

Thus, in the VS case, i.e., when  the parameters of the scattering configuration obey $\theta^{i}=0^{\circ}$ and (\ref{4-270}), the field admits the closed-form expressions: (\ref{4-300}) for $u^{[2]}(\mathbf{x})$, (\ref{4-330}) for $u^{[1]}(\mathbf{x})$, and (\ref{4-410}) for $u^{[0]}(\mathbf{x})$, which are such as to satisfy the four continuity conditions across the various interfaces and the stress-free boundary conditions along the the stress-free boundary of the scatterer.

A last remark: the VS case provides a useful testing ground for the numerical methods and solutions described and offered further on in sect. \ref{num}. This is demonstrated, e.g. in fig. \ref{fm15}.
%%%%%%%%%%%%%%%%%%%%%%%%%%%%%%%%%%%%%%%%%%%%%%%%%%%%%%%%%%%%%%%%%%%%%%%%%%%%%%%%%
\subsection{Coping with $J_{lm}$}
Anticipating the issue of numerics we can expect that the principal problem will be how to cope with $J_{lm}$. If, for the moment, we exclude resorting to a purely-numerical double-integral quadrature scheme, we must dispose of another  more-mathematical method. The one we devised (see also \cite{wi73}) for a brief description) enables the reduction of the evaluation of the double integral to that of  at most  three single-integrals.

Inserting the expressions (\ref{7-090}) for $I_{lm}^{\pm}$ into (\ref{10-040}) gives rise, after interchanges of the orders of integration:
\begin{equation}\label{5-020}
J_{lm}=\frac{\pi}{w^{2}}\int_{-w/2}^{w/2}dx'\cos[k_{xl}(x'+w/2)]\int_{-w/2}^{w/2}dx\cos[k_{xm}(x+w/2)]
\int_{-\infty}^{\infty}\frac{dk_{x}}{\pi k_{y}}\exp^[ik_{x}(x'-x)]~,
\end{equation}
or, using the fact \cite{mf53} that
\begin{equation}\label{5-030}
\int_{-\infty}^{\infty}\frac{dk_{x}}{\pi k_{y}}\exp^[ik_{x}(x'-x)]=H_{0}^{(1)}(k^{(0)}|x'-x|)~,
\end{equation}
wherein $H_{j}^{(1)}(~)$ is the $j$th-order Hankel function of the first kind, we find
\begin{equation}\label{5-040}
J_{lm}=\frac{\pi}{w^{2}}\int_{-w/2}^{w/2}dx'\cos[k_{xl}(x'+w/2)]\int_{-w/2}^{w/2}dx\cos[k_{xm}(x+w/2)]H_{0}^{(1)}(k^{[0]}|x'-x|)
~,
\end{equation}
which becomes, with the changes of variables $\eta'=k^{[0]}(x'+w/2)$, $\eta=k^{[0]}(x+w/2)$, $\alpha_{l}=k_{xl}/k^{[0]}$
\begin{equation}\label{5-050}
J_{lm}=\frac{\pi}{(k^{[0]}w)^{2}}\int_{0}^{k^{[0]}w}d\eta'\cos(\alpha_{l}\eta')\int_{0}^{k^{[0]}w}d\eta\cos(\alpha_{m}\eta)
H_{0}^{(1)}(|\eta'-\eta|)
~.
\end{equation}
From the fact that
\begin{equation}\label{5-060}
\cos(\alpha_{l}\eta')\cos(\alpha_{m}\eta)=
\frac{1}{2}\left\{\cos[\alpha_{-l}(\eta'-\eta)+(\alpha_{m}+\alpha_{-l})\eta]+
\cos[\alpha_{l}(\eta'-\eta)+(\alpha_{m}+\alpha_{l})\eta]\right\}
~.
\end{equation}
we obtain
\begin{equation}\label{5-070}
\frac{(k^{[0]}w)^{2}}{\pi}J_{lm}=\mathcal{J}_{-lm}+\mathcal{J}_{lm}
~,
\end{equation}
wherein
\begin{equation}\label{5-080}
\mathcal{J}_{lm}=\frac{1}{2}\int_{0}^{k^{[0]}w}d\eta'\int_{0}^{k^{[0]}w}d\eta H_{0}^{(1)}(|\eta'-\eta|)\cos[\alpha_{l}(\eta'-\eta)+(\alpha_{m}+\alpha_{l})\eta]
~.
\end{equation}
We make the change of variables $\mu=\eta$ and $\nu=\eta'-\eta$ so as to convert the original rectangular $\eta'-\eta$ integration domain into a parallelogram $\mu-\nu$ domain which is decomposable into a triangular domain for negative $\nu$  and another triangular domain for positive $\nu$, both of which correspond to positive $\mu$. Consequently
\begin{equation}\label{5-090}
\mathcal{J}_{lm}=\int_{0}^{\kappa}d\nu\int_{0}^{\kappa-\nu}d\mu \mathcal{P}(\mu,\nu)+
\int_{-\kappa}^{0}d\nu\int_{-\nu}^{\kappa}d\mu \mathcal{P}(\mu,\nu)=\mathcal{J}^{+}_{lm}+\mathcal{J}^{-}_{lm}
~,
\end{equation}
wherein:
\begin{equation}\label{5-100}
\mathcal{P}(\mu,\nu)=\frac{1}{2}H_{0}^{(1)}(|\nu|)\cos[\alpha_{l}\nu+(\alpha_{m}+\alpha_{l})\mu]~~,~~\kappa=k^{[0]}w
~,
\end{equation}
\begin{equation}\label{5-110}
\mathcal{J}^{+}_{lm}=\int_{0}^{\kappa}d\nu\int_{0}^{\kappa-\nu}d\mu \mathcal{P}(\mu,\nu)~~,~~
\mathcal{J}^{-}_{lm}=\int_{-\kappa}^{0}d\nu\int_{-\nu}^{\kappa}d\mu \mathcal{P}(\mu,\nu)
~.
\end{equation}
We easily find
\begin{multline}\label{5-120}
\mathcal{J}^{+}_{lm}=
\frac{\sin[(\alpha_{l}+\alpha_{m})\kappa]}{2(\alpha_{m}+\alpha_{l})}\int_{0}^{\kappa}d\nu H_{0}^{(1)}(|\nu|)\cos(\alpha_{m}v)-\\
\frac{\cos[(\alpha_{l}+\alpha_{m})\kappa]}{2(\alpha_{m}+\alpha_{l})}\int_{0}^{\kappa}d\nu H_{0}^{(1)}(|\nu|)\sin(\alpha_{m}\nu)-
\frac{1}{2(\alpha_{m}+\alpha_{l})}\int_{0}^{\kappa}d\nu H_{0}^{(1)}(|\nu|)\sin(\alpha_{l}\nu)
~.
\end{multline}
\begin{multline}\label{5-125}
\mathcal{J}^{-}_{lm}=
-\frac{\cos[(\alpha_{l}+\alpha_{m})\kappa]}{2(\alpha_{m}+\alpha_{l})}\int_{0}^{\kappa}d\nu H_{0}^{(1)}(|\nu|)\sin(\alpha_{l}v)+\\
\frac{\sin[(\alpha_{l}+\alpha_{m})\kappa]}{2(\alpha_{m}+\alpha_{l})}\int_{0}^{\kappa}d\nu H_{0}^{(1)}(|\nu|)\cos(\alpha_{l}\nu)-
\frac{1}{2(\alpha_{m}+\alpha_{l})}\int_{0}^{\kappa}d\nu H_{0}^{(1)}(|\nu|)\sin(\alpha_{m}\nu)
~,
\end{multline}
whence
\begin{multline}\label{5-130}
\mathcal{J}_{lm}=
\frac{\kappa}{2}\text {sinc}[(\alpha_{l}+\alpha_{m})\kappa]\int_{0}^{\kappa}d\nu H_{0}^{(1)}(|\nu|)[\cos(\alpha_{m}\nu)+\cos(\alpha_{l}\nu)]-\\
\frac{[1+\cos[(\alpha_{l}+\alpha_{m})\kappa]}{2(\alpha_{m}+\alpha_{l})}
\int_{0}^{\kappa}d\nu H_{0}^{(1)}(|\nu|)[\sin(\alpha_{m}\nu)+\sin(\alpha_{m}\nu)]
~,
\end{multline}
or
\begin{multline}\label{5-135}
\mathcal{J}_{lm}=
\frac{\kappa}{2}\delta_{l,-m}\int_{0}^{\kappa}d\nu H_{0}^{(1)}(|\nu|)[\cos(\alpha_{m}\nu)+\cos(\alpha_{l}\nu)]-\\
\frac{[1+\cos[(\alpha_{l}+\alpha_{m})\kappa]}{2}
\int_{0}^{\kappa}d\nu H_{0}^{(1)}(|\nu|)\nu~
\text {sinc}[(\alpha_{m}+\alpha_{l})\nu/2)]\cos[(\alpha_{m}-\alpha_{l})\nu/2)]
~.
\end{multline}
In the same way we find
\begin{multline}\label{5-140}
\mathcal{J}_{-l,m}=\mathcal{J}_{l,-m}=
\frac{\kappa}{2}\delta_{l,m}\int_{0}^{\kappa}d\nu H_{0}^{(1)}(|\nu|)[\cos(\alpha_{m}\nu)+\cos(\alpha_{l}\nu)]-\\
\frac{[1+\cos[(\alpha_{l}-\alpha_{m})\kappa]}{2}
\int_{0}^{\kappa}d\nu H_{0}^{(1)}(|\nu|)\nu~
\text {sinc}[(\alpha_{m}-\alpha_{l})\nu/2)]\cos[(\alpha_{m}+\alpha_{l})\nu/2)]
~,
\end{multline}
so that
\begin{multline}\label{5-145}
\frac{\kappa^{2}}{\pi}J_{lm}=\mathcal{J}_{l,m}+\mathcal{J}_{-l,m}+
\frac{\kappa}{2}\left(\delta_{l,-m}+\delta_{l,-m}\right)\int_{0}^{\kappa}d\nu H_{0}^{(1)}(|\nu|)[\cos(\alpha_{m}\nu)+\cos(\alpha_{l}\nu)]-\\
\frac{[1+\cos[(\alpha_{l}+\alpha_{m})\kappa]}{2}
\int_{0}^{\kappa}d\nu H_{0}^{(1)}(|\nu|)\nu~
\text {sinc}[(\alpha_{m}+\alpha_{l})\nu/2)]\cos[(\alpha_{m}-\alpha_{l})\nu/2)]+\\
\frac{[1+\cos[(\alpha_{l}-\alpha_{m})\kappa]}{2}
\int_{0}^{\kappa}d\nu H_{0}^{(1)}(|\nu|)\nu~
\text {sinc}[(\alpha_{m}-\alpha_{l})\nu/2)]\cos[(\alpha_{m}+\alpha_{l})\nu/2)]
~,
\end{multline}
from which we deduce, recalling that $l,m=0,1,2,....$:
\begin{equation}\label{5-150}
\frac{\kappa^{2}}{\pi}J_{l,m\ne\pm l}=\left(\frac{1+(-1)^{l+m}}{2}\right)
\left[\frac{\alpha_{m}\mathcal{S}_{m}(\kappa)-\alpha_{l}\mathcal{S}_{l}(\kappa)}{(\alpha_{l})^{2}-(\alpha_{m})^{2}}\right]
~,
\end{equation}
\begin{equation}\label{5-160}
\frac{\kappa^{2}}{\pi}J_{l,l\ne0}=\kappa \mathcal{C}_{l}(\kappa)-\frac{1}{\alpha_{l}}\mathcal{S}_{l}(\kappa)-\mathcal{A}_{l}(\kappa)
~,
\end{equation}
\begin{equation}\label{5-170}
\frac{\kappa^{2}}{\pi}J_{0,0}=2\kappa \mathcal{C}_{0}(\kappa)-2\mathcal{A}_{0}(\kappa)
~,
\end{equation}
with:
\begin{multline}\label{5-140}
\mathcal{S}_{l}(\kappa)=\int_{0}^{\kappa}H_{0}^{(1)}(\nu)\sin(\alpha_{l}\nu)d\nu~,\\
\mathcal{C}_{l}(\kappa)=\int_{0}^{\kappa}H_{0}^{(1)}(\nu)\cos(\alpha_{l}\nu)d\nu~,\\
\mathcal{A}_{l}(\kappa)=\int_{0}^{\kappa}H_{0}^{(1)}(\nu)\nu~\cos(\alpha_{l}\nu)d\nu~.
\end{multline}
Thus, we have reduced the evaluation of the double-integrals in the primitive form of $J_{lm}$, for each $l,m$,  to that of three single-integrals (i.e., $\mathcal{S}_{l}$, $\mathcal{C}_{l}$ and $\mathcal{A}_{l}$).
%%%%%%%%%%%%%%%%%%%%%%%%%%%%%%%%%%%%%%%%%%%%%%%%%%%%%%%%%%%%%%%%%%%%%%%%%%%%%%%%%%%%%%%%%%%%%%%
\section{On the origin of building, hill and mountain resonances}\label{reson}
 The  issue that must be discussed, in theoretical terms before going into the numerics, is that of resonances, which, as stated in the Introduction, is at the  core of our contribution.
%%%%%%%%%%%%%%%%%%%%%%%%%%%%%%%%%%%%%%%%%%%%%%%%%%%%%%%%%%%%%%%%%%
\subsection{Introductory remarks}
We saw that the problem of the response of a convex, rectangular-shaped feature emerging from flat ground leads to a matrix equation (see (\ref{10-030}))
\begin{equation}\label{6-000}
\mathbf{E}\mathbf{d}=\mathbf{c}~,
\end{equation}
wherein the elements of the square, infinite-order  matrix $\mathbf{E}$ are the $E_{lm}$, and the elements of the infinite-order vectors $\mathbf{d}$ and $\mathbf{c}$ are $d_{m}$ and $c_{l}$ respectively. An important feature of this equation is that the amplitude $a^{i}$ and incident angle $\theta^{i}$ of the plane-wave solicitation do not appear in $\mathbf{E}$ but only in $\mathbf{c}$.

The formal solution of the matrix equation is
\begin{equation}\label{6-010}
\mathbf{d}=\mathbf{E}^{-1}\mathbf{c}~,
\end{equation}
wherein $\mathbf{E}^{-1}$ designates the (formal) inverse of $\mathbf{E}$.

Lest it be forgotten, the principal motivation of this study is to find out, as is suggested by empirical evidence and other, former theoretical and numerical studies, why the response at locations within and on a building or mountain (this list could just as well include a building) to a seismic wave can be larger, or even much larger, than the so-called free-field response (i.e., the response on, and below, the ground in the absence of the convex free-surface feature). Our working hypothesis is that such amplifications are due to {\it resonances}.

Loosely-speaking, a resonance designates the moment at which all the parameters of the scattering configuration except $a^{i}$ and $\theta^{i}$ are such that one or several terms in $\mathbf{d}$ are large, the consequence of which is (due to the fact that $\mathbf{d}$ is what largely conditions the amplitude of the displacement field in the convex rectangular feature) that at resonance the displacement field might become large in certain subregions within or below the rectangular region occupied by the convex  feature.

Note that in the seismic engineering community, one often speaks of resonances (typically of a site with flat stress-free boundary underlain by one or several media arranged as layers) and even of a (convex) topographic feature (in \cite{bb96} the authors speak of a "transverse oscillatory resonance mode of a hill from 3 to $5~Hz$"), but the physical origin of these convex topographic resonances are not really explained. They are rather named 'resonances' because they occur (both empirically and in numerical studies) in rather narrow ranges of frequencies and are characterized by response peaks, often qualified as 'amplifications', particularly at the top of the convex surface feature. Some authors \cite{mb14} also underline (but do not explain) the fact that the ('resonance') frequencies of occurrence of these amplifications do not seem to vary with the solicitation (notably the distance and azimuth of the source), but the level of amplifications depend on the shape (notably aspect ratio) and composition of the convex surface feature.
%%%%%%%%%%%%%%%%%%%%%%%%%%%%%%%%%%%%%%%%%%%%%%%%%%%%%%%%%%%%%%%%%%
\subsection{A first definition of (surface shape) resonances}
Let $D$=det $\mathbf{E}$ denote the formal, complex determinant of $\mathbf{E}$. At (or near) resonance, $D$ is equal (or nearly-) equal to zero since this is what makes $\mathbf{E}^{-1}$, and thus one or several elements of $\mathbf{d}$, large. Since there exist attenuation mechanisms in the scattering problem at hand, such as radiation damping (energy that escapes to the outer confines of the bottom half-space \cite{ws96}) and material damping related to the lossy nature of the material(s) within the convex feature, the real and imaginary parts of  $D$ will not vanish entirely and simultaneously (at real frequencies). Thus, we modify our definition of resonance as the moment when $\|D\|$ is very small or $1/\|D\|$ is very large (large meaning compared to the off-resonance situations). Note that the resonances are easily spotted when the attenuations are small, but possibly hard to spot when the attenuations are large (this will be demonstrated further on in the computed transfer functions, and will, in fact constitute the method by which we shall determine the resonance frequencies). Again, note that our definition of resonance is independent of $a^{i}$ and $\theta^{i}$, whatever be the degree of attenuation.

We shall define the term 'coupling to a resonance' as the moment (for fixed $h_{1},~h_{2},~w,$ $\mu^{[0]},~\mu^{[1]},~\mu^{[2]},~\beta^{[0]},~\beta^{[1]},~\beta^{[2]},~a^{i},~\theta^{i}$) at which the frequency $f$ of the seismic  solicitation equals a resonant frequency (whose meaning will emerge in what follows).

As we shall see in the numerical results, coupling to a resonance results in a large value of at least one term in the SOV representations of the field within the protuberance, which fact should not be interpreted as the amplification of the field {\it at all locations within the protuberance}. The configuration of the  field, within and outside  the protuberance, at a resonance frequency is termed the  mode  at this frequency.
%%%%%%%%%%%%%%%%%%%%%%%%%%%%%%%%%%%%%%%%%%%%%%%%%%%%%%%%%%%%%%%
\subsection{A second definition of (surface shape) resonances}
 Consider the determinant $D(\omega)$ (which we have written as a function of $\omega=2\pi f$ instead of $f$), with $\omega$ now thought of as being a complex variable. In the complex $\omega$ plane, a zero of $D(\omega)$ occurs at that (generally-complex) value $\omega^{R}$ of $\omega$  for which
\begin{equation}\label{6-012}
D(\omega)=0~.
\end{equation}
 To this zero corresponds a pole, synonymous with infinite $1/D$ and infinite $\mathbf{d}$.  As we know that the zero generally does not occur for real $\omega^{R}$ we write $\omega^{R}=\omega^{R'}-i\omega^{R''}$ wherein $\omega^{R'}\ge 0$ and $\omega^{R''}\ge 0$. The 'physical' angular frequency is $\omega^{'}$ so that we can expect $\|1/D(\omega^{'})\|$ to be largest when $\omega^{'}\simeq \omega^{R'}$, which fact can be seen as follows. By means of a Taylor series expansion in the neighborhood of $\omega=\omega^{R}$ we find
\begin{equation}\label{6-014}
\|D(\omega)\|^{-2}\simeq\|D_{,\omega}\big(\omega^{R}\big)\|^{-2}\|\omega-\omega^{R}\|^{-2}=
\|D_{,\omega}\big(\omega^{R}\big)\|^{-2}\Big[\|\omega\|^{2}+\|\omega\|^{2}-2\Re(\omega^{R}\omega^{*}\big]^{-1}~,
\end{equation}
so that
\begin{equation}\label{6-016}
\|D(\omega^{'})\|^{-2}\simeq
\|D_{,\omega}\big(\omega^{R}\big)\|^{-2}\Big[\big(\omega^{'}-\omega^{R'}\big)^{2}+\big(\omega^{R''}\big)^{2}\big]^{-1}~.
\end{equation}
This function (as well as $\|D(\omega^{'})\|^{-1}$ which we shall depict in many of our numerical results hereafter), has the shape of a lorentzian,  indicative of $resonant~response$, whose maximum is situated at $\omega^{'}=\omega^{R'}$, which is called the 'resonance frequency', and whose width at half height is $2\omega^{R''}$. Note that more than one resonance can occur in the function $\|D(\omega^{'})\|^{-1}$ in which case what was offered until now remains true for each such resonance as long as the various resonance frequencies are well-separated in the complex $\omega$ plane. This constitutes our second definition of surface shape resonances.

This definition of (surface shape) resonances applies equally-well to the seismic response of a below-ground structure (BGS), and, in fact, to the vibratory or wave-like (electromagnetic, acoustic, elastic, hydrodynamic, etc.) response of any (including inhomogeneous, such as multilayered) object \cite{ej57,wi73,pe80,ma82,dg82,ma88,mr88,mv85,wi88,wi90b,wi95,lw84,lm98,zz08}.
%%%%%%%%%%%%%%%%%%%%%%%%%%%%%%%%%%%%%%%%%%%%%%%%%%%%%%%%%
\subsection{An alternate system of equations}\label{num}
As it stands,  the matrix equation $\mathbf{E}\mathbf{d}=\mathbf{c}$ is not particularly-appropriate for the determination of the diffraction coefficients $\mathbf{d}$, particularly as concerns the  issue of the dependence of the resonances on  $h_{1},~h_{2}$ on the one hand, and on $w$ on the other hand. The reason for this is that the matrix $\mathbf{E}$ is the sum of two matrices both of which involve all the configurational parameters $h_{1},~h_{2},~w$.

The way to resolve this problem is actually quite simple: in (\ref{10-030}), divide $E_{lm}$ by $\sigma_{m}$ and multiply $d_{m}$ by $\sigma_{m}$ so as to obtain
\begin{equation}\label{5-010}
\sum_{m=0}^{\infty}\mathcal{E}_{lm}\mathcal{F}_{m}=\mathcal{G}_{l}~;~l=0,1,2,...~,
\end{equation}
in which:
\begin{equation}\label{5-015}
\mathcal{E}_{lm}=E_{lm}\frac{\epsilon_{l}}{\sigma_{m}}=\delta_{lm}\frac{\kappa_{l}}{\sigma_{l}}-
\epsilon_{l}\frac{iw}{2\pi}\frac{\mu^{[1]}}{\mu^{[0]}}k_{ym}^{[1]}J_{lm}~~,~~
\mathcal{G}_{l}=c_{l}\epsilon_{l}=2a^{i}\epsilon_{l}I_{l}^{+}(k_{x}^{i})
~~,~~\mathcal{F}_{m}=d_{m}\sigma_{m}~.
\end{equation}
As expected, only the first term (i.e., the one with $\delta_{lm}$) involves $h_{1},~h_{2}$, and only the second term (i.e., the one with $J_{lm}$) involves  $w$.
%%%%%%%%%%%%%%%%%%%%%%%%%%%%%%%%%%%%%%%%%%%%%%%%%%%%%%%%%%%%%%%%%%
\subsection{The issue of the infinite dimensions of the matrix equation for the diffraction amplitudes}
The matrix equation (\ref{5-010}) can be written symbolically as $\boldsymbol{\mathcal{\mathcal{E}}}\boldsymbol{\mathcal{F}}=\boldsymbol{\mathcal{G}}$. An issue that should not be avoided, notably in connection with resonances, is that of the infinite dimensions of the matrix $\boldsymbol{\mathcal{E}}$. In any numerical study (such as the one undertaken in sect. \ref{num} hereafter), $\boldsymbol{\mathcal{E}}$ must be treated as if it had finite dimensions, i.e., $\boldsymbol{\mathcal{E}}$ being an $M$-by $M$ matrix, with  $M$ a finite integer. Similarly to the matrix equation $\mathbf{E}\mathbf{d}=\mathbf{c}$ in which we denoted the determinant of $\mathbf{E}$ by $D$, now we denote the determinant of $\boldsymbol{\mathcal{E}}$ by $\mathcal{D}$.

The first point to underline is that if the resonance is to occur in the $n$-th coefficient of $\boldsymbol{\mathcal{F}}$, but $M$ is chosen to be inferior $n$, it will be impossible to detect the said resonance. Thus, for instance, if $M$ is chosen to be 0, then only an approximation of the resonant coupling to $\mathcal{F}_{0}$ is possible, this meaning that accounting for the possible resonant coupling to $\mathcal{F}_{m>0}$ is impossible by this means.

Since it has been hypothesized in previous publications that  coupling to the so-called fundamental $m=0$ component (called 'fundamental mode') of the field in the topographic feature is the dominant mechanism for explaining the seismic response of the said feature, we shall first pay attention to this component, as well as  to the  $M=0$ approximation of this response, notably to find out whether it can really be of resonant nature.
%%%%%%%%%%%%%%%%%%%%%%%%%%%%%%%%%%%%%%%%%%%%%%%%%%%%%%%%%%%%%%%%%%%%%
\subsection{Resonances from the point of view of the $M=0$ approximation}\label{reson0}
What the last line of the preceding section means is that we first examine the equation
\begin{equation}\label{6-020}
\mathcal{E}_{00}\mathcal{F}_{0}^{(0)}=\mathcal{G}_{0}\Longleftrightarrow E_{00}d_{0}^{(0)}=c_{0}~,
\end{equation}
wherein the superscript $(0)$ means the zeroth-order approximation, from which it is immediately-evident (as underlined previously in the $M\rightarrow\infty$ context) that $d_{0}^{(0)}$ is all the larger, the smaller is $D^{(0)}=E_{00}$. So let us take a close look at $\mathcal{E}_{00}$ via (\ref{5-090}) and the fact that $k_{y0}^{[1]}=k^{[1]}=2\pi f/\beta^{[1]}$,
\begin{equation}\label{6-030}
\mathcal{D}^{(0)}=\mathcal{E}_{00}=\frac{\kappa_{0}}{\sigma_{0}}-
i\frac{\mu^{[1]}\beta^{[0]}}{\mu^{[0]}\beta^{[1]}}
\frac{k^{[0]}w}{2\pi}J_{00}~.
\end{equation}
whose first term depends only on $h_{1}$, $h_{2}$ and whose second term depends only on $k^{[0]}w$ since (see (\ref{5-170}))
\begin{equation}\label{6-035}
K_{00}(k^{[0]}w)=\frac{k^{[0]}w}{2\pi}J_{00}(k^{[0]}w)=\mathcal{C}_{0}(k^{[0]}w)-\frac{1}{k^{[0]}w}\mathcal{D}_{0}(k^{[0]}w)~.
\end{equation}
We are now in a position to find out, especially when  $k^{[0]}w$ is small (which affects only $\frac{k^{[0]}w}{2\pi}J_{00}$),  for what frequencies (i.e., the resonant frequencies) $\|\mathcal{E}_{00}\|$ can be small.
%%%%%%%%%%%%%%%%%%%%%%%%%%%%%%%%%%%%%%%%%%%%%%%%%%%%%%%%%%%%%%%%%%%%%
\subsubsection{Origin of the so-called shear-wall resonance (SWR)}\label{SWR}
Employing (11.3.20) and (11.3.24) in \cite{as68} gives
\begin{equation}\label{6-040}
\mathcal{A}_{0}(k^{[0]}w)=k^{[0]}w_{1}H^{(1)}(k^{[0]}w)+\frac{2i}{\pi}~,
\end{equation}
so that
\begin{equation}\label{6-045}
K_{00}(k^{[0]}w)=\mathcal{C}_{0}(k^{[0]}w)-H^{(1)}(k^{[0]}w)-\frac{1}{k^{[0]}w}\frac{2i}{\pi}~,
\end{equation}

In \cite{as68}, p. 360 we find the asymptotic forms:
\begin{equation}\label{6-060}
H_{0}^{(1)}(\zeta)\sim \frac{2}{i\pi}\ln\zeta~~,~~H_{1}^{(1)}(\zeta)\sim\frac{2}{i\pi\zeta}~;~\zeta\rightarrow 0,
\end{equation}
so that $-H_{1}^{(1)}(k^{[0]}w)-\frac{1}{k^{[0]}w}\frac{2i}{\pi}\sim 0~;~k^{[0]}w\rightarrow 0$. On the other hand,
\begin{equation}\label{6-070}
\mathcal{C}_{0}(k^{[0]}w)=\int_{0}^{k^{[0]}w}H_{0}^{(1)}(\zeta)d\zeta\sim \frac{2}{i\pi}\int_{0}^{k^{[0]}w}\ln\zeta d\zeta=k^{[0]}w \ln(k^{[0]}w)-k^{[0]}w=0  ~;~k^{[0]}w\rightarrow 0~,
\end{equation}
so that
\begin{equation}\label{6-075}
K_{00}(k^{[0]}w)=0  ~;~k^{[0]}w\rightarrow 0~,
\end{equation}
whence
\begin{equation}\label{6-080}
\mathcal{E}_{00}\sim \frac{\kappa_{0}}{\sigma_{0}}~;~k^{[0]}w\rightarrow 0~.
\end{equation}
Recall, via (\ref{4-240})-(\ref{4-250}), that:
\begin{equation}\label{6-090}
\kappa_{0}=\cos(k^{[1]}h_{1}+k^{[2]}h_{2})+\left(\frac{\mu^{[2]}\beta^{[1]}}{\mu^{[1]}\beta^{[2]}}-1\right)
\sin(k^{[1]}h_{1})\sin(k^{[2]}h_{2})
~,
\end{equation}
\begin{equation}\label{6-100}
\sigma_{0}=\sin(k^{[1]}h_{1}+k^{[2]}h_{2})+\left(\frac{\mu^{[2]}\beta^{[1]}}{\mu^{[1]}\beta^{[2]}}-1\right)
\cos(k^{[1]}h_{1})\sin(k^{[2]}h_{2})
~,
\end{equation}
and assume that
\begin{equation}\label{6-110}
\left\|\frac{\mu^{[2]}\beta^{[1]}}{\mu^{[1]}\beta^{[2]}}-1\right\|<<1
~,
\end{equation}
whence
\begin{equation}\label{6-120}
\kappa_{0}\approx \cos(k^{[1]}h_{1}+k^{[2]}h_{2})~~,~~\sigma_{0}\approx\sin(k^{[1]}h_{1}+k^{[2]}h_{2})
~.
\end{equation}
Thus, $\mathcal{E}_{00}$ is minimal when $\cos(k^{[1]}h_{1}+k^{[2]}h_{2})=0$ which occurs for
\begin{equation}\label{6-130}
k^{[1]}h_{1}+k^{[2]}h_{2}=(2L+1)\frac{\pi}{2}~;~L=0,1,2,....
~,
\end{equation}
or for
\begin{equation}\label{6-140}
f_{L}=\frac{2L+1}{4\left(\frac{h_{1}}{\beta^{[1]}}+\frac{h_{2}}{\beta^{[2]}}\right)}~;~L=0,1,2,....
~,
\end{equation}
which are the $M=0$ approximation of the resonance frequencies for the scattering configuration: (a) under the assumption (\ref{6-110}), and (b) at low driving frequency (i.e., $2\pi f/\beta^{[0}<<1$) and/or narrow AGS's (i.e., $w<<1$). The condition $w<<1$ means that the convex rectangular protuberance resembles a (thin) wall, and since we are dealing with shear motion, the $f_{L}$ are often termed the shear wall resonance frequencies, usually, in connection with a homogeneous (with respect to the wavespeed and shear modulus) feature such that $\frac{h_{1}}{\beta^{[1]}}+\frac{h_{2}}{\beta^{[2]}}=\frac{h}{\beta^{[1]}}$ and $\frac{\mu^{[2]}\beta^{[1]}}{\mu^{[1]}\beta^{[2]}}-1=0$ whose consequences are
\begin{equation}\label{6-150}
f_{L}^{SWR}=\frac{(2L+1)\beta^{[1]}}{4h}~;~L=0,1,2,....
~,
\end{equation}
which are more properly termed the 'homogeneous shear wall resonance (HSWR) frequencies'. Finally, $f_{0}=\frac{\beta^{[1]}}{4h}$ is often termed the fundamental resonance frequency (or just the fundamental frequency) of the protuberance.

Some comments are in order concerning these resonances. Strictly speaking, they can occur only when the media in the protuberance are non-lossy (i.e., the wavespeeds therein are real), in which case $\mathcal{E}_{00}$ is strictly =0 at $f=f_{L}$, this meaning that $\mathcal{F}_{0}$, and therefore $d_{0}$ blow up at the resonance frequencies $f=f_{L}$. There are two reasons why this never occurs. The first is that all real materials are lossy, even slightly-so, so that $\mathcal{E}_{00}\ne 0$ at $f=f_{L}$ or even near $f_{L}$. The second reason is that the diffraction coefficient $\mathcal{F}_{0}^{(0)}$ defined in (\ref{6-020}), with $\mathcal{E}_{00}$ therein replaced by its asymptotic form $\kappa_{0}/\sigma_{0}$, is only an approximation of $\mathcal{F}_{0}$ that takes no account of radiation damping. In fact, the replacement of $\mathcal{E}_{00}$  by its $k^{[0]}w\rightarrow 0$ asymptotic form has (as we shall see further on in the numerical results) important consequences, notably concerning the location of the resonance frequencies. The reason why we invoked the asymptotic analysis was simply to establish the connection of our investigation with the more-traditional ones based on shear-wall, or mass-spring type resonator paradigms \cite{tr72,jb73,wt75,tt92,jl19}. This means, that the correct equation for finding the zeroth-order approximation of the resonance frequencies is  $D^{(0)}(\omega)=0$ (for complex $\omega$) or $\omega=$arg min $D^{(0)}$ (for real $\omega$) wherein $E_{00}$ is given by its exact expression $E_{00}=\kappa_{0}-\frac{iw}{2\pi}\frac{\mu^{[1]}}{\mu^{[0]}}k_{y0}^{[1]}\sigma_{0}J_{00}$.

Moreover, the same principle holds for higher-order (i.e., $M>0$) approximations of the resonance frequencies: the matrix equation to deal with is $\boldsymbol{\mathcal{E}}^{(M)}\boldsymbol{\mathcal{F}}^{(M)}=\boldsymbol{\mathcal{G}}$ with the understanding that the  $M$-th order approximation of the resonance frequencies  are those that correspond to minima of $D^{(M)}=\text {det }\mathbf{E}^{(M)}$. As stated briefly previously, by taking $M>0$, we expect to find a whole new set of resonances that might show up as large values of the diffraction coefficients of order greater than zero. This will be demonstrated numerically further on, after treating in more detail the issue of coupling to a resonance.
%%%%%%%%%%%%%%%%%%%%%%%%%%%%%%%%%%%%%%%%%%%%%%%%
\subsubsection{Coupling to a HSWR resonance}
Coupling to a resonance means: (a) obtaining the vector $\mathbf{d}$ of diffraction coefficients at the resonance frequency and (b) determining the field in and on the convex surface feature (and eventually on and beneath the flat portions of ground) from  $\mathbf{d}$ at the resonant frequency. We shall first examine coupling to the HSWR resonance because it encompasses some of the features of coupling to more complicated resonances.

The point of departure is the relation $\mathcal{F}_{0}^{(0)}(f)=\frac{\mathcal{G}(f)}{\mathcal{E}_{00}(f)}$, and because $\mathcal{F}_{0}^{(0)}(f)=d_{0}^{(0)}(f)\sigma_{0}$

\begin{equation}\label{6-160}
d_{0}^{(0)}(f)=\frac{\mathcal{G}(f)}{\sigma_{0}(f)\mathcal{E}_{00}(f)}
~.
\end{equation}
We found previously that  $\sigma_{0}(f)\mathcal{E}_{00}(f)\sim \kappa_{0}(f)~;~k^{[0]}w\rightarrow 0$ and $\mathcal{G}(f)=2a^{i}I_{0}^{+}(k_{x}^{i})=2a^{i}$sinc$\left(k_{x}^{i}w/2\right)$$\sim 2a^{i}~;~k^{[0]}w\rightarrow 0$, so that
\begin{equation}\label{6-170}
d_{0}^{(0)}(f)\sim\frac{2a^{i}(f)}{\kappa_{0}(f)}=\frac{2a^{i}(f)}{\cos\big(\frac{2\pi fh}{\beta^{[1]}}\big)}~;~k^{[0]}w\rightarrow 0
~,
\end{equation}
which shows that the diffraction amplitude $d_{0}^{(0)}(f)$ is not only conditioned by the resonant factor $1/\cos\big(\frac{2\pi fh}{\beta^{[1]}}\big)$, but also by the spectral amplitude $a^{i}(f)$ of the solicitation, so that
\begin{equation}\label{6-180}
d_{0}^{(0)}(f_{L}^{HSWR})=\frac{2a^{i}(f_{L}^{HSWR})}{\cos\Big((2L+1)\pi/2\Big)}~;~L=0,1,2,...
~,
\end{equation}
which is all the larger (it is infinite as it stands, but, because of the material losses and/or radiation damping, actually finite) the larger is $a^{i}(f_{L}^{HSWR})$. Thus, coupling to the HSWR resonance is conditioned by the spectral amplitude $a_{i}(f)$ at the resonant frequency.

We now examine coupling to the fields at resonance. Within the rectangular-shape protuberance the zeroth-order approximation to the field is
\begin{equation}\label{6-190}
u^{[1](0)}(x,y;f)=d_{0}^{(0)}(f)\cos\Big(\frac{2\pi f}{\beta^{[1]}}(y-h)\Big)
~,
\end{equation}
which shows that: (1) the field does not depend on $x$ within the protuberance whatever the frequency $f$, and (2) the field is maximal at the top ($z=h$) of the surface feature whatever be $f$. It follows that
\begin{equation}\label{6-200}
u^{[1](0)}(x,0;f)=d_{0}^{(0)}(f)\kappa_{0}\sim 2a^{i}(f)~;~k^{[0]}w/2\rightarrow 0
~,
\end{equation}
which shows that the field at the base of the protuberance is asymptotically the same as what it would be in the absence of the said feature, this being true for all (low frequency) $f$. This result is paradoxical  because if the base of the protuberance is impervious to the incident wave, it is impossible for it to penetrate into the said surface feature. As we shall see hereafter, this paradox disappears as soon as the asymptotic analysis is dropped. In any case, with the adoption of the approximations embodied in (\ref{6-200}), we see that
\begin{equation}\label{6-210}
\|u^{[1](0)}(x,h;f_{L}^{HSWR})\|=\|2a_{i}(f_{L}^{HSWR})\|~;~L=0,1,2,....
~,
\end{equation}
which tells us that coupling to the HSWR does not result in any amplification or deamplification of the top displacement field of the convex rectangular-shape surface feature. This also means that the field (actually its modulus) within the homogeneous shear wall is not amplified at the HSWR resonance frequency since this field is inferior or equal to the field at the top of the HSW.
%%%%%%%%%%%%%%%%%%%%%%%%%%%%%%%%%%%%%%%%%%%%%%%%%%%%%%
\subsubsection{Coupling to resonances when the $k^{[0]}w\rightarrow 0$ asymptoticity is not assumed}
The first task is to determine $d_{0}^{(0)}(f)$ at, or near resonance, starting from (\ref{6-020}):
\begin{equation}\label{6-220}
d_{0}^{(0)}(f)=\frac{\mathcal{G}_{0}(f)}{\mathcal{E}_{00}(f)\sigma_{0}(f)}=\frac{c_{0}(f)}{E_{00}(f)}~,
\end{equation}
\begin{multline}\label{6-230}
\mathcal{E}_{00}(f)=\frac{\kappa_{0}(f)}{\sigma_{0}(f)}-
i\frac{\mu^{[1]}\beta^{[0]}}{\mu^{[0]}\beta^{[1]}}\left[\mathcal{C}_{0}(k^{[0]}w)-\frac{1}{k^{[0]}w}\mathcal{A}_{0}(k^{[0]}w)\right]~~,~~\\
\mathcal{G}_{0}(f)=2a^{i}(f)I_{0}^{+}(k_{x}^{i})=2a^{i}(f)\text{sinc}\left(k_{x}^{i}w/2\right)~.
\end{multline}
As stated previously, the resonance frequencies $f_{L}$ are those frequencies for which $\|E_{00}(f)\|$ is minimal. These frequencies are all the closer to to $f_{L}^{SWR}$ the smaller (in absolute value) is the second term  $i\frac{\mu^{[1]}\beta^{[0]}}{\mu^{[0]}\beta^{[1]}}\left[C_{0}(k^{[0]}w)-\frac{1}{k^{[0]}w}D_{0}(k^{[0]}w)\right]$ relative to the first term $\frac{\kappa_{0}(f)}{\sigma_{0}(f)}$ in $\mathcal{E}_{00}$. Moreover, the presence of this second term (which is complex) in $\mathcal{E}_{00}$ is the reason why the so-obtained $d_{0}^{(0)}(f_{L});~L=0,1,...$ are not infinite. Thus, resonant coupling of   $d_{0}^{(0)}(f)$  manifests itself by a relatively-large, but finite, value of this diffraction coefficient at the resonant frequency $f_{L}$. We note also that not only is this coupling more efficient, due to the proximity of the resonance frequency to the frequency of the maximum of $a^{i}(f)$, but also to the proximity of $f_{L}$ to $f=0$  (due to the fact that the sinc function is maximum when its argument is nil).

The second aspect of resonant coupling has to do with the field at resonance within the protuberance. As previously, we have
\begin{equation}\label{6-240}
u^{[1](0)}(x,y;f)=d_{0}^{(0)}(f)\cos\Big(\frac{2\pi f}{\beta^{[1]}}(y-h)\Big)
~,
\end{equation}
which again shows that: (1) the field does not depend on $x$ within the protuberance whatever the frequency $f$, and (2) the field is maximal at the top ($z=h$) of the protuberance whatever be $f$. It follows that
\begin{equation}\label{6-250}
u^{[1](0)}(x,0;f)=d_{0}^{(0)}(f)\kappa_{0}
~,
\end{equation}
which indicates that the field at the base of the protuberance is no longer the same as what it would be in the absence of the said protuberance, this being true for all  $f$ including the resonant frequencies. This result means, as one would expect, that the base segment of the protuberance is no longer impervious to the incident wave, thus making it possible for the incident wave  to penetrate into the said protuberance.

The last, all-important, feature of (\ref{6-240}), is that
\begin{equation}\label{6-260}
\|u^{[1](0)}(x,h;f_{L})\|\geq \|2a_{i}(f_{L})\|~;~L=0,1,2,....
~,
\end{equation}
due to the fact that $\|d_{0}^{(0)}(f_{L})\|\geq \|2a^{i}(f_{L})\|$. This means that coupling to a resonance generally results in amplification of the top displacement field of the  rectangular-shape protuberance (with respect to its value on  flat ground). However, this amplification does not necessarily occur at other heights within the protuberance due to the presence of the $\cos$ term in (\ref{6-240}).
%%%%%%%%%%%%%%%%%%%%%%%%%%%%%%%%%%%%%%%%%%%%%%%%%%%%%%%%%%%%%%%%%%%%%
\subsection{Resonances from the point of view of the $M=1$ approximation}
The $M=1$ approximation of the diffraction coefficient vectors $\boldsymbol{\mathcal{F}}$ and $\mathbf{d}$ originates in the linear system(s):
\begin{equation}\label{6-300}
\begin{array}{c}
\mathcal{E}_{00}\mathcal{F}_{0}^{(1)}+\mathcal{E}_{01}\mathcal{F}_{1}^{(1)}=\mathcal{G}_{0}\\
\mathcal{E}_{10}\mathcal{F}_{0}^{(1)}+\mathcal{E}_{11}\mathcal{F}_{1}^{(1)}=\mathcal{G}_{1}
\end{array}
\Longleftrightarrow
\begin{array}{c}
E_{00}d_{0}^{(1)}+E_{01}d_{1}^{(1)}=c_{0}\\
E_{10}d_{0}^{(1)}+E_{11}d_{1}^{(1)}=c_{1}
\end{array}
~,
\end{equation}
the latter of whose solution is
\begin{equation}\label{6-310}
d_{0}^{(1)}=\frac{c_{0}E_{11}-c_{1}E_{01}}
{E_{00}E_{11}-E_{10}E_{01}}~~,~~d_{1}^{(1)}=
\frac{c_{1}E_{00}-c_{0}E_{10}}
{E_{00}E_{11}-E_{10}E_{01}}
~,
\end{equation}
from which we see that resonant coupling to two diffraction coefficients is possible due to the fact that both are affected by the resonances(s) resulting from the minima of the function $\|D^{(1)}\|$ wherein
\begin{equation}\label{6-320}
D^{(1)}=E_{00}E_{11}-E_{10}E_{01}
~.
\end{equation}
However, the fact that the numerators in (\ref{6-310}) are different, the coupling to $d_{0}^{(1)}$ is not necessarily                                                                as efficient as the coupling to $d_{1}^{(1)}$, which means that one or the other, or even both, of these coefficients are not necessarily large at resonance. Another feature of (\ref{6-310}), which will be demonstrated numerically further on, is that due to the fact that $D^{(1)}$ is a more complicated function than $D^{(0)}=E_{00}$, notably at relatively-high frequencies, the number of resonant frequencies associated with  $D^{(1)}$ is larger than those associated with $D^{(1)}$. Finally, (even) the lower-frequency resonances associated with $D^{(1)}$ occur at frequencies that are different from the resonant frequencies associated with $D^{(0)}$, which fact shows that generally, {\it one cannot correctly describe the resonances of the configuration by basing the description on the sole $M=0$ approximation}.

Another important aspect of resonant coupling to $\mathbf{d}^{(1)}$ is the fact that it depends not only on the driving term $c_{0}$ but also on $c_{1}$. Both of these are functions of $a^{i}(f)$ so that the previous comments concerning the influence of this factor remain true, but they also depend on the sinc functions contained in $I_{l}^{+}(k_{x}^{i})$ and it is possible, for non-normal incidence, that $\|I_{1}^{+}(k_{x}^{i}\|>\|I_{0}^{+}(k_{x}^{i}\|$ so as to make the coupling to $\mathbf{d}^{(1)}$ become larger at non-normal incidence than at normal incidence, contrary to the case of resonant coupling to $\mathbf{d}^{(0)}$ in which the envelope of $\|I_{0}^{+}(k_{x}^{i}\|$ diminishes  with increasing incident angle. This possibility will be illustrated numerically further on.

Next, consider the resonant coupling to the field within the protuberance. To simplify the message that we want to bring across, we  choose the case of a homogeneous protuberance, i.e., $M^{[2]}=M^{[1]}$. The $M=1$ approximation of the field within the protuberance is then
\begin{equation}\label{6-320}
u^{[2](1)}(x,y;f)=d_{0}^{(1)}(f)\cos\big(k_{y0}(y-h)\big)+d_{1}^{(1)}(f)\cos\big(k_{x1}(x+w/2)\big)\cos\big(k_{y1}(y-h)\big)
~,
\end{equation}
 from which we observe that, contrary to what occurs in the zeroth-order approximation of the field, now the $M=1$ approximation thereof: (1) depends on the $x$ coordinate, 2) is not necessarily-maximal at $y=h$, and 3) the $x-y$ pattern of resonant response can be dominated by the second term in (\ref{6-320}) if, as is possible (see the comments a few lines back), $d_{1}(f)$ dominates $d_{0}(f)$ at a resonant frequency. The $x-y$ pattern of  response can be  even more involved when both $d_{1}(f)$  and $d_{0}(f)$ are influential at a resonant frequency and even be such that the field is not amplified at most of the locations within the protuberance. These observations, which will be illustrated by the numerical results offered further on, again underline the absolute necessity of going beyond $M=0$ to correctly predict the seismic response of the hill or mountain, this being especially so if one wants to account for coupling to  other than the fundamental mode resonance (i.e., the one that occurs at the lowest frequency, corresponding to the $L=0$ HSWR when the convex feature resembles a homogeneous shear wall).
%%%%%%%%%%%%%%%%%%%%%%%%%%%%%%%%%%%%%%%%%%%%%%%%%%%%%%%%%%%%%%%%%%%%%
\subsection{Resonances from the point of view of the $M>1$ approximations}
Needless to say, essentially everything that was written for the $M=1$ approximation holds for the $M>1$ approximations of resonant response. This will be illustrated in the numerical examples which follow. The latter will show that the possibility of a resonance, manifested by a small value of $\|D(f)\|$ at a so-called resonant frequency $f^{R}$, does not guarantee that the response will be amplified considerably and/or at all locations of the hill or mountain because of the interplay of the resonant diffraction coefficients (themselves depending on the spectrum of the solicitation) with the geometric factors (i.e., that depend on the $x,y$ coordinates) contained in the SOV representation of the displacement field. In other words: significant amplification of this field at $f^{R}$ requires (i.e., is a necessary condition for) the existence of a resonance at $f^{R}$, but is not a sufficient condition for this amplification to manifest itself (this manifestation depending heavily on the spectral attributes and incident angle of the solicitation, as well as on the location at which the field is sensed.
%%%%%%%%%%%%%%%%%%%%%%%%%%%%%%%%%%%%%%%%%%%%%%%%%%%%%%%%%%%%%%%%%%%%%%%%%%%%%%%%%
\section{Numerical resolution of the linear system of equations}
 The first task is to obtain numerically the set $\{\mathcal{F}_{m}\}$  from the linear system of equations (\ref{5-010}). Once this set is found, it is introduced into $d_{m}=\frac{\mathcal{F}_{m}}{\sigma_{m}}$ to get $\{d_{m}\}$, and into  (\ref{8-070}), (\ref{8-073}), (\ref{8-075}) to obtain the sets  $\{a_{m}\}$, $\{b_{m}\}$, $\{\mathcal{B}(k_{x})\}$ .
When all these coefficients (we mean those whose values depart significantly from zero) are found, they enable the computation of the seismic response (i.e., the displacement field) in all the subdomains of the site and city via (\ref{6-010}), (\ref{6-014}), (\ref{6-030}) and (\ref{6-040})  (in the last two expressions, the sums are taken from $m=0$ to $m=M$, $M$ a finite, relatively-small integer defined hereafter).

Concerning the resolution of the infinite system of linear equations (\ref{5-010}), the approach is basically to replace it by the finite system of linear equations
\begin{equation}\label{7-010}
\sum_{m=0}^{M}\mathcal{E}_{lm}\mathcal{F}_{m}^{(M)}=\mathcal{G}_{l}~;~l=0,1,2,...M~,
\end{equation}
in which the superscript  $(M)$ signifies the $M$-th order approximation of $\mathcal{F}_{m}$ obtained via (\ref{7-010}), the  procedure being to increase $M$ so as to generate the sequence of numerical solutions $\{F_{0}^{(0)}\}$, $\{F_{0}^{(1)},F_{1}^{(1)}\}$,....until the values of the first few members of  these sets stabilize and the remaining members become very small. This is usually obtained for reasonably-small values of $M$, especially in the low frequency regime of interest in our seismic response problem.

The so-obtained numerical solutions (which are henceforth based on the assumption $a^{i}(\omega)=1~;~\forall \omega\ge 0$) were found to: i) reproduce the theoretical solution for  VS configurations, ii)  satisfy the conservation of flux relation \cite{wi20} with an error of less than a half percent for all $M\ge 0$, and iii) be in agreement with numerical results obtained by a finite element method \cite{gr05,gw08},  so that they can be considered,  for all practical purposes, to be 'exact'. This issue will benefit from supplementary comments further on.
%\clearpage
%\newpage
%%%%%%%%%%%%%%%%%%%%%%%%%%%%%%%%%%%%%%%%%%%%%%%%%%%%%%%%%%%%%%%%%%%%%%%%%%%%%%%%%%%%%%%%%%%%%%%%%%%%%%
\section{Seismic response from stress-free boundary irregularities in the sense of Sills and beyond}\label{Sills}
%%%%%%%%%%%%%%%%%%%%%%%%%%%%%%%%%%%%%%%%%%%%%%%%%%%%%%%%%%%%%%%%%%%
\subsection{Comparison with the numerical results of Sills}
The publication \cite{si78} represents one of the earliest efforts to compute, as exactly as possible, the seismic response of a hill, assumed by Sills to be of semi-circular shape, radius $h$ and (homogeneous) composition identical to that of the underlying half space. This was achieved by a boundary integral equation numerical scheme to show that the so-obtained amplifications are the result of stress-free boundary irregularities. Due to their quasi-rigorous nature, these numerical results  provide a means of comparison with are own results, especially by giving evidence of  the universal nature of the fundamental hill resonance  (in the sense that it occurs for a variety of hill shapes).
\begin{figure}[ht]
\begin{center}
\includegraphics[width=0.65\textwidth]{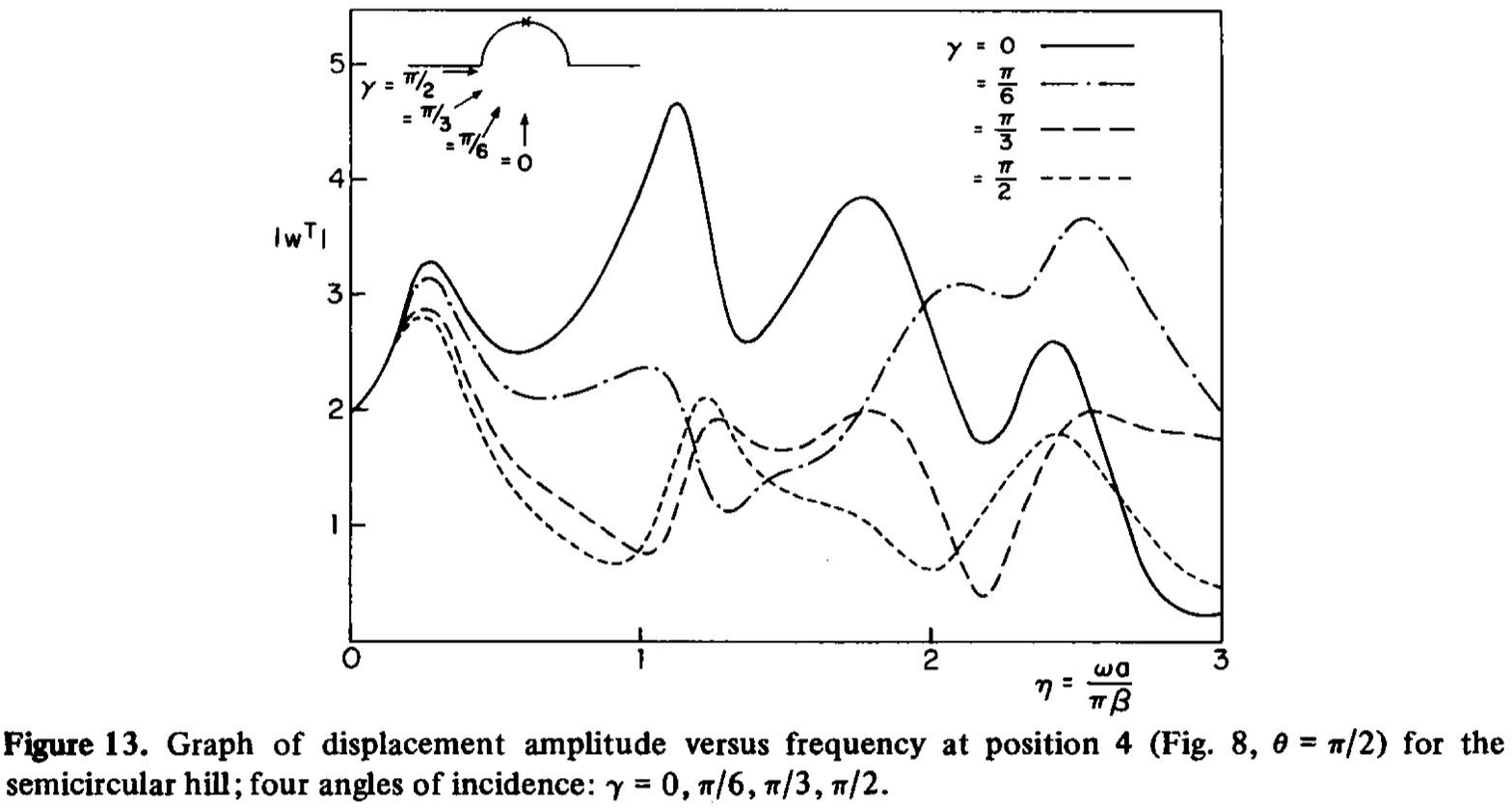}
\caption{Sills' results for the total displacement field $\|u\|$ at the top of a semi-circular hill for various incident angles.}
\label{sills-13}
\end{center}
\end{figure}
\begin{figure}[ptb]
\begin{center}
\includegraphics[width=0.64\textwidth]{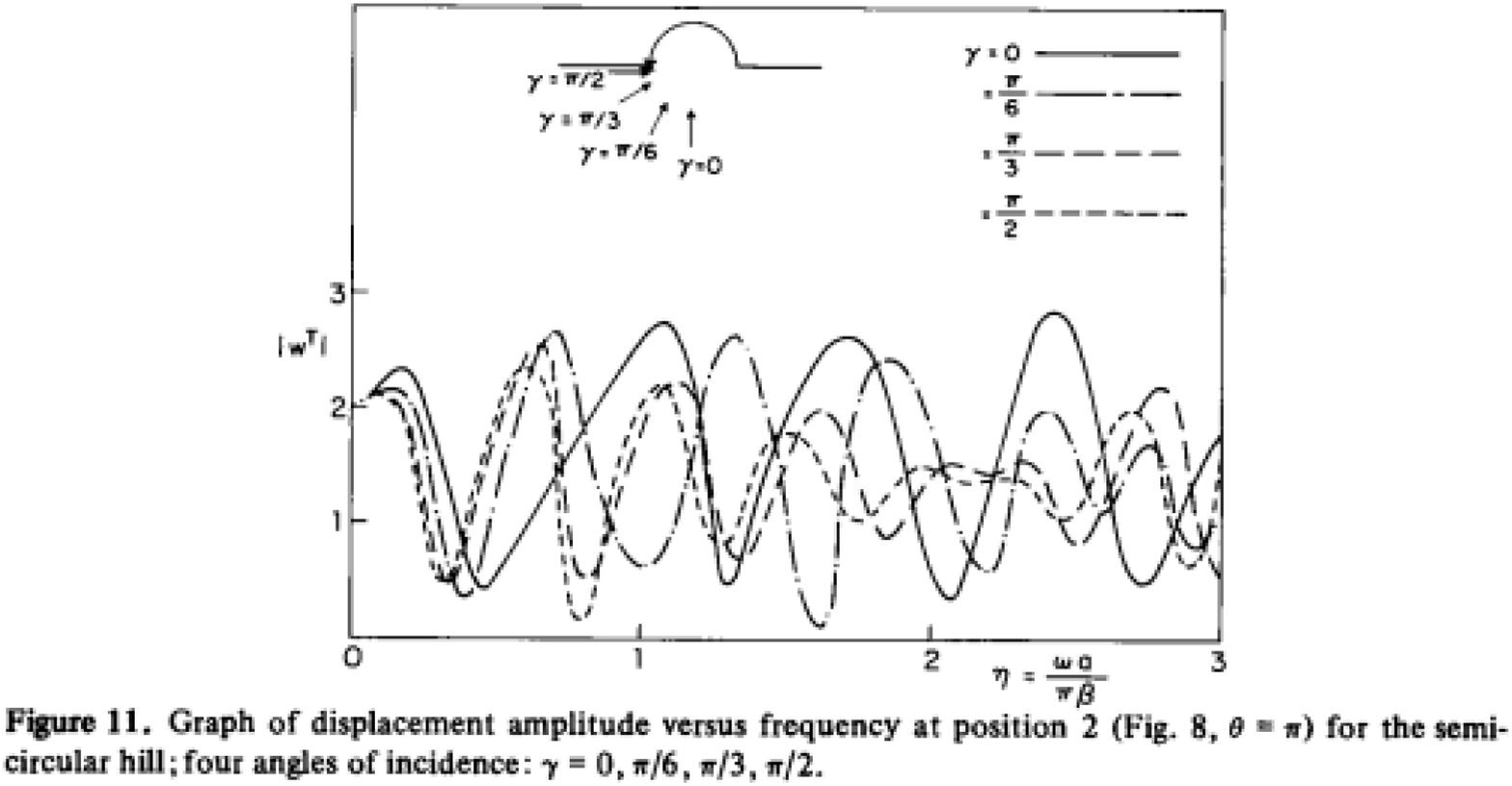}
\caption{Sills' results for the total displacement field $\|u\|$ at the left-hand corner of a semi-circular hill for various incident angles.}
\label{sills-11}
\end{center}
\end{figure}

It is interesting to cite what is written by Sills in \cite{si78} about these graphs (our figs. \ref{sills-13} and \ref{sills-11} corresponding to his figs. 13 and 11 respectively), the first of which applies to the responses, for $\theta^{i}=30,~60,~90^{\circ}$, at the midpoint of the top segment of the hill ($x=0,~y=h$ in our notation) and the second to the responses, for the same incident angles, at the bottom left-hand corner of the protuberance ($x=-h,~y=0$ in our notation):\\\\
"The  rapidly  varying  displacement  amplitude  indicates a complex constructive  and destructive  interference  pattern  which  results  from  the presence of the irregularity.... On  the  other hand,  for  vertical incidence  the  energy  appears  to have been  focused toward  the top of the hill (Fig.  13) resulting in rather high amplifications. In
particular,  for a frequency  of approximately  1 ,  the  displacement is almost 2.5 times greater
than  that  which  is  expected  in  the  case of a featureless topography.... It is quite clear from these graphs that the displacement amplitude is highly dependent upon
angle of incidence and frequency, as well as position along the boundary. Hence one cannot
conclude  on  the  basis  of  these  graphs  that  either  amplification  or  de-amplification  will
always occur  at  a certain  position  on this type of irregularity.  One can conclude however,
that  the  irregularity  does have  an  effect, which appears profound  in this case, and should
be accounted for when analyzing earthquakes in regions of irregular topography."\\\\
What Sills means  by 'vertical incidence' is (in our notation) $\theta^{i}=0^{\circ}$, and by 'frequency'  actually $\frac{k^{[0]}h}{\pi}=\frac{\omega h}{\beta^{[0]}\pi}$. Note that he makes {\it no evocation of resonances}, but rather of "rapidly varying displacement" which he attributes to "complex constructive and destructive interference" nevertheless associated with "the presence of the irregularity".

  The fact that the location of the first peak in Sills' two figures does not seem to depend on the incident angle is an indication that we are actually in presence of what we previously termed the fundamental mode resonance at this frequency, this not being in contradiction with the fact that their corner peak appears at a frequency that is lower than their summit  peak (or what amounts to the same, the response at the corner is lower than at the summit, at the frequency of maximal summit response, in agreement with our prediction for the resonant response of a rectangular hill).

  Let us attempt a more quantitative comparison with our own numerical results for a rectangular-shaped hill. Actually, it appeared to us to be most appropriate for this sake to choose our hill to be of height $h=250~m$ and width $w=2h=500~m$ (which minimally- circumscribes Sills' semi-circular hill. Since the latter is homogeneous, lossless and composed of the same material as that of the underground, we chose $\beta^{[2]}=\beta^{[1]}=\beta^{[0]}$ and $\mu^{[2]}=\mu^{[1]}=\mu^{[0]}$ and somewhat arbitrarily $\beta^{[0]}=1629.4~ms^{-1}$ and $\mu^{[0]}=6.85~MPa$ corresponding to a density $\rho^{[0]}=2580.1~Kgm^{-3}$ which is close to the Bouguer density $2670~Kgm^{-3}$ (itself in the range $2650-2750 Kgm^{-3}$ of granite density).
\clearpage
\newpage
%%%%%%%%%%%%%%%%%%%%%%%%%%%%%%%%%%%%%%%%%
\subsubsection{Our computed transfer functions and $1/D$ for $\theta^{i}=0^{\circ}$}
\begin{figure}[ht]
\begin{center}
\includegraphics[width=0.64\textwidth]{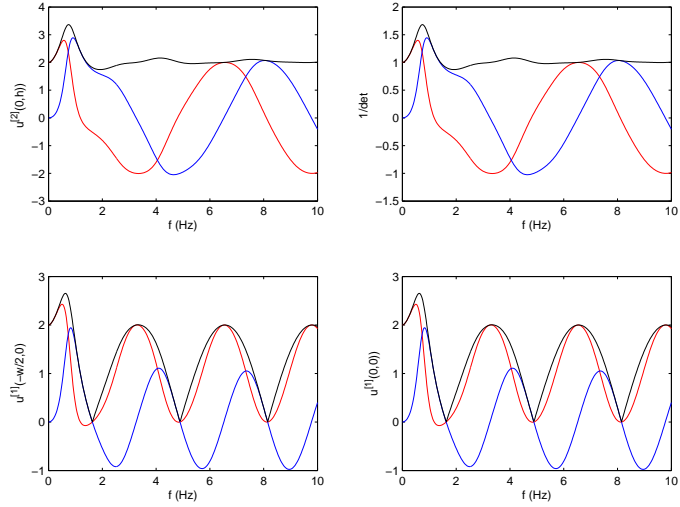}
\caption{The lower right panel is for $T^{(M)}(0,0;f)$, the lower left panel for $T^{(M)}(-w/2,0;f)$ and the upper left panel for $T^{(M)}0,h=h_{2};f)$ whereas the upper right panel depicts $1/D^{(M)}(\omega)$, with $D^{(M)}$ the determinant of the $(M+1)-$by$-(M+1)$ matrix equation involved in the computation of the modal coefficient vector $\mathbf{d}^{(M)}$. The red curves are relative to the real part, the blue curves to the imaginary part and the black curves to the absolute value. $\beta^{[1]}=\beta^{[0]}=1629.4~ms^{-1}$ and $\mu^{[1]}=\mu^{[0]}=6.85~MPa$. Case $h_{1}=250~m$,  $h_{2}=0~m$,  $w=500~m$. $\theta^{i}=0^{\circ}$,  $\mathbf{M=0}$.}
\label{sillsD-010}
\end{center}
\end{figure}

\begin{figure}[ptb]
\begin{center}
\includegraphics[width=0.64\textwidth]{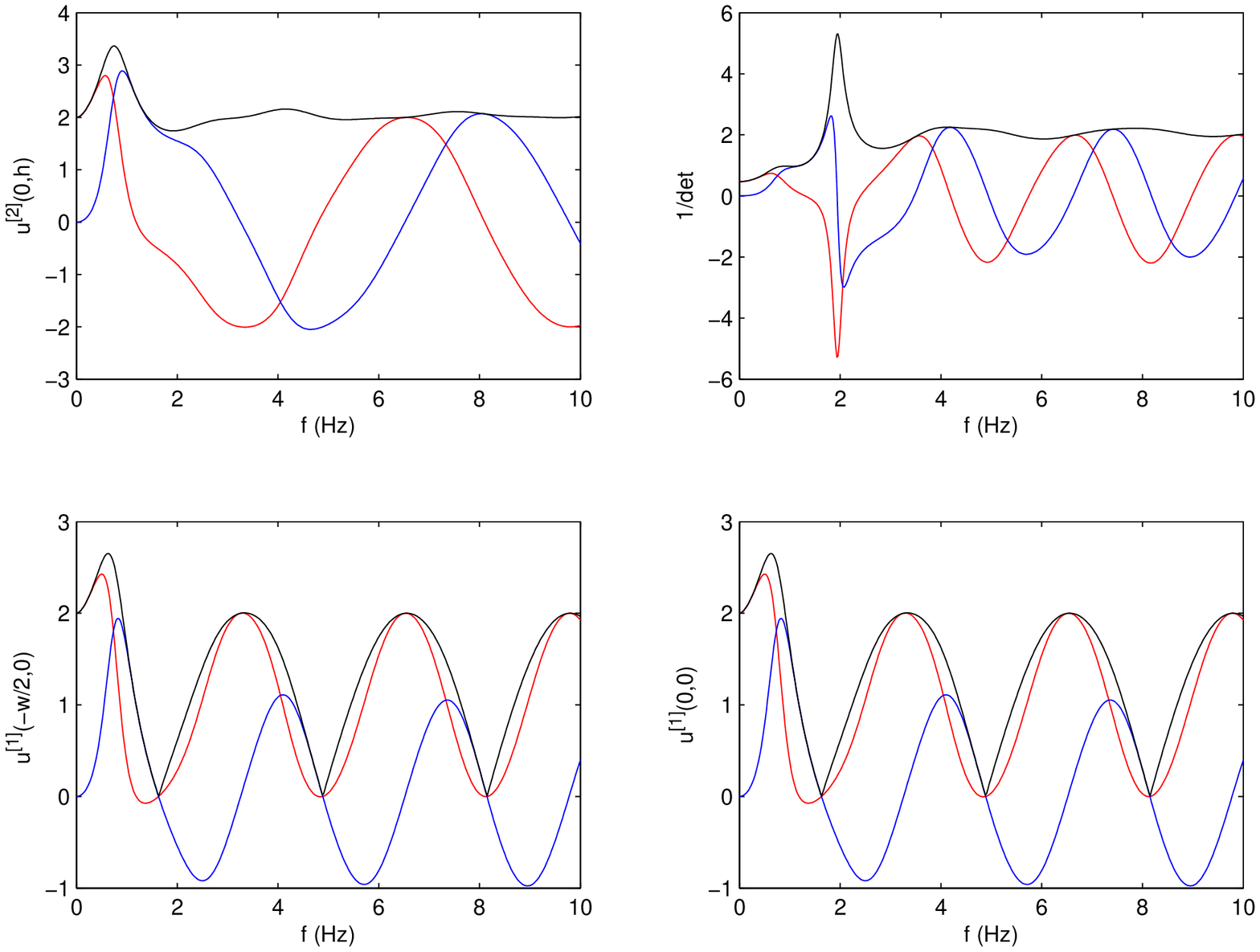}
\caption{$\theta^{i}=0^{\circ}$. Same as fig. \ref{sillsD-010} except that  $\mathbf{M=1}$.}
\label{sillsD-020}
\end{center}
\end{figure}

\begin{figure}[ptb]
\begin{center}
\includegraphics[width=0.64\textwidth]{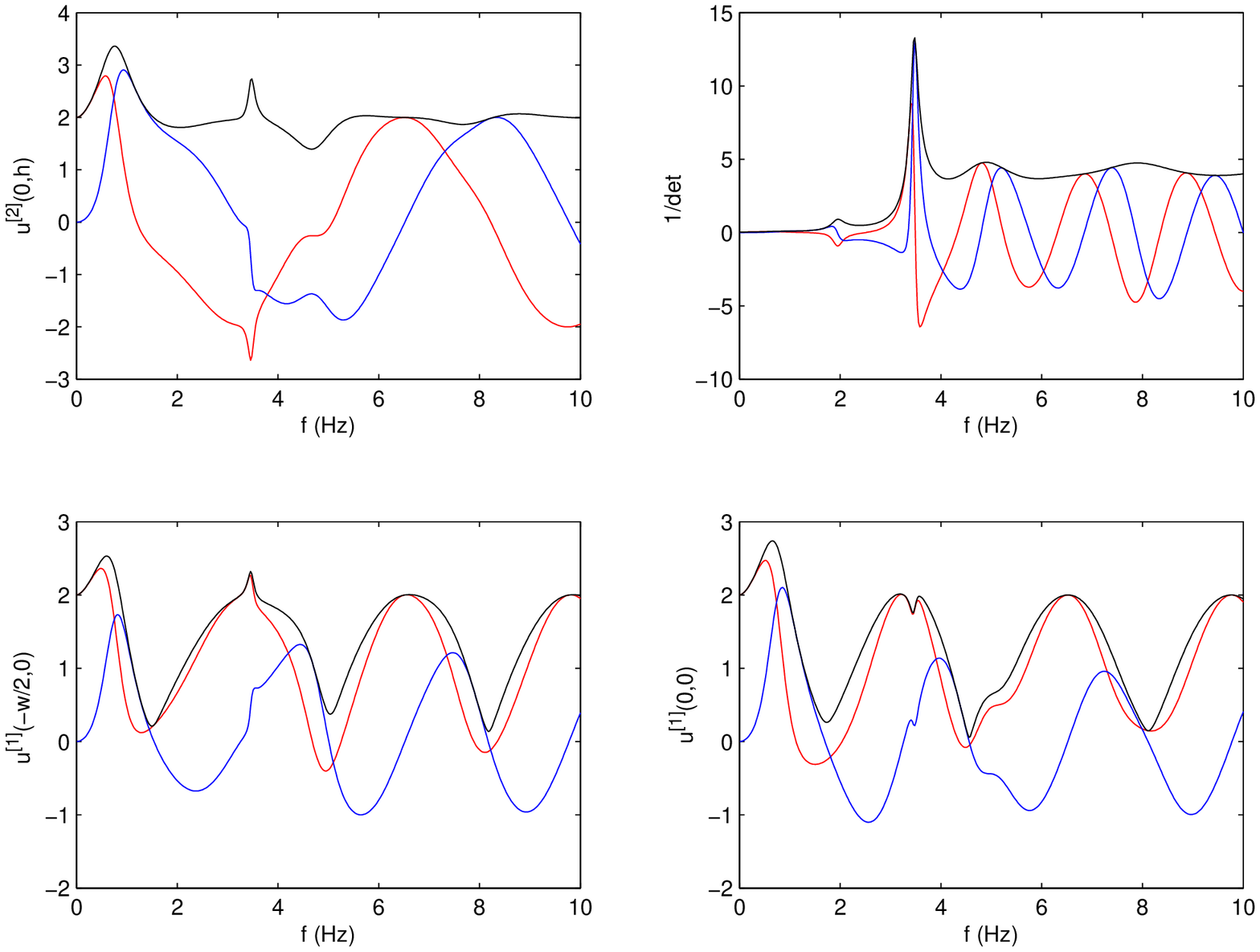}
\caption{$\theta^{i}=0^{\circ}$. Same as fig. \ref{sillsD-010} except that    $\mathbf{M=2}$.}
\label{sillsD-030}
\end{center}
\end{figure}

\begin{figure}[ptb]
\begin{center}
\includegraphics[width=0.64\textwidth]{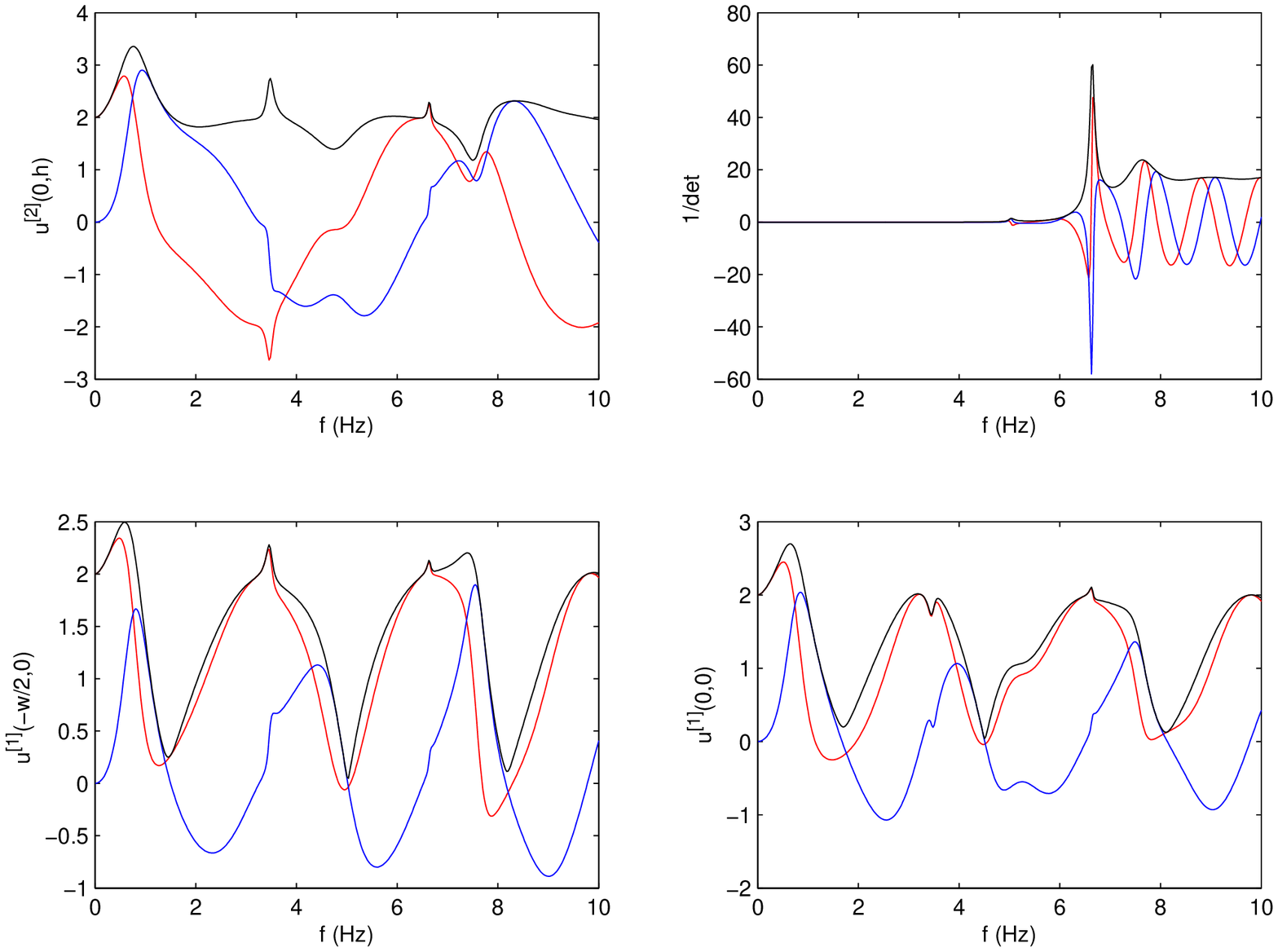}
\caption{$\theta^{i}=0^{\circ}$. Same as fig. \ref{sillsD-010} except that   $\mathbf{M=4}$.}
\label{sillsD-040}
\end{center}
\end{figure}

Figs. \ref{sillsD-010}-\ref{sillsD-040} depict our computed transfer functions $T^{(M)}(x,y;f)=u(x,y;f)/a^{i}(\omega)$ and $1/D^{(M)}(\omega)$ for  incident angle $\theta^{i}=0^{\circ}$ and for various approximation orders $M$.
\clearpage
\newpage
%%%%%%%%%%%%%%%%%%%%%%%%%%%%%%%%%%%%%%%%%
\subsubsection{Our computed transfer functions and $1/D$ for $\theta^{i}=30^{\circ}$}
\begin{figure}[ht]
\begin{center}
\includegraphics[width=0.64\textwidth]{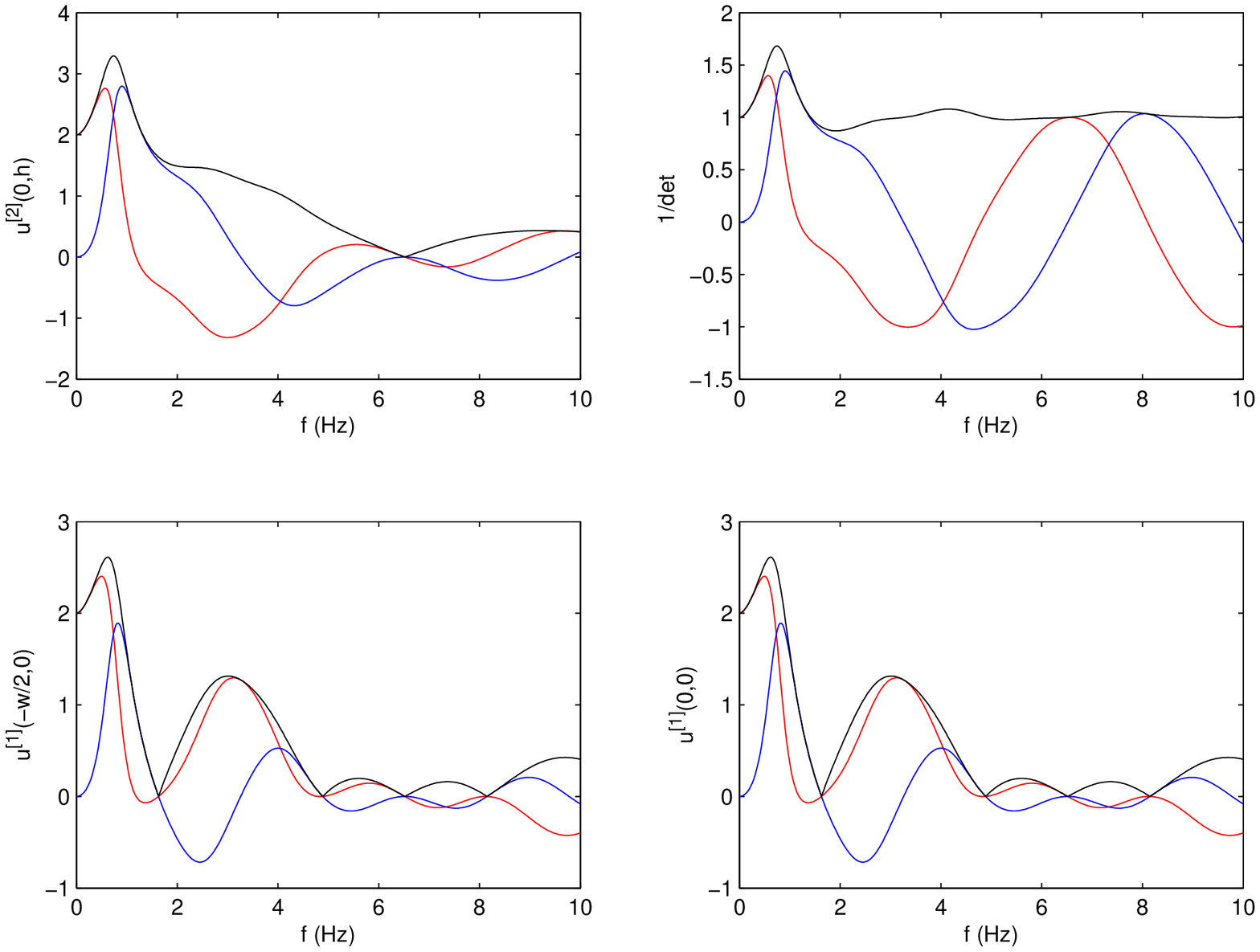}
\caption{$\theta^{i}=30^{\circ}$.  Otherwise, same as fig. \ref{sillsD-010} except that    $\mathbf{M=0}$.}
\label{sillsD-050}
\end{center}
\end{figure}

\begin{figure}[ptb]
\begin{center}
\includegraphics[width=0.64\textwidth]{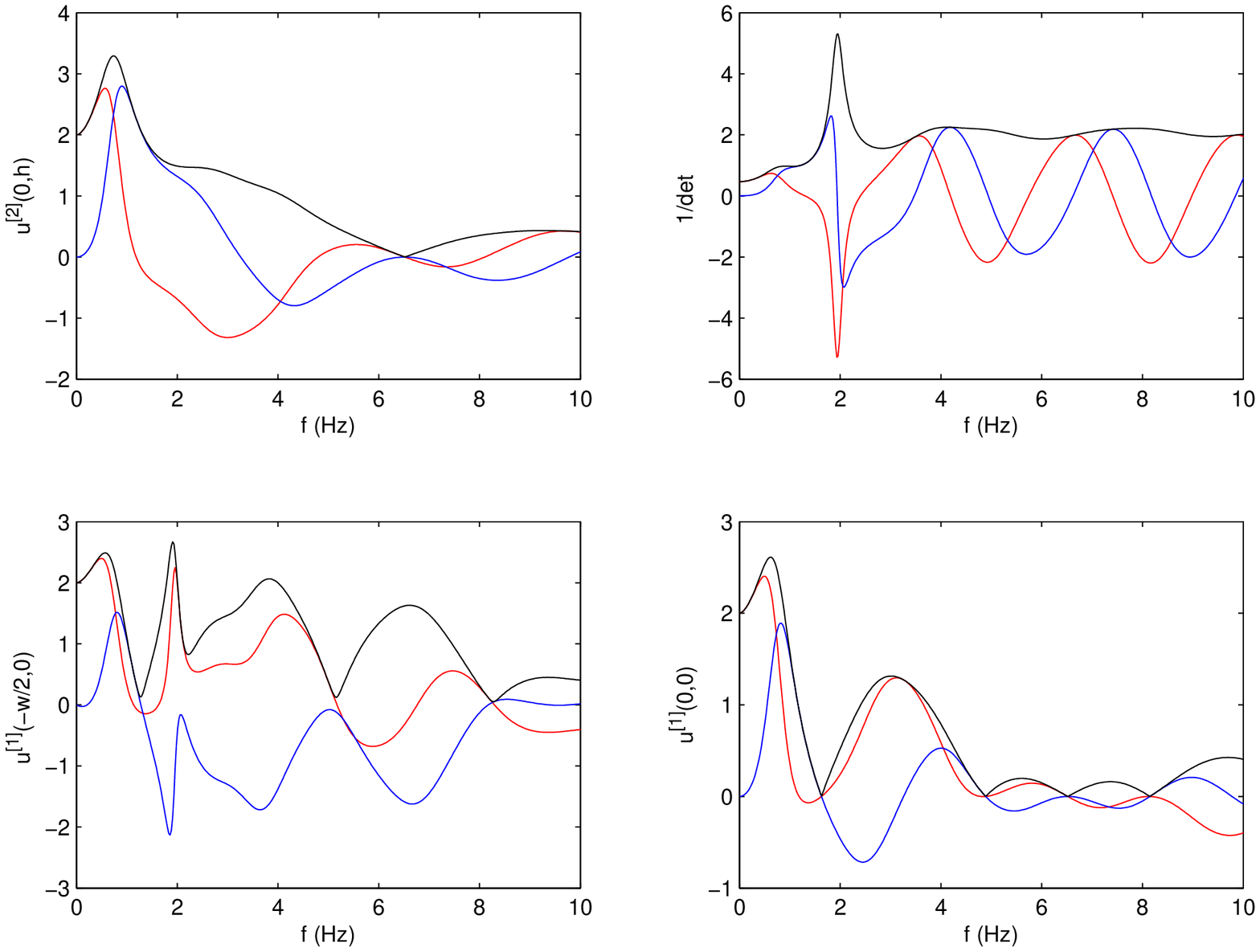}
\caption{$\theta^{i}=30^{\circ}$. Otherwise, same as fig. \ref{sillsD-010} except that    $\mathbf{M=1}$.}
\label{sillsD-060}
\end{center}
\end{figure}

\begin{figure}[ptb]
\begin{center}
\includegraphics[width=0.64\textwidth]{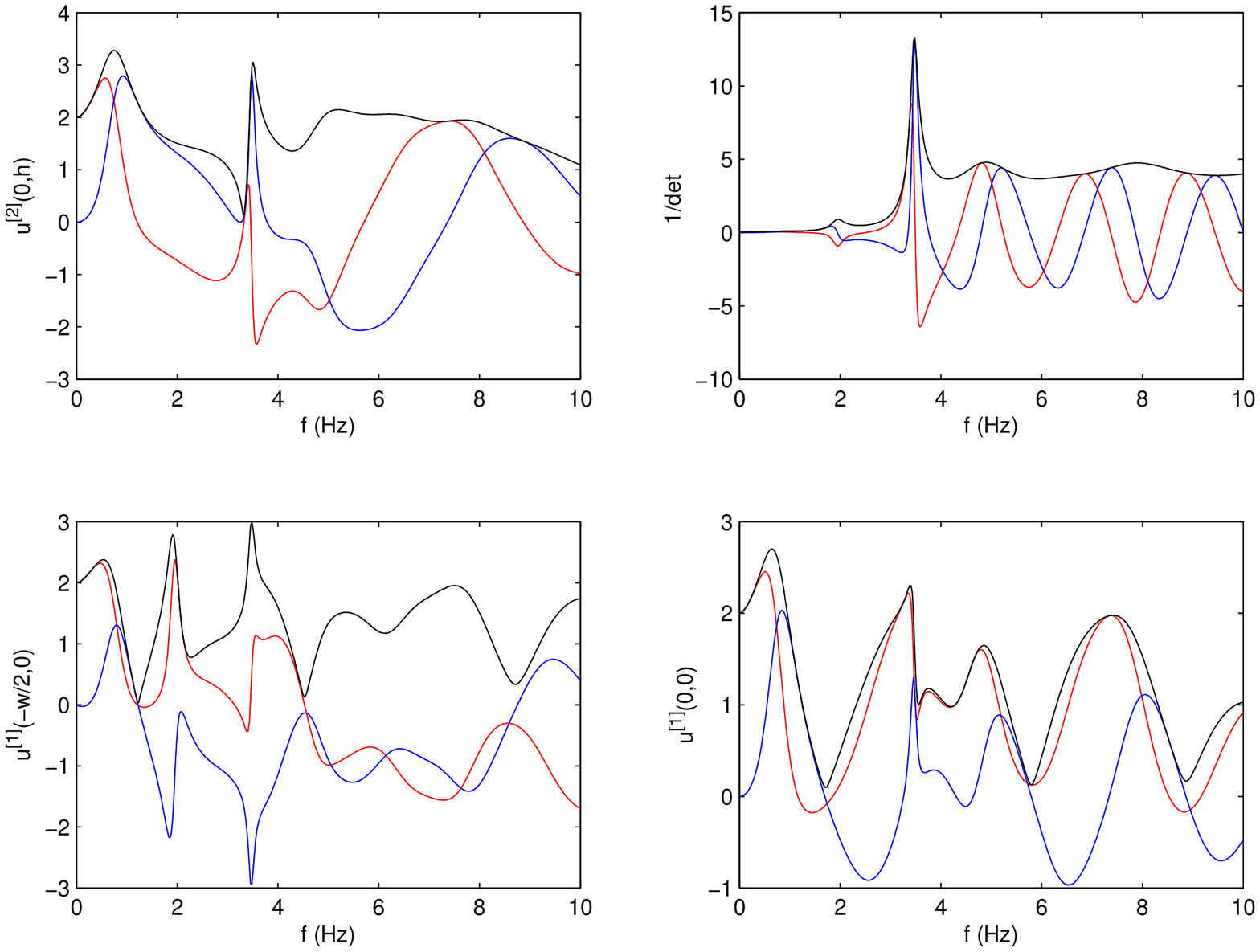}
\caption{$\theta^{i}=30^{\circ}$. Otherwise, same as fig. \ref{sillsD-010} except that   $\mathbf{M=2}$.}
\label{sillsD-070}
\end{center}
\end{figure}

\begin{figure}[ptb]
\begin{center}
\includegraphics[width=0.64\textwidth]{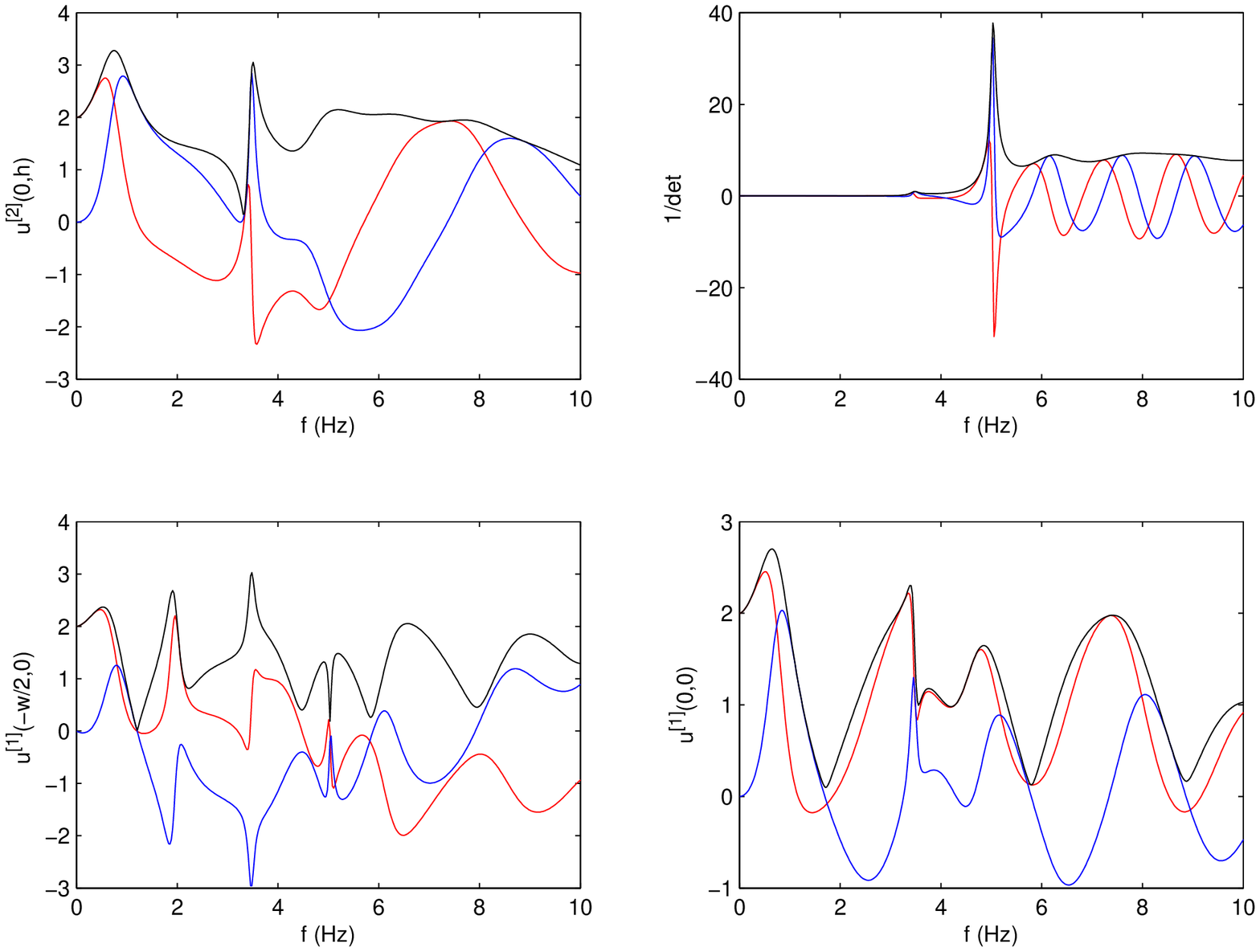}
\caption{$\theta^{i}=30^{\circ}$. Otherwise same as fig. \ref{sillsD-010} except that    $\mathbf{M=3}$.}
\label{sillsD-080}
\end{center}
\end{figure}

\begin{figure}[ptb]
\begin{center}
\includegraphics[width=0.64\textwidth]{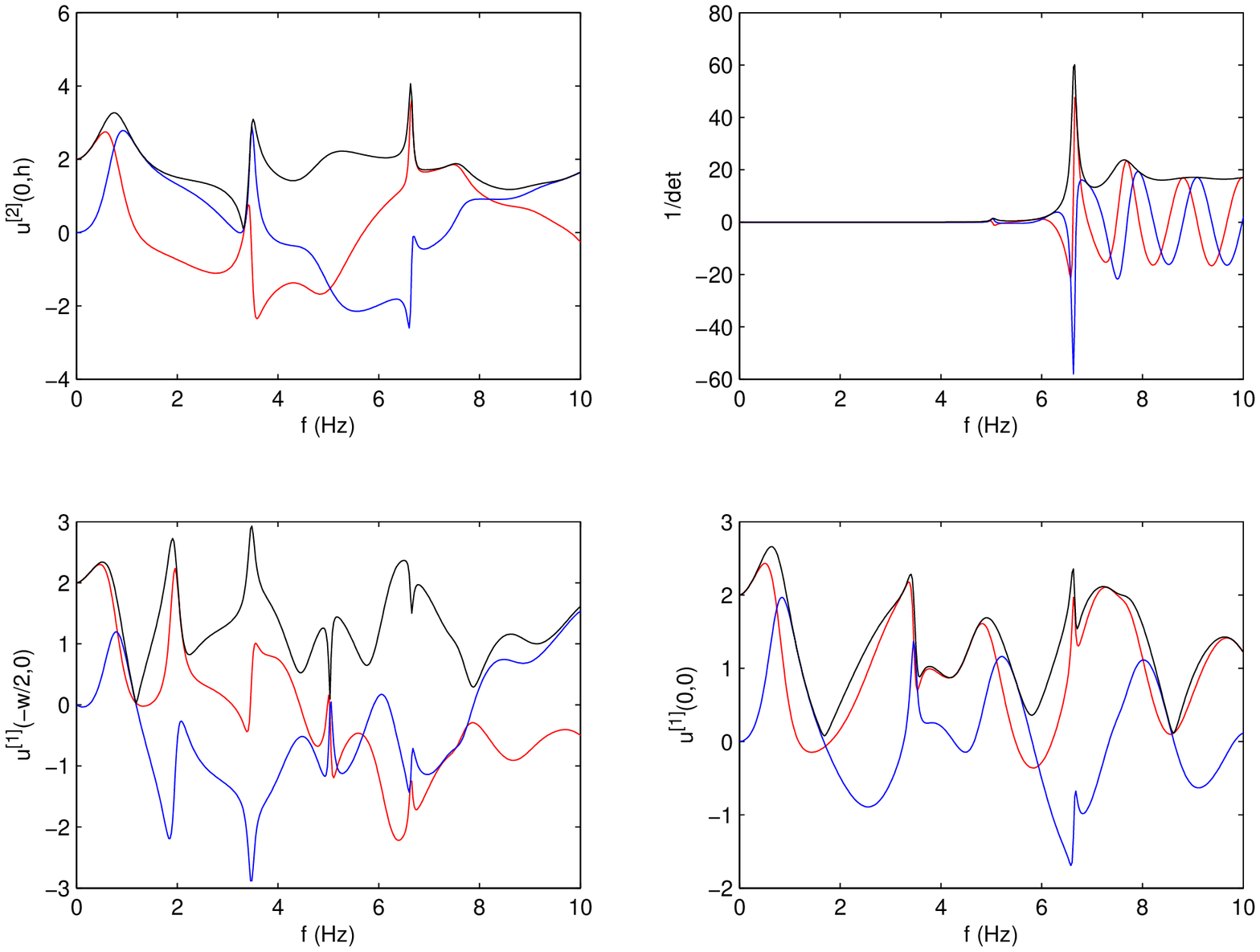}
\caption{$\theta^{i}=30^{\circ}$. Otherwise same as fig. \ref{sillsD-010} except that    $\mathbf{M=4}$.}
\label{sillsD-090}
\end{center}
\end{figure}

Figs. \ref{sillsD-050}-\ref{sillsD-090} depict our computed transfer functions $T^{(M)}(x,y;f)=u(x,y;f)/a^{i}(\omega)$ and $1/D^{(M)}(\omega)$ for  incident angle $\theta^{i}=30^{\circ}$ and for various approximation orders $M$.
\clearpage
\newpage
%%%%%%%%%%%%%%%%%%%%%%%%%%%%%%%%%%%%%%%%%
\subsubsection{Our computed transfer functions and $1/D$ for $\theta^{i}=60^{\circ}$}
\begin{figure}[ht]
\begin{center}
\includegraphics[width=0.64\textwidth]{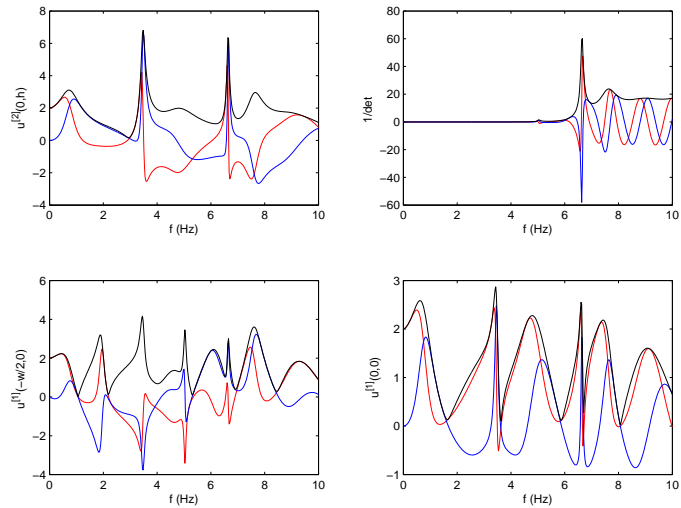}
\caption{$\theta^{i}=60^{\circ}$. Otherwise, same as fig. \ref{sillsD-010} except that    $\mathbf{M=4}$.}
\label{sillsD-100}
\end{center}
\end{figure}

Figs. \ref{sillsD-100}-\ref{sillsD-090} depict our computed transfer functions $T^{(M)}(x,y;f)=u(x,y;f)/a^{i}(\omega)$ and $1/D^{(M)}(\omega)$ for  incident angle $\theta^{i}=60^{\circ}$ and  approximation order $M=4$.
\clearpage
\newpage
%%%%%%%%%%%%%%%%%%%%%%%%%%%%%%%%%%%%%%%%%
\subsubsection{Discussion}

Note the rapid stabilization of the transfer functions as $M$ is increased, this setting in more rapidly at low than at higher frequencies. Note the resemblance (position and height) of our first peak of the summit response to the one of Sills. This agreement is somewhat less true concerning the first peak of the left corner response, and the disagreement increases for both the summit and left corner  succeeding peaks.
Note that the shape of the $\|1/D(f)\|$ (particularly evident for the first and second) peaks  is lorentzian and thus evocative of resonances. Moreover, the positions of the peaks of $\|1/D(f)\|$ in these figures do not vary with incident angle, which fact is another characteristic of surface shape resonances. We also find that the $m$-th peak does not appear  in $\|1/D^{(M)}(f)\|$ for $M<m$ and its position does not vary for $M\ge m$. Thus, it seems reasonable to assume that Sills' peaks of summit and corner response are due to the excitation of surface shape resonances. This will be further substantiated (for our rectangular Sills-like hill) further on.
%%%%%%%%%%%%%%%%%%%%%%%%%%%%%%%%%%%%%%%%%%%%%%%%%%%%%%%%%%%%%%%%%%%%%%%%%%%%%%%%%%%%%%%%%%%%%%%%%%%%%%%%%%%%%%%%%%%%%%%%%
\subsection{Beyond Sills: seismic response within the rectangular version of Sills' hill at the first five resonant frequencies}
Recall that Sills' graphs relate to the response   at locations exclusively {\it on the stress-free boundary} of the hill and ground. In fact, practically all studies (both theoretical/numerical and empirical) of seismic response of hills and mountains are of the same nature (i.e., they do not pertain to the field {\it  within} the convex feature, which is much harder to measure than the field on the protuberance surface) and they either ignore, or do not make clear, the relation of amplified response to the occurrence of resonances. As we shall discover hereafter, a look at the field within the convex feature, at various frequencies that we are sure are resonance frequencies, will make it clear that {\it amplified response is indeed associated with coupling to surface shape resonances}.

\begin{figure}[ht]
\begin{center}
\includegraphics[width=0.64\textwidth]{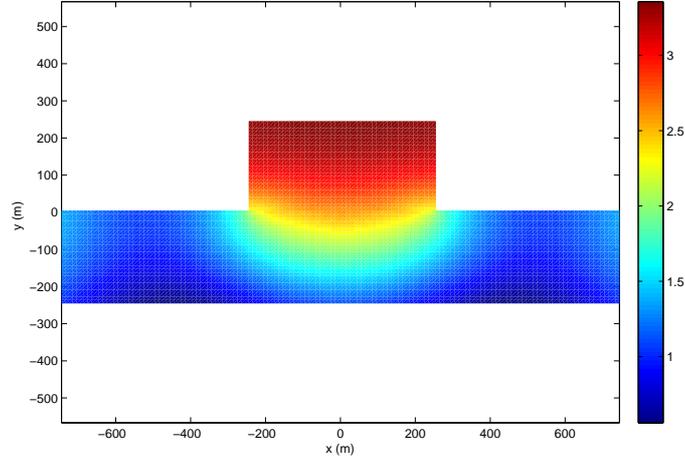}
\caption{Map of the modulus of $T^{(5)}(x,y;\mathbf{f=0.7592~Hz})$ for $\theta^{i}=0^{\circ}$. $\mathbf{d}^{(5)}=\{
   2.2835 + 2.4412i,~
   0.0000 - 0.0000i,~
  -0.0072 - 0.0136i,~
  -0.0000 + 0.0000i,~
  -0.0001 - 0.0002i,~
   0.0000 - 0.0000i\}$.}
\label{silldisp-010}
\end{center}
\end{figure}

\begin{figure}[ptb]
\begin{center}
\includegraphics[width=0.64\textwidth]{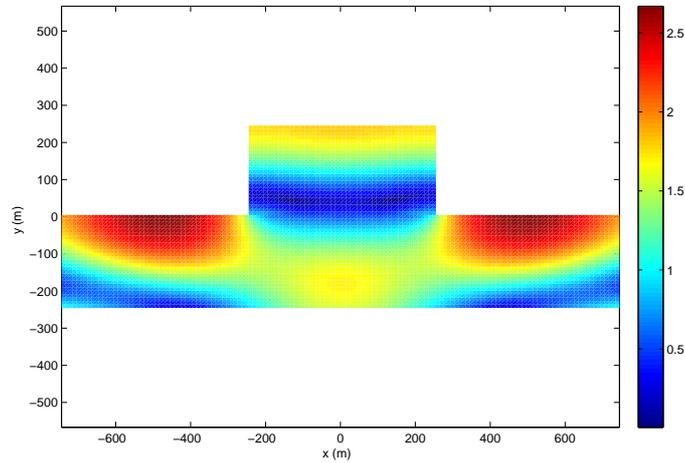}
\caption{Map  of the modulus of $T^{(5)}(x,y;\mathbf{f=1.958~Hz})$ for $\theta^{i}=0^{\circ}$. $\mathbf{d}^{(5)}=\{-0.8614 + 1.5748i,~
  -0.0000 + 0.0000i,~
   0.0575 - 0.0027i,~
  -0.0000 - 0.0000i,~
   0.0004 - 0.0001i,~
   0.0000 - 0.0000i\}$.}
\label{silldisp-020}
\end{center}
\end{figure}

\begin{figure}[ptb]
\begin{center}
\includegraphics[width=0.64\textwidth]{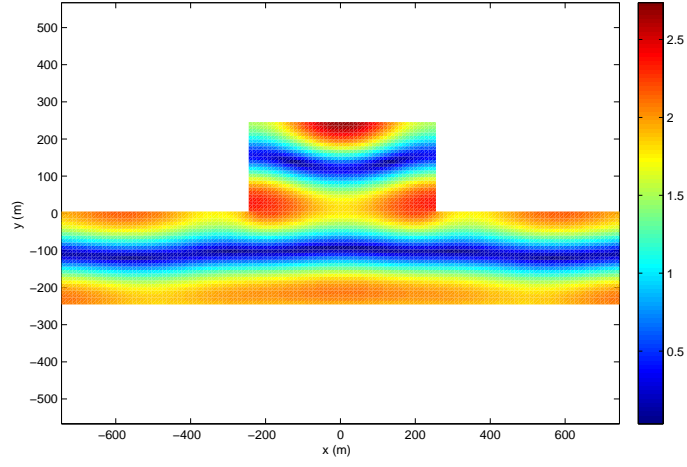}
\caption{Map  of the modulus of $T^{(5)}(x,y;\mathbf{f=3.482~Hz})$ for $\theta^{i}=0^{\circ}$. $\mathbf{d}^{(5)}=\{-2.0516 - 0.4222i,~
  -0.0000 - 0.0000i,~
   0.5326 + 0.4933i,~
  -0.0000 - 0.0000i,~
  -0.0003 - 0.0003i,~
  -0.0000 - 0.0000i\}$.}
\label{silldisp-030}
\end{center}
\end{figure}

\begin{figure}[ptb]
\begin{center}
\includegraphics[width=0.64\textwidth]{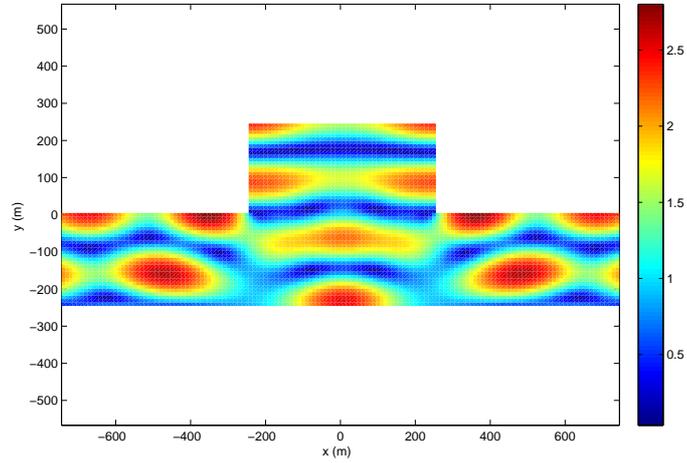}
\caption{Map  of the modulus of $T^{(5)}(x,y;\mathbf{f=5.030~Hz})$ for $\theta^{i}=0^{\circ}$. $\mathbf{d}^{(5)}=\{
0.4552 - 1.9221i,~
   0.0000 + 0.0000i,~
   0.4901 - 0.3430i,~
  -0.0000 - 0.0000i,~
   0.0118 - 0.0024i,~
   0.0000 + 0.0000i\}$.}
\label{silldisp-040}
\end{center}
\end{figure}

The graphs (figs. \ref{sillsD-010}-\ref{sillsD-100}) of $\|1/D^{(4)}(f)\|$ showed that the first five resonant frequencies for this hill are: $0.7592~Hz$, $1.958~Hz$, $3.482~Hz$, $5.030~Hz$, $6.628~Hz$.
In figs. \ref{silldisp-010}-\ref{silldisp-160} we display the computed displacement field within this hill at these four frequencies for incident angles $\theta^{i}=0,~30,~60,~80^{\circ}$. In the caption of each figure, we also give the corresponding first six entries of the diffraction coefficient vector $\mathbf{d}$.
Recall that: $\beta^{[1]}=\beta^{[0]}=1629.4~ms^{-1}$ and $\mu^{[1]}=\mu^{[0]}=6.85~MPa$. $h_{1}=250~m$,  $h_{2}=0~m$,  $w=500~m$, and we shall take $M=5$.
%%%%%%%%%%%%%%%%%%%%%%%%%%%%%%%%%%%%%%%%%%%%%%%%%%%%%%%
\subsubsection{Displacement field graphs for $\theta^{i}=0^{\circ}$}
\begin{figure}[ptb]
\begin{center}
\includegraphics[width=0.64\textwidth]{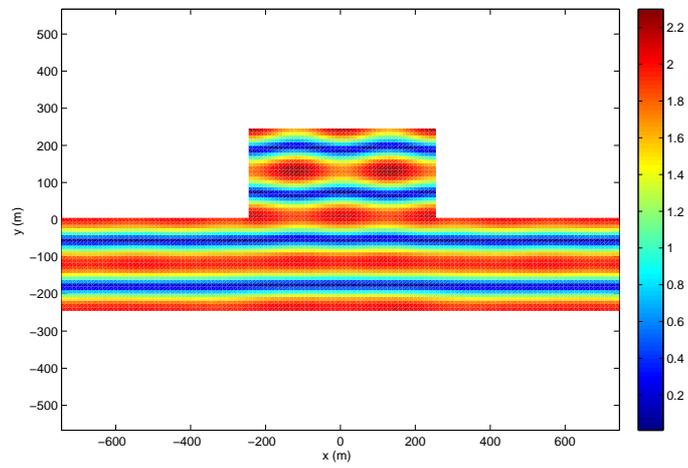}
\caption{Map  of the modulus of $T^{(5)}(x,y;\mathbf{f=6.628~Hz})$ for $\theta^{i}=0^{\circ}$. $\mathbf{d}^{(5)}=\{
   2.0058 + 0.2015i,~
   0.0000 - 0.0000i,~
   0.0197 - 0.0380i,~
   0.0000 + 0.0000i,~
   0.2787 + 0.0818i,~
  -0.0000 - 0.0000i\}$.}
\label{silldisp-050}
\end{center}
\end{figure}
\clearpage
\newpage
\noindent These figures show that the resonant coupling to (and therefore the amplification of) the fields within the hill is rather weak at all five resonant frequencies when the incident angle is $\theta^{i}=0^{\circ}$. This can be appreciated by the recollection of the fact  that the value of the transfer function on the stress-free boundary of flat ground (i.e., in the absence of the protuberance) is $2$.

Other features of these figures are that: 1) the field is maximal on the top, and independent of $x$ throughout most of the protuberance, at the lowest-frequency resonance, this being in agreement with the previous discussion on the $M=0$ approximation, 2) the field is progressively-more inhomogeneous as the resonance frequency increases, except (see fig. \ref{silldisp-050}) when something similar to the VS conditions are  satisfied, 3) at all but the second and fourth of these figures, the field can be greater at locations outside of the protuberance, than within and on the top of, the protuberance. This is in agreement with some empirically-observed results \cite{cc02} in environments wherein the hill is actually a building.
%%%%%%%%%%%%%%%%%%%%%%%%%%%%%%%%%%%%%%%%%%%%%%%%%%%%%%%
\subsubsection{Displacement field graphs for $\theta^{i}=30^{\circ}$}
\begin{figure}[ht]
\begin{center}
\includegraphics[width=0.64\textwidth]{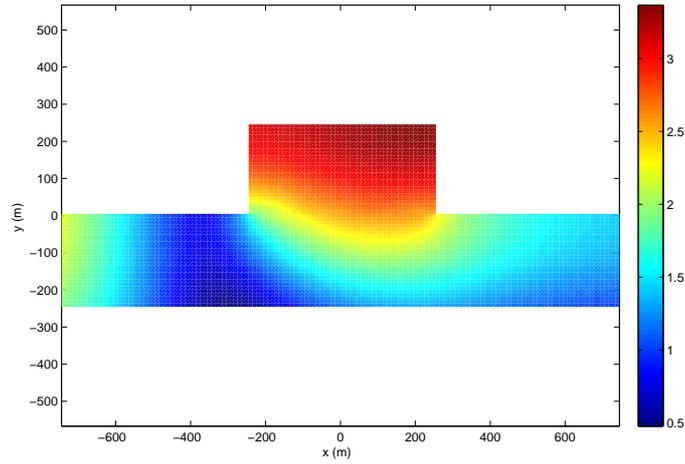}
\caption{Map  of the modulus of $T^{(5)}(x,y;\mathbf{f=0.7592~Hz})$ for $\theta^{i}=30^{\circ}$. $\mathbf{d}^{(5)}=\{
   2.2254 + 2.3731i,~
  -0.0101 - 0.1672i,~
  -0.0097 - 0.0133i,~
  -0.0000 - 0.0008i,~
  -0.0001 - 0.0002i,~
  -0.0000 - 0.0000i\}$.}
\label{silldisp-060}
\end{center}
\end{figure}
\begin{figure}[ptb]
\begin{center}
\includegraphics[width=0.64\textwidth]{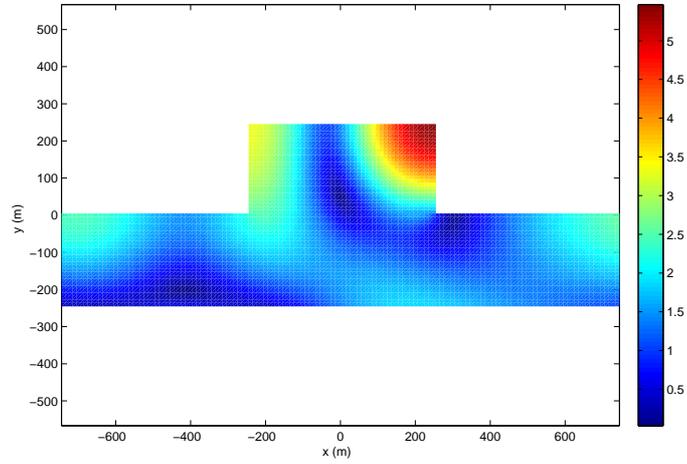}
\caption{Map  of the modulus of $T^{(5)}(x,y;\mathbf{f=1.958~Hz})$ for $\theta^{i}=30^{\circ}$. $\mathbf{d}^{(5)}=\{
-0.6914 + 1.3477i,~
   4.0190 - 1.4750i,~
   0.0195 - 0.0010i,~
  -0.0041 - 0.0000i,~
   0.0001 - 0.0001i,~
  -0.0001 + 0.0000i\}$.}
\label{silldisp-070}
\end{center}
\end{figure}
\begin{figure}[ptb]
\begin{center}
\includegraphics[width=0.64\textwidth]{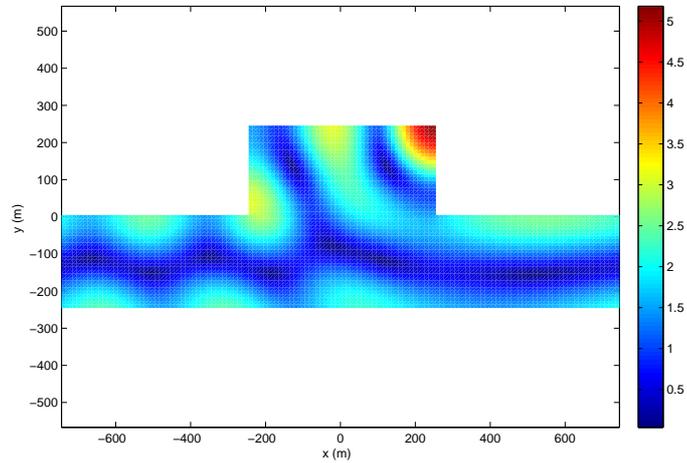}
\caption{Map  of the modulus of $T^{(5)}(x,y;\mathbf{f=3.482~Hz})$ for $\theta^{i}=30^{\circ}$. $\mathbf{d}^{(5)}=\{
-1.0119 - 0.0611i,~
   0.3314 + 1.9177i,~
  -0.2862 - 2.9526i,~
   0.0033 - 0.0018i,~
  -0.0007 + 0.0009i,~
   0.0000 - 0.0000i\}$.}
\label{silldisp-080}
\end{center}
\end{figure}
\begin{figure}[ptb]
\begin{center}
\includegraphics[width=0.64\textwidth]{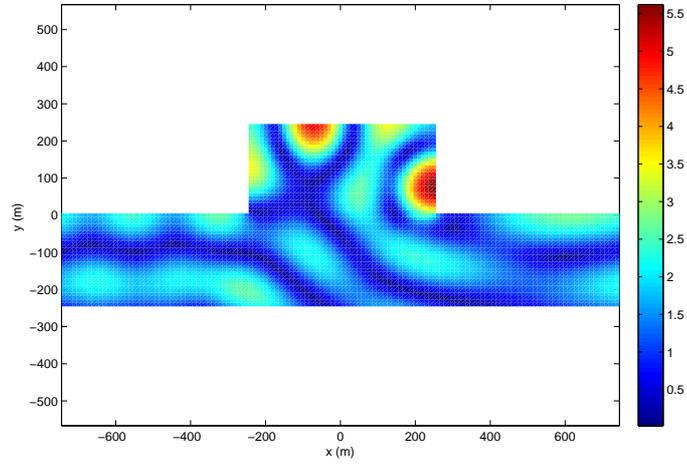}
\caption{Map  of the modulus of $T^{(5)}(x,y;\mathbf{f=5.030~Hz})$ for $\theta^{i}=30^{\circ}$. $\mathbf{d}^{(5)}=\{
0.0318 - 0.4632i,~
  -2.4289 - 0.2308i,~
   1.5591 + 1.0492i,~
   2.3066 + 1.5719i,~
  -0.0008 + 0.0054i,~
  -0.0006 - 0.0002i\}$.}
\label{silldisp-090}
\end{center}
\end{figure}
\begin{figure}[ptb]
\begin{center}
\includegraphics[width=0.64\textwidth]{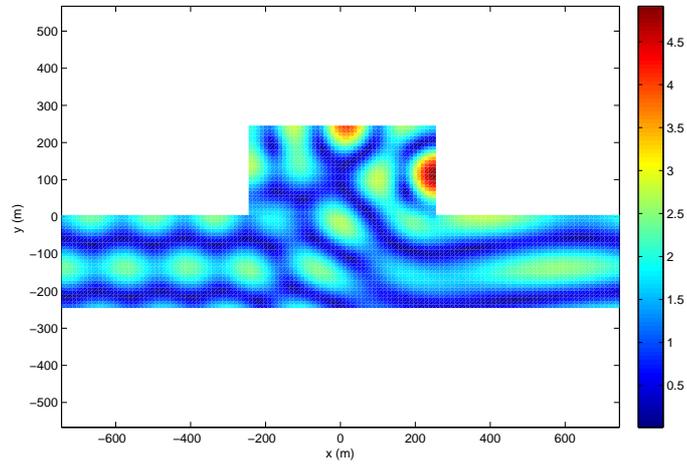}
\caption{Map  of the modulus of $T^{(5)}(x,y;\mathbf{f=6.628~Hz})$ for $\theta^{i}=30^{\circ}$. $\mathbf{d}^{(5)}=\{
0.0229 - 0.1060i,~
  -0.3427 - 1.5837i,~
  -1.6540 + 1.0562i,~
   1.2068 - 0.1370i,~
   1.7564 - 1.0051i,~
   0.0033 + 0.0049i\}$.}
\label{silldisp-100}
\end{center}
\end{figure}
\clearpage
\newpage
\noindent These figures show that the resonant coupling to the fields at certain 'hot spots' within the hill is fairly-strong at all five resonant frequencies when the incident angle is $\theta^{i}=30^{\circ}$. However, at other locations within the protuberance, the displacement field can be smaller than at certain locations in the underground. Thus, when we speak of amplified motion (with respect to the ground motion) in the protuberance at resonance, it should be understood that this amplification does not systematically occur at all locations within the protuberance and for all locations on the ground. Moreover, at the higher resonant frequencies, the field at most locations within the protuberance is much greater than that at most locations on the ground and within the underground.
%%%%%%%%%%%%%%%%%%%%%%%%%%%%%%%%%%%%%%%%%%%%%%%%%%%%%%%%%%%%%%%%%%%%%%%%%%
\subsubsection{Displacement field graphs for $\theta^{i}=60^{\circ}$}
\begin{figure}[ht]
\begin{center}
\includegraphics[width=0.64\textwidth]{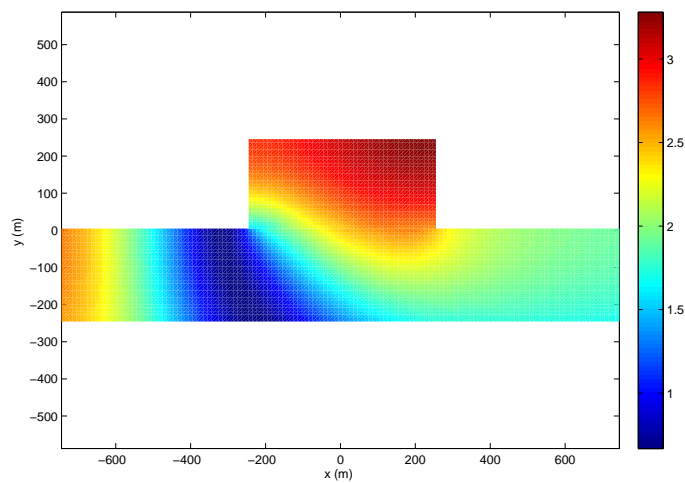}
\caption{Map  of the modulus of $T^{(5)}(x,y;\mathbf{f=0.7592~Hz})$ for $\theta^{i}=60^{\circ}$. $\mathbf{d}^{(5)}=\{
2.1117 + 2.2398i,~
  -0.0169 - 0.2818i,~
  -0.0145 - 0.0126i,~
  -0.0001 - 0.0012i,~
  -0.0002 - 0.0002i,~
  -0.0000 - 0.0000i\}$.}
\label{silldisp-110}
\end{center}
\end{figure}
\begin{figure}[ptb]
\begin{center}
\includegraphics[width=0.64\textwidth]{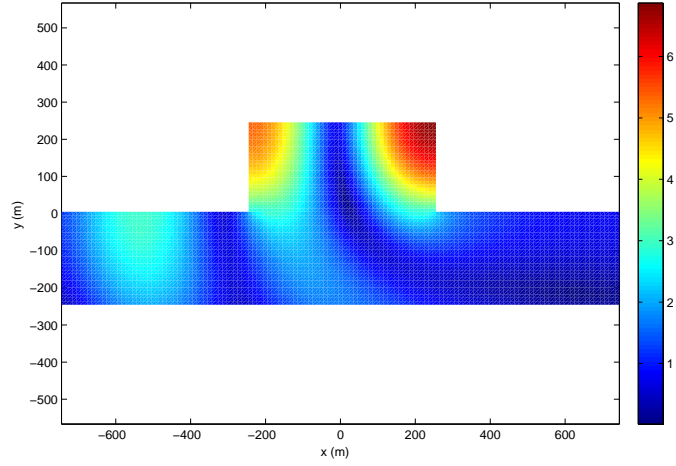}
\caption{Map  of the modulus of $T^{(5)}(x,y;\mathbf{f=1.958~Hz})$ for $\theta^{i}=60^{\circ}$. $\mathbf{d}^{(5)}=\{
  -0.4013 + 0.9526i,~
   5.7206 - 2.1027i,~
  -0.0441 + 0.0019i,~
  -0.0059 + 0.0023i,~
  -0.0002 - 0.0001i,~
  -0.0001 + 0.0000i\}$.}
\label{silldisp-120}
\end{center}
\end{figure}
\begin{figure}[ptb]
\begin{center}
\includegraphics[width=0.64\textwidth]{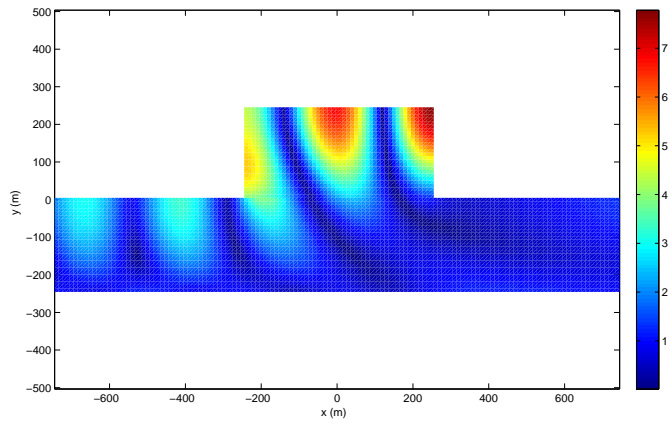}
\caption{Map  of the modulus of $T^{(5)}(x,y;\mathbf{f=3.482~Hz})$ for $\theta^{i}=60^{\circ}$. $\mathbf{d}^{(5)}=\{
0.2285 + 0.3301i
   0.4251 + 1.6318i,~
  -1.1174 - 6.3388i,~
   0.0057 + 0.0276i,~
   0.0002 + 0.0021i,~
   0.0000 + 0.0002i\}$.}
\label{silldisp-130}
\end{center}
\end{figure}
\begin{figure}[ptb]
\begin{center}
\includegraphics[width=0.64\textwidth]{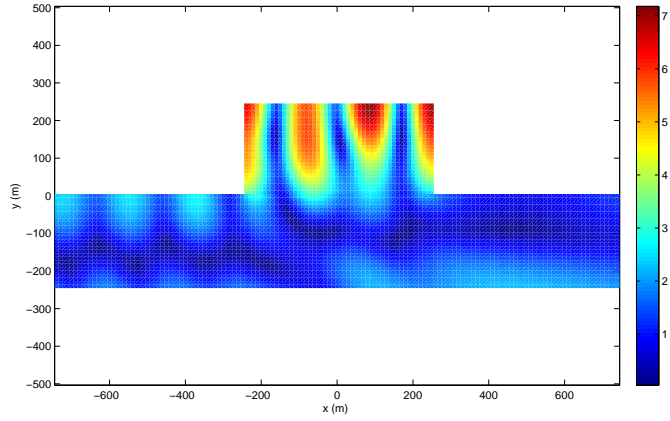}
\caption{Map  of the modulus of $T^{(5)}(x,y;\mathbf{f=5.030~Hz})$ for $\theta^{i}=60^{\circ}$. $\mathbf{d}^{(5)}=\{
 -0.1428 + 0.4549i,~
   0.1120 + 0.0833i,~
   1.5944 + 1.2412i,~
  -5.4215 + 3.4319i,~
   0.0099 + 0.0060i,~
   0.0005 - 0.0001i\}$.}
\label{silldisp-140}
\end{center}
\end{figure}
\begin{figure}[ptb]
\begin{center}
\includegraphics[width=0.64\textwidth]{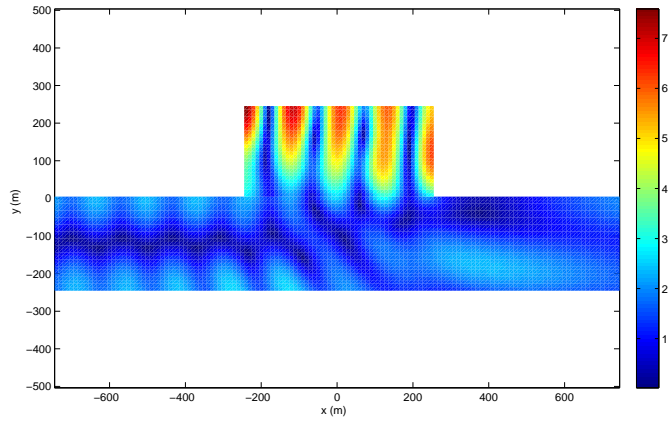}
\caption{Map  of the modulus of $T^{(5)}(x,y;\mathbf{f=6.628~Hz})$ for  $\theta^{i}=60^{\circ}$. $\mathbf{d}^{(5)}=\{
-0.0414 - 0.0074i,~
  -0.2068 + 0.5995i,~
  -0.1802 + 0.1227i,~
   1.9697 - 0.3537i,~
   4.1149 + 4.8242i,~
   0.0042 - 0.0013i\}$.}
\label{silldisp-150}
\end{center}
\end{figure}
\clearpage
\newpage
\noindent These figures show that the resonant coupling to the fields at  'hot columns' within the hill is  rather strong at all five resonant frequencies when the incident angle is $\theta^{i}=60^{\circ}$. Otherwise, the same comments as previously apply to this case.
%%%%%%%%%%%%%%%%%%%%%%%%%%%%%%%%%%%%%%%%%%%%%%%%%%%%%%%%%%%%%%%%%%%%%%%%%%
\subsubsection{Displacement field graphs for $\theta^{i}=80^{\circ}$}
\begin{figure}[ht]
\begin{center}
\includegraphics[width=0.64\textwidth]{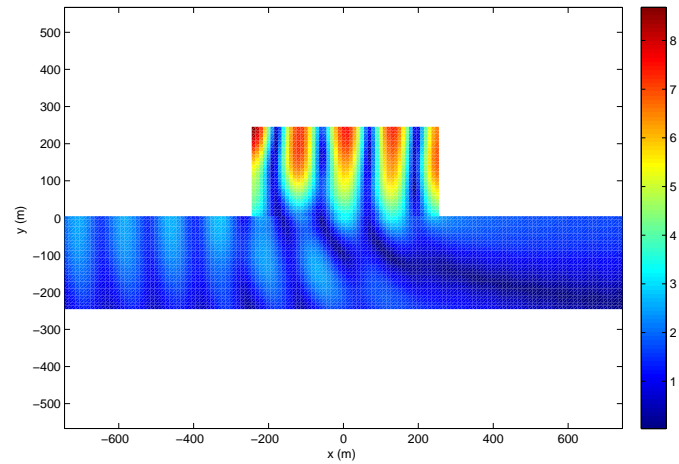}
\caption{Map  of the modulus of $T^{(5)}(x,y;\mathbf{f=6.628~Hz})$ for $\theta^{i}=80^{\circ}$.}
\label{silldisp-160}
\end{center}
\end{figure}
\clearpage
\newpage
\noindent This figure, and previous ones, show that the resonant coupling to the fields within the hill is  usually strongest at the highest resonant frequency (but this conclusion may change, as we shall see further on, in the presence of lossy media) and when the incident angle is the largest, i.e., $\theta^{i}=80^{\circ}$. Moreover, the field within the protuberance attains values, at the resonance frequencies, that are much larger than those found by Sills for a similar hill, this being probably due to the fact that the midpoint of the top segment and the left hand bottom corner are not usually the locations at which the field is at its maximum.
%%%%%%%%%%%%%%%%%%%%%%%%%%%%%%%%%%%%%%%%%%%%%%%%%%%%%%%%%%%%%%%%%%%%%%%%%%%%%%%%%%%%%%%%%%%%%%%%%%%%%%%%%%%%%%%%%
%Config of the Civita di Bagnoregio hill (Paolucci \cite{pa02})
%%%%%%%%%%%%%%%%%%%%%%%%%%%%%%%%%%%%%%%%%%%%%%%%%%%%%%%%%%%%%%%%%%%%%%%%%%%%%%%%%%%%%%%%%%%%%%%%%%%%%%%%%%%%%%%%%
\section{Simulation of the seismic response of the Civita di Bagnoregio hill}
The Civita di Bagnoregio hill \cite{ca13} provides a testing ground for our rectangular ridge model, all the more so than the seismic response of a somewhat similar (to our) protuberance has been previously simulated, via a spectral element numerical scheme, by Paolucci \cite{pa02}.

The first problem in connection with SH wave simulations is to assign values to the shear modulus $\mu$ (or the mass density $\rho$) and the real and imaginary parts of shear wavespeed $\beta$. In \cite{pa02}, Paolucci writes, concerning his computations for the Civita di Bagnoregio hill, that the constitutive parameters of the tuff material of which the hill, as well as of the half-space basement, are supposedly-composed,   are:
$\beta=600~ms^{-1}$ and $\nu=0.25$, for the shear wavespeed and the Poisson ratio respectively. Unfortunately, he does not give the  other key parameter, $\rho$ or  $\mu$, so that we were obliged to refer to another publication \cite{hb14} treating the subject of the mechanical properties of tuff. If we suppose that this tuff is dry,  the parameters given in \cite{hb14} are in the ranges: $\rho=1270-1330~Kgm^{-3}$, $\beta=1250-1280~ms^{-1}$, $\nu=0.28$, $\mu=1.99-2.19~GPa$. If, on the one hand, we retain the Paolucci value for $\beta$ and the  value for $\mu$ provided by the authors (Heap et al.) of \cite{hb14} then it follows that $\rho=5777.8~Kgm^{-3}$ which is clearly too large. If, on the other hand, we retain the Heap et al. values for $\rho$ and the Paolucci value for $\beta$ then it follows that $\mu=0.468~GPa$ which appears to be somewhat small.

Thus, our first choice was: $\beta=600~ms^{-1}$ and $\mu=0.668~GPa$. But this does not resolve the problem of the attenuation in the tuff. Since neither Paolucci nor Heap deal with this issue, we chose $\beta^{''}=-40~ms^{-1}$ so that $\beta=600-i40~ms^{-1}$. Actually, with smaller attenuations, the amplifications turned out to be larger than those of Paolucci.

Our second choice appealed essentially to the mean values of the parameters provided by Heap et al.: $\beta=1265~ms^{-1}$, $\mu=2.08~GPa$. Again, this does not resolve the attenuation issue, so we adopted the same relative attenuation as in the first choice: $\beta^{''}=-84~ms^{-1}$, whence $\beta=1265-i84~ms^{-1}$.

The second problem in connection with the SH seismic wave response of a structure such as a hill (assumed to be a protrusive body emerging from flat ground) is to define the shape and dimensions of the structure. Since our study has to do with ridge-like hills of rectangular shape (in the cross-section), the shape parameters are the width $w$ and height $h$ of the rectangle. The shape of the ridge-like Civita di Bagnoregio hill in \cite{pa02} is much more complicated than a simple rectangle, but there appears to exist a nearly-rectangular region near the top whose dimensions are $w=275~m$ and $h=50~m$, these being the geometrical parameters we adopted in our initial (i.e., first and second model) computations.

Last but not least, in conformity with what was assumed by Paolucci, we chose the seismic disturbance to be  a normally-incident plane wave (i.e., $\theta^{i}=0^{\circ}$), and both the basement and hill to be occupied by the same (macroscopically-homogeneous) material so that the aforementioned parameters apply throughout the scattering configuration (except for $\beta^{[0]''}$ which is taken to be zero).

Thus, we shall assume, for the first model, that: $\theta^{i}=0^{\circ}$, $h_{1}=0~m$, $h_{2}=50~m$, $w=275~m$, $\mu^{[0]}=0.668~GPa$, $\beta^{[0]}=600~ms^{-1}$,  $\mu^{[2]}=0.668~GPa$, $\beta^{[2]}=600-i40~ms^{-1}$.
%%%%%%%%%%%%%%%%%%%%%%%%%%%%%%%%%%%%%%%%
\subsection{First model results}
\begin{figure}[ht]
\begin{center}
\includegraphics[width=0.65\textwidth]{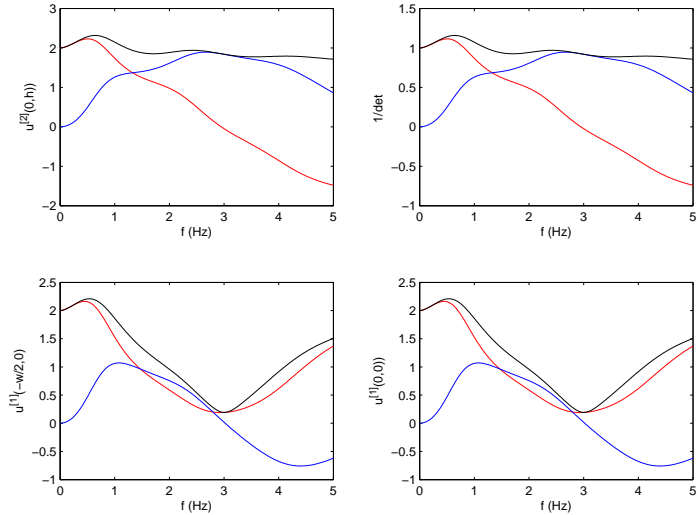}
\caption{The lower right panel is for $T^{(M)}(0,0;f)$, the lower left panel for $T^{(M)}(-w/2,0;f)$ and the upper left panel for $T^{(M)}0,h=h_{2};f)$ whereas the upper right panel depicts $1/D^{(M)}(\omega)$. The red curves are relative to the real part, the blue curves to the imaginary part and the black curves to the absolute value. Case $\theta^{i}=0^{\circ}$, $h_{1}=0~m$, $h_{2}=50~m$, $w=275~m$, $\mu^{[0]}=0.668~GPa$, $\beta^{[0]}=600~ms^{-1}$,  $\mu^{[2]}=0.668~GPa$, $\beta^{[2]}=600-i40~ms^{-1}$.  $\mathbf{M=0}$.}
\label{tf010}
\end{center}
\end{figure}

\begin{figure}[ptb]
\begin{center}
\includegraphics[width=0.65\textwidth]{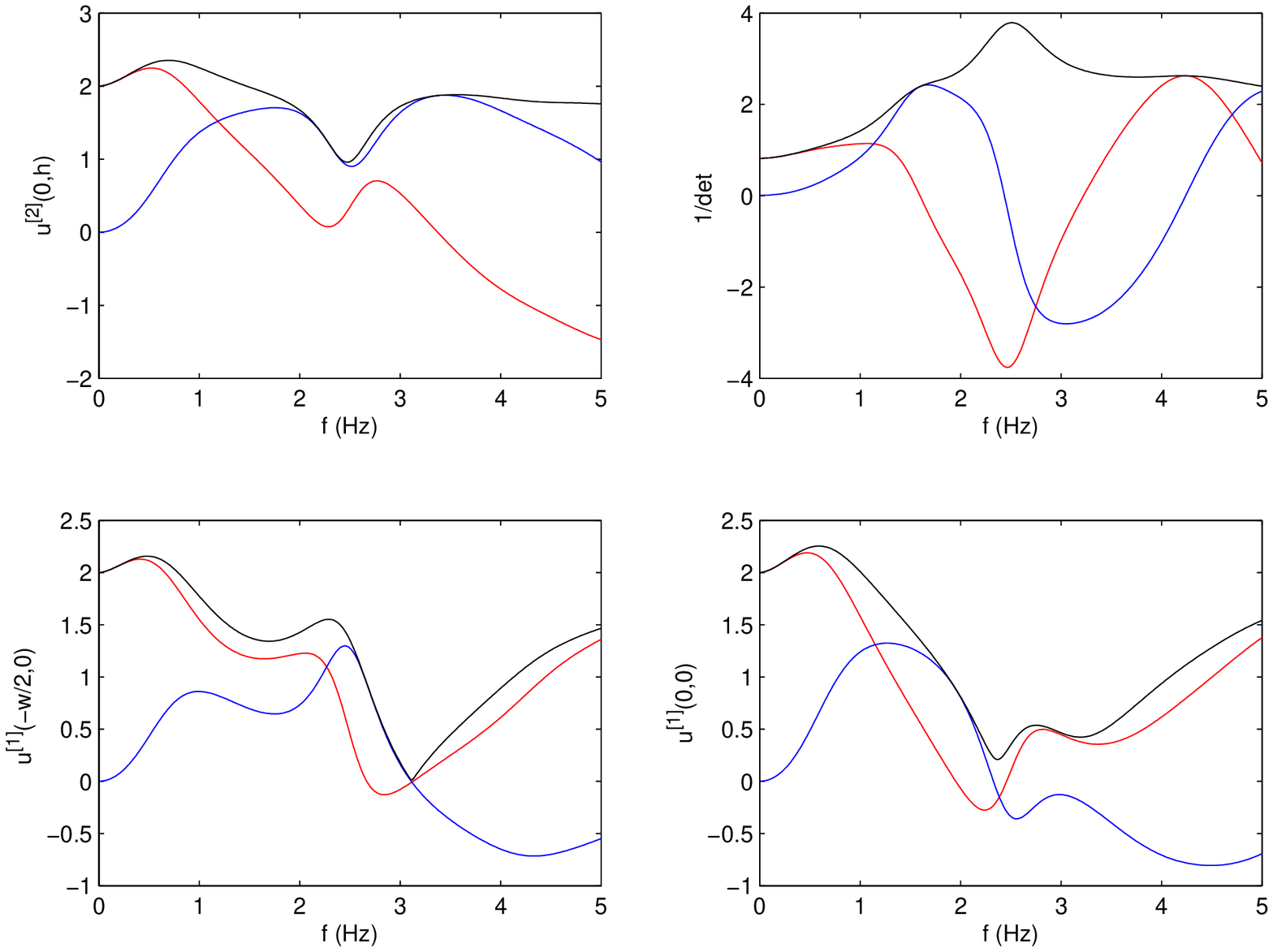}
\caption{$\theta^{i}=0^{\circ}$, $h_{1}=0~m$, $h_{2}=50~m$,  $w=275~m$, $\mu^{[0]}=0.668~GPa$, $\beta^{[0]}=600~ms^{-1}$, $\mu^{[1]}=0.668~GPa$, $\beta^{[1]}=600-i40~ms^{-1}$, $\mu^{[2]}=0.668~GPa$, $\beta^{[2]}=600-i40~ms^{-1}$. Same as fig. \ref{tf010} except that $\mathbf{M=2}$.}
\label{tf020}
\end{center}
\end{figure}

\begin{figure}[ptb]
\begin{center}
\includegraphics[width=0.65\textwidth]{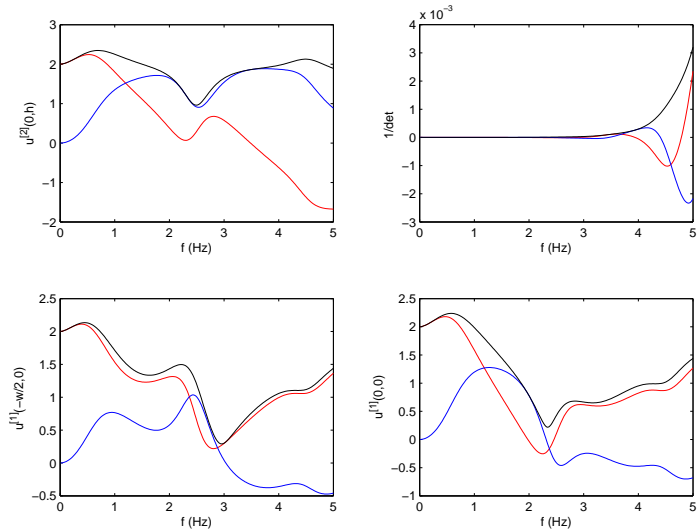}
\caption{$\theta^{i}=0^{\circ}$, $h_{1}=0~m$, $h_{2}=50~m$,  $w=275~m$, $\mu^{[0]}=0.668~GPa$, $\beta^{[0]}=600~ms^{-1}$, $\mu^{[1]}=0.668~GPa$, $\beta^{[1]}=600-i40~ms^{-1}$. Same as fig. \ref{tf010} except that  $\mu^{[2]}=0.668~GPa$, $\beta^{[2]}=600-i40~ms^{-1}$, $\mathbf{M=8}$.}
\label{tf030}
\end{center}
\end{figure}

The next step is to choose the number of terms $M+1$ in the modal representation of the hill. We found that the choice $M=8$ guaranteed that neither the first nine modal coefficients nor the displacement field  changed significantly for larger $M$ when the frequencies are near $1~Hz$. These coefficients (i.e., the first nine (for the first model at $f=1~Hz$) are:\\
$\mathbf{d} ^{(8)}=\{
   1.7924 + 1.2365i,~
   0.0000 - 0.0000i,~
  -0.0112 - 0.1229i,~
  -0.0000 + 0.0000i,~
  -0.0025 - 0.0102i,~
  -0.0000 - 0.0000i,~
  -0.0005 - 0.0016i,~
  -0.0000 + 0.0000i,~
  -0.0001 - 0.0003i\}$
which shows that the dominant mode is the $m=0$ mode, the odd-order mode coefficients are unsurprisingly nil (due to the assumption of normal incidence) and the other even-order modes die out rather rapidly as $m$ increases. The transfer functions, defined (in the same manner as in \cite{pa02}) as $T^{(M)}(x,y;f)=u^{(M)}(x,y;f)/a^{i}(\omega)$, are depicted in our figs. \ref{tf010}-\ref{tf030}  for $M=0,~2,~8$ respectively in which the lower right panel is for $T^{(M)}(0,0;f)$, the lower left panel for $T^{(M)}(-w/2,0;f)$ and the upper left panel for $T^{(M)}(0,h=h_{2};f)$. The upper right panel depicts $1/D^{(M)}(\omega)$, with $D^{(M)}$ the determinant of the $(M+1)-$by$-(M+1)$ matrix equation involved in the computation of the modal coefficient vector $\mathbf{d}^{(M)}$.
\clearpage
\newpage
In the upper right panels of figs. \ref{tf010}-\ref{tf030} we observe lorenztian behavior in $1/D(f)$, whose maxima indicate the frequencies of resonance (at which  a mode of the basement/hill configuration is excited) and we can notice that $f=1~Hz$ does not coincide with any such resonant frequency. Moreover,  contrary to what was written in \cite{pa02}, no obviously-resonant (i.e., maximal) response (notably at the summit) appears at $1~Hz$. Finally, our top transfer function has only a vague resemblance to those in fig. 5 of \cite{pa02}.

\begin{figure}[ht]
\begin{center}
\includegraphics[width=0.45\textwidth]{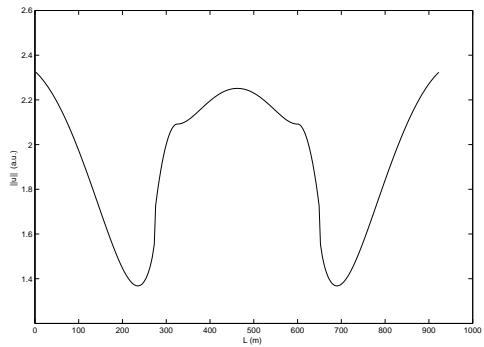}
\caption{Normalized total displacement field $\|u(x,y)/a^{i}\|$ along the stress-free boundary. $\theta^{i}=0^{\circ}$, $h_{1}=0~m$, $h_{2}=50~m$,  $w=275~m$, $\mu^{[0]}=0.668~GPa$, $\beta^{[0]}=600~ms^{-1}$, $\mu^{[1]}=0.668~GPa$, $\beta^{[1]}=600-i40~ms^{-1}$, $\mu^{[2]}=0.668~GPa$, $\beta^{[2]}=600-i40~ms^{-1}$, $\mathbf{M=8}$.}
\label{fsf010}
\end{center}
\end{figure}

In spite of this, we chose to depict in our fig.  \ref{fsf010}  the graph of the displacement field along the stress-free boundary at $f=1~Hz$,
\clearpage
\newpage
\begin{flushleft}wherein it can be observed that the field takes on values that do not exceed 2.3 (recall that this value is 2 in the absence of the hill), has the cosine-modulated shape of the zeroth-order mode on the upper face of the hill as one expects from the fact that this mode was found previously to be dominant at $1~Hz$, i.e., $u(x,h)\approx d_{0}+d_{2}\cos\big(k_{x2}(x+w/2)\big)~;~\|d_{0}\|>>\|d_{2}\|$,  and is even larger on portions of the ground than on the top of the hill (contrary to what is found in \cite{pa02}).\end{flushleft}
%\newpage
%%%%%%%%%%%%%%%%%%%%%%%%%%%%%%%%%%%%%%%%
\subsection{Second model results}
The parameters of the second model are: $\theta^{i}=0^{\circ}$, $h_{1}=0~m$, $h_{2}=50~m$,  $w=275~m$, $\mu^{[0]}=2.080~GPa$, $\beta^{[0]}=1265~ms^{-1}$, $\mu^{[2]}=2.080~GPa$, $\beta^{[2]}=1265-i84~ms^{-1}$.

Again, we found that $M=8$ provided stable numerical results for the modal coefficients and to be such, at $f=1~Hz$,  that the dominant mode is again the $m=0$ mode, the odd-order mode coefficients are nil (due to the assumption of normal incidence) and the other even-order modes die out  rapidly as $m$ increases.

\begin{figure}[ht]
\begin{center}
\includegraphics[width=0.65\textwidth]{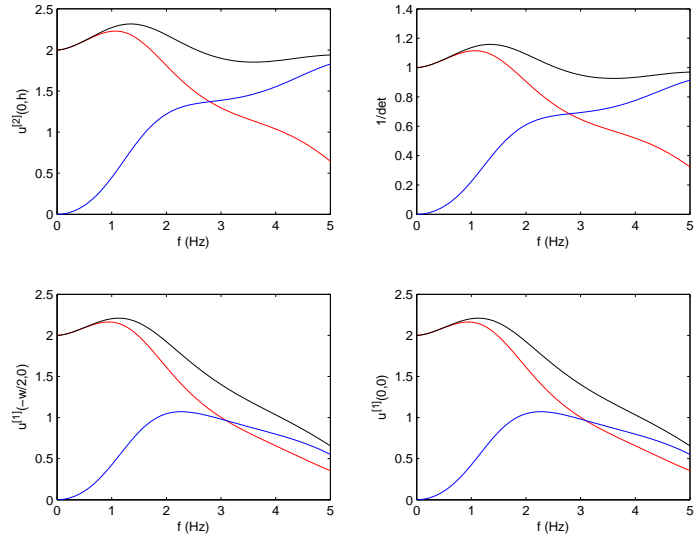}
\caption{The lower right panel is for $T^{(M)}(0,0;f)$, the lower left panel for $T^{(M)}(-w/2,0;f)$ and the upper left panel for $T^{(M)}0,h=h_{2};f)$ whereas the upper right panel depicts $1/D^{(M)}(\omega)$, with $D^{(M)}$ the determinant of the $(M+1)-$by$-(M+1)$ matrix equation involved in the computation of the modal coefficient vector $\mathbf{d}^{(M)}$. The red curves are relative to the real part, the blue curves to the imaginary part and the black curves to the absolute value. $\theta^{i}=0^{\circ}$, $h_{1}=0~m$, $h_{2}=50~m$,  $w=275~m$, $\mu^{[0]}=2.080~GPa$, $\beta^{[0]}=1265~ms^{-1}$, $\mu^{[2]}=2.080~GPa$, $\beta^{[2]}=1265-i84~ms^{-1}$. Same as fig. \ref{tf010} and  $\mathbf{M=0}$.}
\label{tf040}
\end{center}
\end{figure}

\begin{figure}[ht]
\begin{center}
\includegraphics[width=0.65\textwidth]{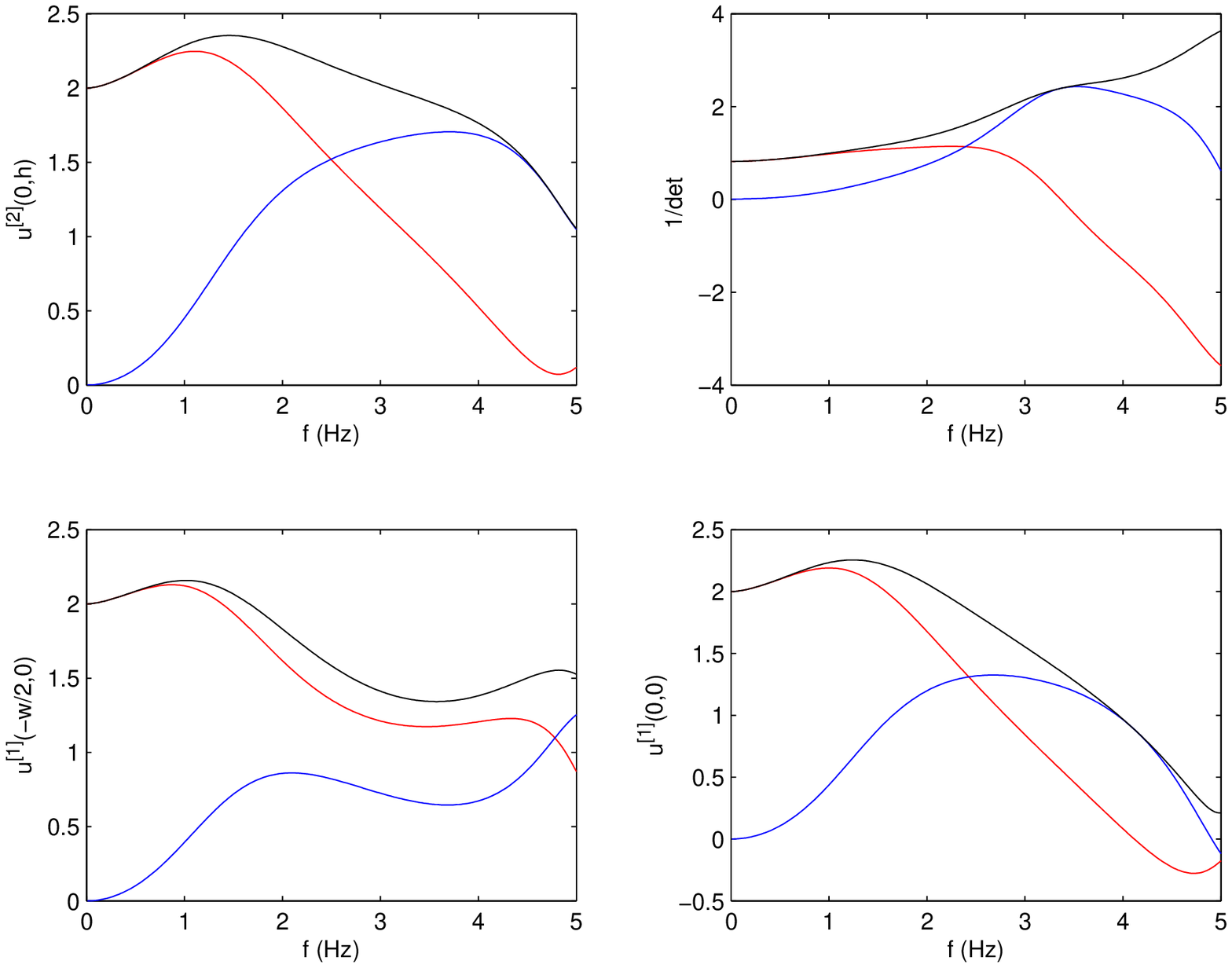}
\caption{$\theta^{i}=0^{\circ}$, $h_{1}=0~m$, $h_{2}=50~m$,  $w=275~m$, $\mu^{[0]}=2.080~GPa$, $\beta^{[0]}=1265~ms^{-1}$, $\mu^{[1]}=2.080~GPa$, $\beta^{[1]}=1265-i84~ms^{-1}$, $\mu^{[2]}=2.080~GPa$, $\beta^{[2]}=1265-i84~ms^{-1}$. Same as fig. \ref{tf040} except that $\mathbf{M=2}$.}
\label{tf050}
\end{center}
\end{figure}

\begin{figure}[ht]
\begin{center}
\includegraphics[width=0.65\textwidth]{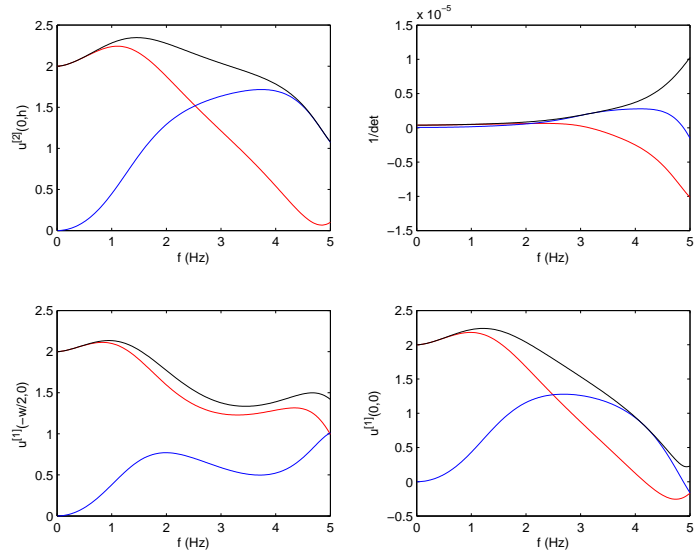}
\caption{$\theta^{i}=0^{\circ}$, $h_{1}=0~m$, $h_{2}=50~m$,  $w=275~m$, $\mu^{[0]}=2.080~GPa$, $\beta^{[0]}=1265~ms^{-1}$, $\mu^{[1]}=2.080~GPa$, $\beta^{[1]}=1265-i84~ms^{-1}$, $\mu^{[2]}=2.080~GPa$, $\beta^{[2]}=1265-i84~ms^{-1}$. Same as fig. \ref{tf040} except that $\mathbf{M=8}$.}
\label{tf060}
\end{center}
\end{figure}

  The $M=0,~2,~8$ transfer functions for this scattering configuration are exhibited in figs. \ref{tf040}-\ref{tf060}
\clearpage
\newpage
\noindent which, as concerns the top $T$, now  more closely resembles (although the maximum of the top $T$, as well as the first maximum of $1/D^{(M)}(\omega)$, are now situated beyond $1~Hz$) the corresponding transfer functions in fig. 5 of \cite{pa02}.

\begin{figure}[ht]
\begin{center}
\includegraphics[width=0.55\textwidth]{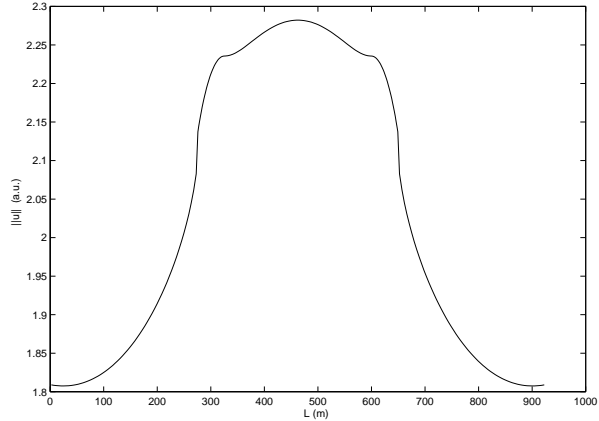}
\caption{Normalized total displacement field $\|u(x,y)/a^{i}\|$ along the stress-free boundary. $\theta^{i}=0^{\circ}$, $h_{1}=0~m$, $h_{2}=50~m$,  $w=275~m$, $\mu^{[0]}=2.080~GPa$, $\beta^{[0]}=1265~ms^{-1}$, $\mu^{[1]}=2.080~GPa$, $\beta^{[1]}=1265-i84~ms^{-1}$, $\mu^{[2]}=2.080~GPa$, $\beta^{[2]}=1265-i84~ms^{-1}$.   $\mathbf{\mathbf{M=0}}$.}
\label{fsf02}
\end{center}
\end{figure}
%
%\clearpage
%\newpage
The displacement field on the stress-free boundary (at $f=1~Hz$) is displayed in fig.  \ref{fsf02}  wherein it can be observed that the field: 1) again takes on values that do not exceed 2.3, 2) again has the cosinus-modulated shape of the zeroth-order mode on the upper face of the hill as one expects from the fact that this mode was found previously to be dominant at $1~Hz$, i.e., $u(x,h)\approx d_{0}+d_{2}\cos\big(k_{x2}(x+w/2)\big)~;~\|d_{0}\|>>\|d_{2}\|$, 3)  is now much smaller on the flanks at ground level than on the top of the hill as is observed to be the case in  \cite{pa02}). Note also that the field on the top is more constant than in the first model, this also being in agreement with what is observed in fig. 5 of \cite{pa02}.

To conclude this discussion, we have found that in order to obtain a result that compares favorably with the spectral element simulation of the seismic response of the somewhat irregularly-shaped hill in fig. 5 of \cite{pa02}, we had to choose, in our own simulation,  a rectangular hill whose base (and top)   is less wide and whose height is smaller than the hill of \cite{pa02}, and (for the second model) a shear wavespeed in the hill that is both complex (to include a fairly-substantial attenuation) and whose real part is double that of the one in \cite{pa02}. This attenuation has the  effect of smoothing out and reducing somewhat the height of the first broad resonance peak in the top transfer function, as well as reducing to an acceptable level the field on the top of the hill at $f=1~Hz$.  Our computation thus seems to substantiate the contention of Paolucci that the (actually modest)  field amplification at this frequency is indeed due to a resonance (broader  and whose maximum is at a higher frequency than Paolucci's) resonance at $f=1~Hz$). Our results provide, in addition, the indication that this resonance is due to the excitation of the zeroth-order mode of the (rectangular) hill.
%%%%%%%%%%%%%%%%%%%%%%%%%%%%%%%%%%%%%%%%
\subsection{Alternative (third) model results}
Paolucci mentions in his article that the  Civita di Bagnoregio hill was the site of a major earthquake on 11 June, 1695 that caused great damage to the town located on its plateau-like summit as well as landslides along its  flank(s), without notable damage having been reported in neighboring towns located beneath the hill (presumably at what we call ground level). With the relatively-modest motion of our second model, it is not easy to explain why this would cause major damage  at $1~Hz$, especially to  what one can guess (in 1695) are small buildings (4 to 8 stories) whose fundamental resonance frequencies are in the  $2.5-5~Hz$ range. Of course, we do not know what the dominant frequency of this earthquake was, but it cannot be excluded that it was in the $2.5-5~Hz$ range rather than near $1~Hz$. This suggests that a not-too-radical modification of our second hill model might give rise to a resonance in the $2.5-5~Hz$ range and thus explain why the buildings on the top of the hill are subjected to strong ground motion when the seismic solicitation is in this same frequency range.

Our modification consists in: a) increasing somewhat the height (this parameter is not easy to ascertain in fig.5 of \cite{pa02}) of the hill  to $h=h_{2}=65~m$, and b) hardening the underground medium without changing the hill medium so that the shear modulus and shear wavespeed in the underground are now $8~GPa$ and $2000~ms^{-1}$ respectively. It is not certain that the latter choice (amounting to a relatively-soft hill overlying a relatively-hard underground) is geologically-admissible, but the supposition that the underground medium be identical to the hill medium is also debatable.

Let us now see what the consequences of these changes are. The parameters of this third model are: $\theta^{i}=0^{\circ}$, $h_{1}=0~m$, $h_{2}=65~m$,  $w=275~m$, $\mu^{[0]}=8~GPa$, $\beta^{[0]}=2000~ms^{-1}$, $\mu^{[2]}=2.080~GPa$, $\beta^{[2]}=1265-i84~ms^{-1}$.

\begin{figure}[ht]
\begin{center}
\includegraphics[width=0.55\textwidth]{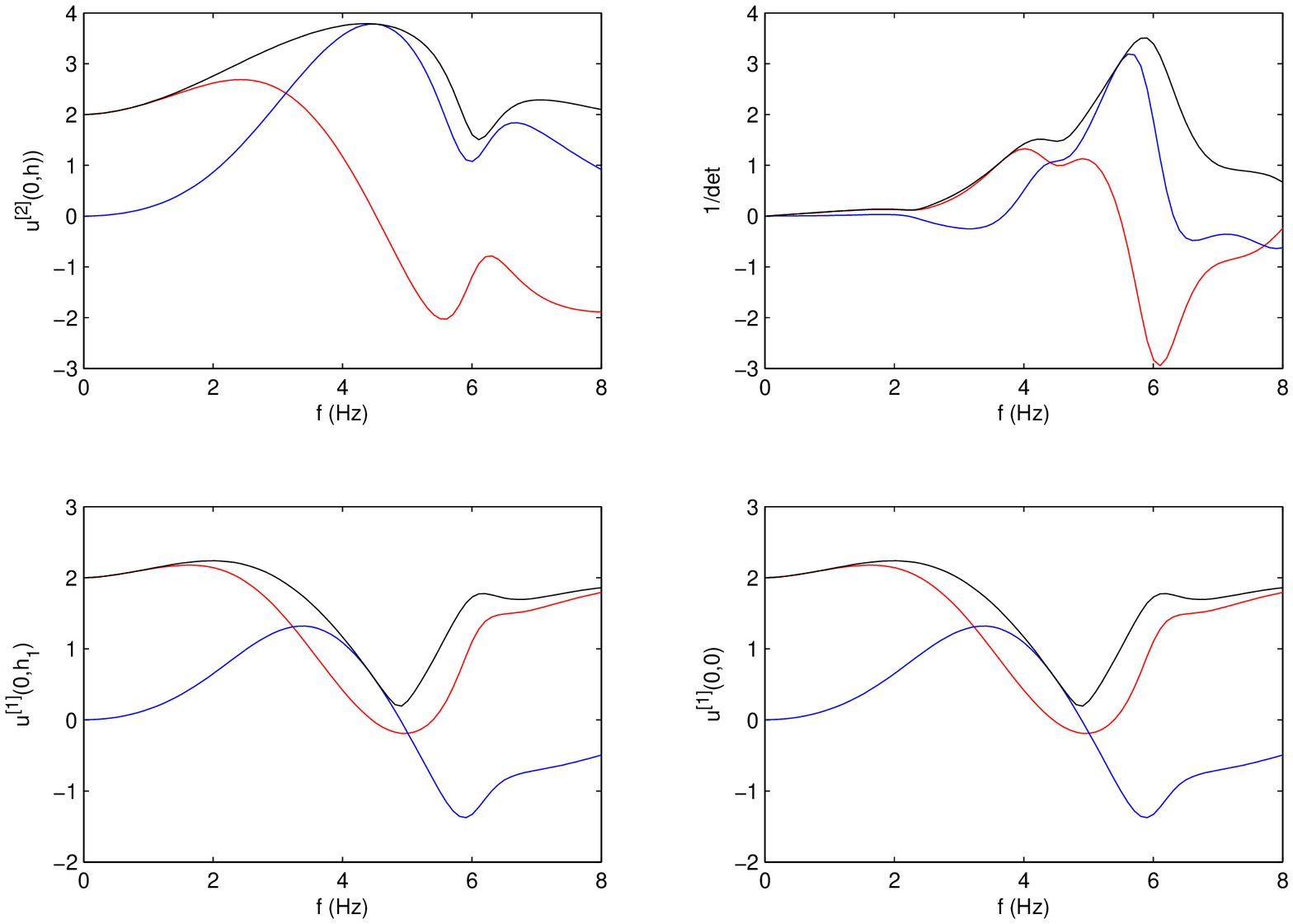}
\caption{The lower right panel is for $T^{(M)}(0,0;f)$, the lower left panel for $T^{(M)}(-w/2,0;f)$ and the upper left panel for $T^{(M)}0,h=h_{2};f)$ whereas the upper right panel depicts $1/\mathcal{D}^{(M)}(\omega)$, with $\mathcal{D}^{(M)}$ the determinant of the $(M+1)-$by$-(M+1)$ matrix equation involved in the computation of the modal coefficient vector $\boldsymbol{\mathcal{F}}^{(M)}$. The red curves are relative to the real part, the blue curves to the imaginary part and the black curves to the absolute value. $\theta^{i}=0^{\circ}$, $h_{1}=0~m$, $h_{2}=65~m$,  $w=275~m$, $\mu^{[0]}=8~GPa$, $\beta^{[0]}=2000~ms^{-1}$, $\mu^{[2]}=2.080~GPa$, $\beta^{[2]}=1265-i84~ms^{-1}$.   $M=4$. The resonances appear at: $f=4.2~Hz$.}
\label{tf070}
\end{center}
\end{figure}

\begin{figure}[ptb]
\begin{center}
\includegraphics[width=0.50\textwidth]{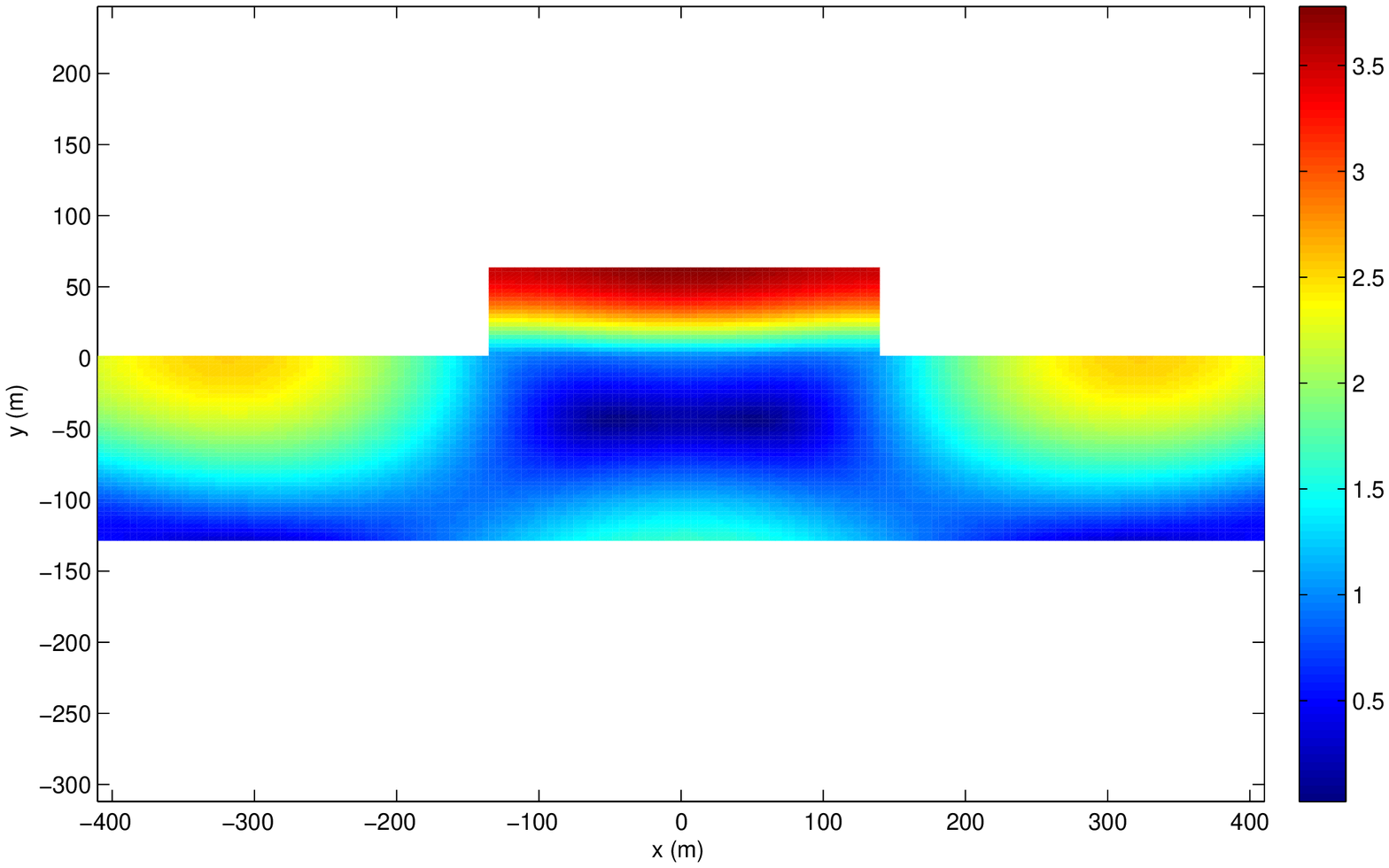}
\caption{Map of $\|T^{(M)}(x,y;4.2~Hz))\|$. $f=4.2~Hz$, $\theta^{i}=0^{\circ}$, $h_{1}=0~m$, $h_{2}=65~m$,  $w=275~m$, $\mu^{[0]}=8~GPa$, $\beta^{[0]}=2000~ms^{-1}$,  $\mu^{[2]}=2.080~GPa$, $\mathbf{\beta^{[2]}=1265-i84~ms^{-1}}$.   $\mathbf{\mathbf{M=4}}$.}
\label{fm03}
\end{center}
\end{figure}

\begin{figure}[ptb]
\begin{center}
\includegraphics[width=0.50\textwidth]{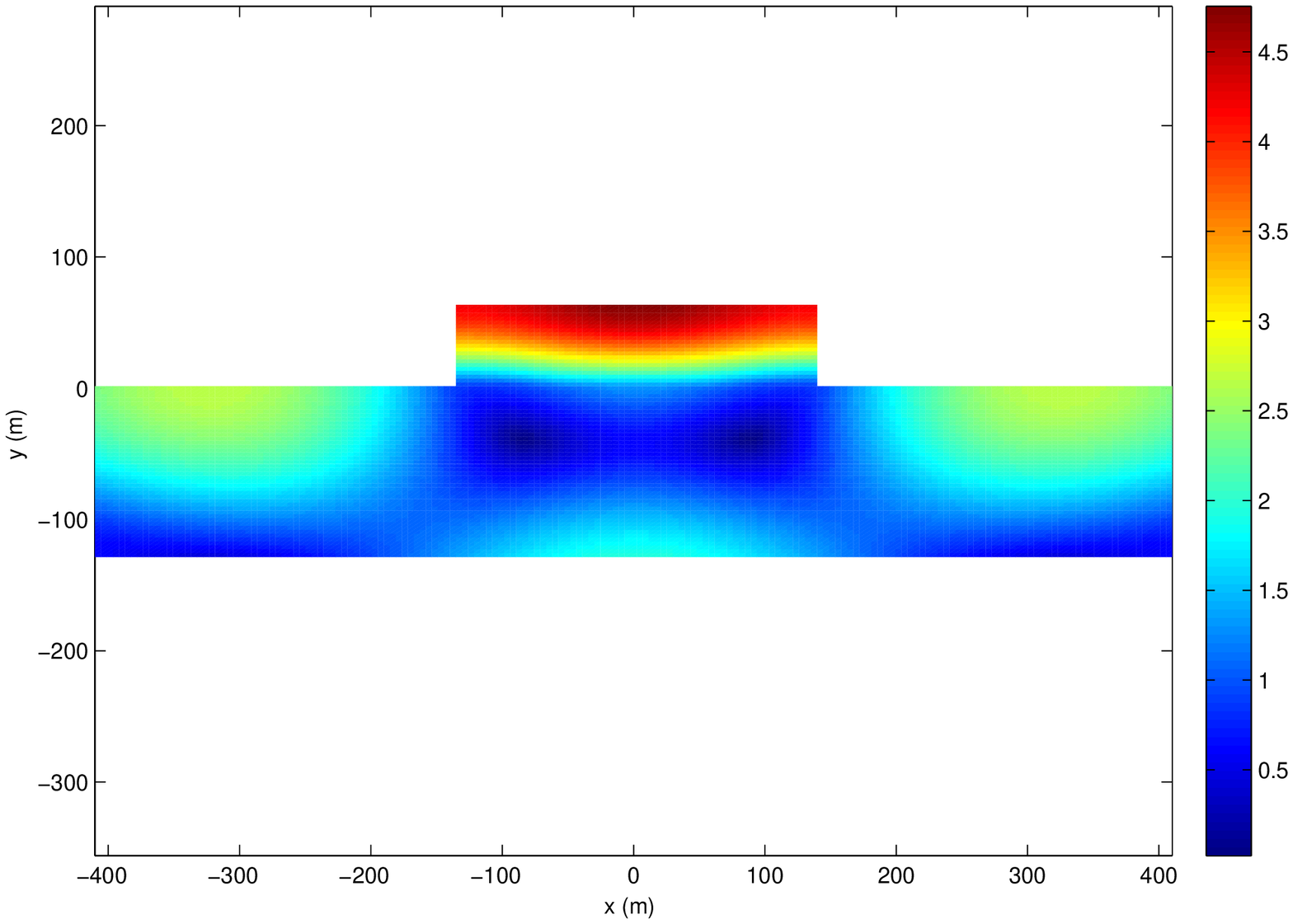}
\caption{Map of $\|T^{(M)}(x,y;4.2~Hz))\|$. $f=4.2~Hz$, $\theta^{i}=0^{\circ}$, $h_{1}=0~m$, $h_{2}=65~m$,  $w=275~m$, $\mu^{[0]}=8~GPa$, $\beta^{[0]}=2000~ms^{-1}$,  $\mu^{[2]}=2.080~GPa$, $\mathbf{\beta^{[2]}=1265-i0~ms^{-1}}$.  $M=4$.}
\label{fm04}
\end{center}
\end{figure}

Fig. \ref{tf070} depicts the transfer functions whereby we observe  that a broad maximum occurs for the top response centered at $f=4.2~Hz$.
\clearpage
\newpage
\clearpage
\newpage
Fig. \ref {fm03} depicts the displacement field map at this frequency wherein we observe that the top field has now increased from $2.3$ to $3.7$ (a.u.) over most of the plateau of the hill. If we suppress the attenuation  in the hill, the displacement field map looks like fig. \ref{fm04}.
\clearpage
\newpage
\noindent in which we observe that the maximum value of the field on the top increases from 3.7 to 4.7 (a.u), both these values  being possibly-sufficient to account for relatively strong motion (generally measured by a quantity proportional to $\|u\|^{2}$ \cite{bg07,wi18a}) and damage to small buildings located on the crest of the Civita hill caused by a seismic wave whose dominant frequency is $4.2~Hz$.

In fig. \ref{tf070} we also note the presence of a strong maximum of $1/\|\mathcal{D}\|$ at $f=5.9~Hz$ which suggests the existence of a resonance at this frequency as well, with the concurrent possibility of significant  amplification of the displacement field within the hill for an incident seismic wave whose dominant frequency is $f=5.9~Hz$.

\begin{figure}[ptb]
\begin{center}
\includegraphics[width=0.50\textwidth]{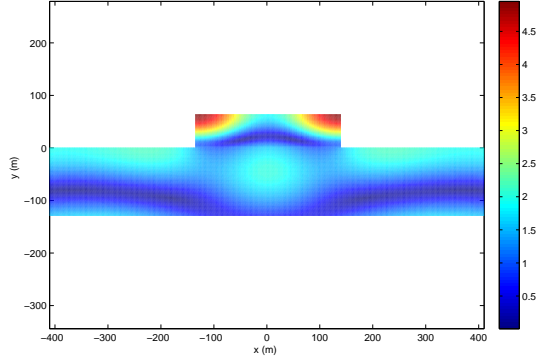}
\caption{Map of $\|T^{(M)}(x,y;5.9~Hz))\|$. $f=5.9~Hz$, $\mathbf{\theta^{i}=0^{\circ}}$, $h_{1}=0~m$, $h_{2}=65~m$,  $w=275~m$, $\mu^{[0]}=8~GPa$, $\beta^{[0]}=2000~ms^{-1}$,  $\mu^{[2]}=2.080~GPa$, $\beta^{[2]}=1265-i84~ms^{-1}$.  $M=8$. }
\label{fm05}
\end{center}
\end{figure}

Fig. \ref{fm05} tells us what the field distribution looks like at this frequency. Now the field is concentrated at the upper edges, which might explain the origin of the landslides observed in 1695, and at various occasions since then, affecting the Civita hill.

This shows that, at $f=5.9~Hz$, both the $m=0$ and $m=2$ modes are being excited (i.e., resonate). Now, instead of $u(x,h)\approx d_{0}+d_{2}\cos\big(k_{x2}(x+w/2)\big)~;~\|d_{0}\|>>\|d_{2}\|$, we have  $u(x,h)\approx d_{0}+d_{2}\cos\big(k_{x2}(x+w/2)\big)~;~\|d_{0}\|\approx\|d_{2}\|$ whence $u(\pm w/2,0)\approx d_{0}+d_{2}$ and $u(0,0)\approx d_{0}-d_{2}$, which on account of the signs of the real and imaginary parts of $d_{0}$ and $d_{2}$, accounts for the fact that the fields at the corners are maximal and the field at the center of the plateau is minimal at $f=5.9~Hz$.
\clearpage
\newpage
\noindent A look at the coefficients of the hill modes helps to understand why the field concentrates at the upper edge.
$\mathbf{d}^{(9)} =\{
  -1.9083 + 2.7632i,~
   0.0000 - 0.0000i,~
  -0.3842 + 1.6453i,~
  -0.0000 - 0.0000i,~
   0.0195 - 0.0017i,~
  -0.0000 - 0.0000i,~
   0.0017 - 0.0005i,~
  -0.0000 - 0.0000i,~
   0.0002 - 0.0001i,~
   0.0000 + 0.0000i\}$.

\begin{figure}[ht]
\begin{center}
\includegraphics[width=0.50\textwidth]{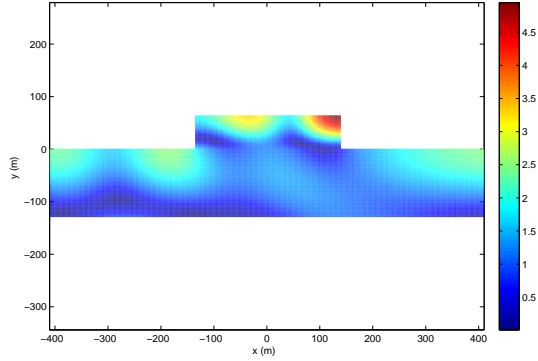}
\caption{Map of $\|T^{(M)}(x,y;5.9~Hz))\|$. $f=5.9~Hz$, $\mathbf{\theta^{i}=40^{\circ}}$, $h_{1}=0~m$, $h_{2}=65~m$,  $w=275~m$, $\mu^{[0]}=8~GPa$, $\beta^{[0]}=2000~ms^{-1}$,  $\mu^{[2]}=2.080~GPa$, $\beta^{[2]}=1265-i84~ms^{-1}$.  $M=8$. $N=200$.}
\label{fm06}
\end{center}
\end{figure}
%
%\clearpage
%\newpage
Actually, the landslides of the Civita hill, which continue until this day,  appear to affect mostly its eastern flank. This led us to hypothesize that most of the seismic disturbances in the Bagnoregio area come from the same source which is not directly (but far) beneath the hill but rather to the left (in our field map figures). If we suppose that this amounts to a seismic solicitation in the form of a plane SH wave having $\theta^{i}=40^{\circ}$ incidence, then the displacement field map has the appearance of fig. \ref{fm06}
\noindent wherein we observe that the region of the strongest motion is now concentrated at the  eastern edge of the plateau, thus possibly explaining why the landslides of this topographic feature take place predominantly on its eastern flank.
%
%%%%%%%%%%%%%%%%%%%%%%%%%%%%%%%%%%%%%%%%%%%%%%%%%%%%%%%%%%%%%%%%%%%%%%%%%%%%%%%%%%%%%%%%%%%%%%%%%%%%%%%%%%%%%%%%%%%%%%%%%%
\section{On the possibility of very-strong seismic response in a hill}\label{vssr}
The previous numerical results seemed to imply that the seismic displacement response in hills, entirely-filled with a solid that is either the same as, or different from, that of the underground, is systematically inferior to about four times the flat-ground response (the latter equals 2). In this section we show, via Figs. \ref{vsc-010}-\ref{fm17}, that this response can, in fact, be much stronger, even for a hill having a relatively-large aspect ratio $w/h=500~m/150~m=3.333$. {\it Now, we assume that the hill is entirely-filled with a solid  that is different from that of the underground}.
%%%%%%%%%%%%%%%%%%%%%%%%%%%%%%%%%%%%%%
\subsection{Without attenuation}
We first examine the case in which there are no material losses within and underneath the  hill. The parameters of the configuration are: $h_{1}=0~m$, $h_{2}=150~m$, $w=500~m$, $\mu^{[0]}=6.85~GPa$, $\beta^{[0]}=1629.4~ms^{-1}$,  $\mu^{[2]}=2~GPa$, $\beta^{[2]}=1000-i0~ms^{-1}$.
%%%%%%%%%%%%%%%%%%%%%%%%%%%%%%%%%%%%%%
\subsubsection{Search for the first four resonant frequencies}
\begin{figure}[ht]
\begin{center}
\includegraphics[width=0.65\textwidth]{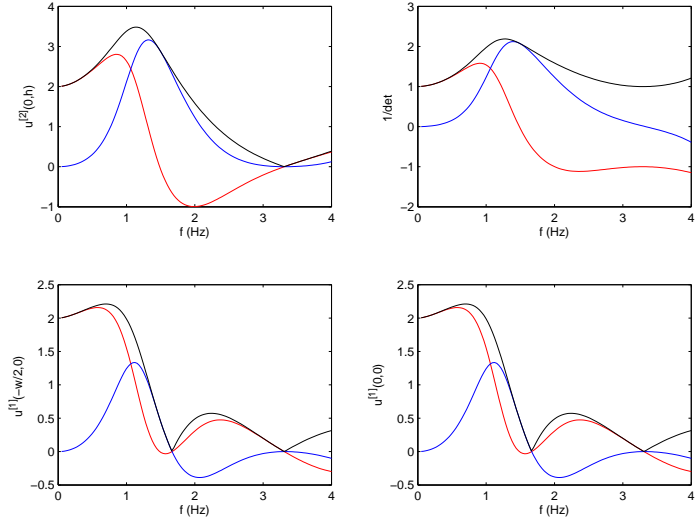}
\caption{The lower right panel is for $T^{(M)}(0,0;f)$, the lower left panel for $T^{(M)}(-w/2,0;f)$ and the upper left panel for $T^{(M)}0,h=h_{2};f)$ whereas the upper right panel depicts $1/D^{(M)}(\omega)$, with $D^{(M)}$ the determinant of the $(M+1)-$by$-(M+1)$ matrix equation involved in the computation of the modal coefficient vector $\mathbf{d}^{(M)}$. The red curves are relative to the real part, the blue curves to the imaginary part and the black curves to the absolute value. $\theta^{i}=80^{\circ}$, $h_{1}=0~m$, $h_{2}=150~m$, $w=500~m$, $\mu^{[0]}=6.85~GPa$, $\beta^{[0]}=1629.4~ms^{-1}$, $\mu^{[2]}=2~GPa$, $\beta^{[2]}=1000-i0~ms^{-1}$. $\mathbf{\mathbf{M=0}}$. }
\label{vsc-010}
\end{center}
\end{figure}

\begin{figure}[ptb]
\begin{center}
\includegraphics[width=0.65\textwidth]{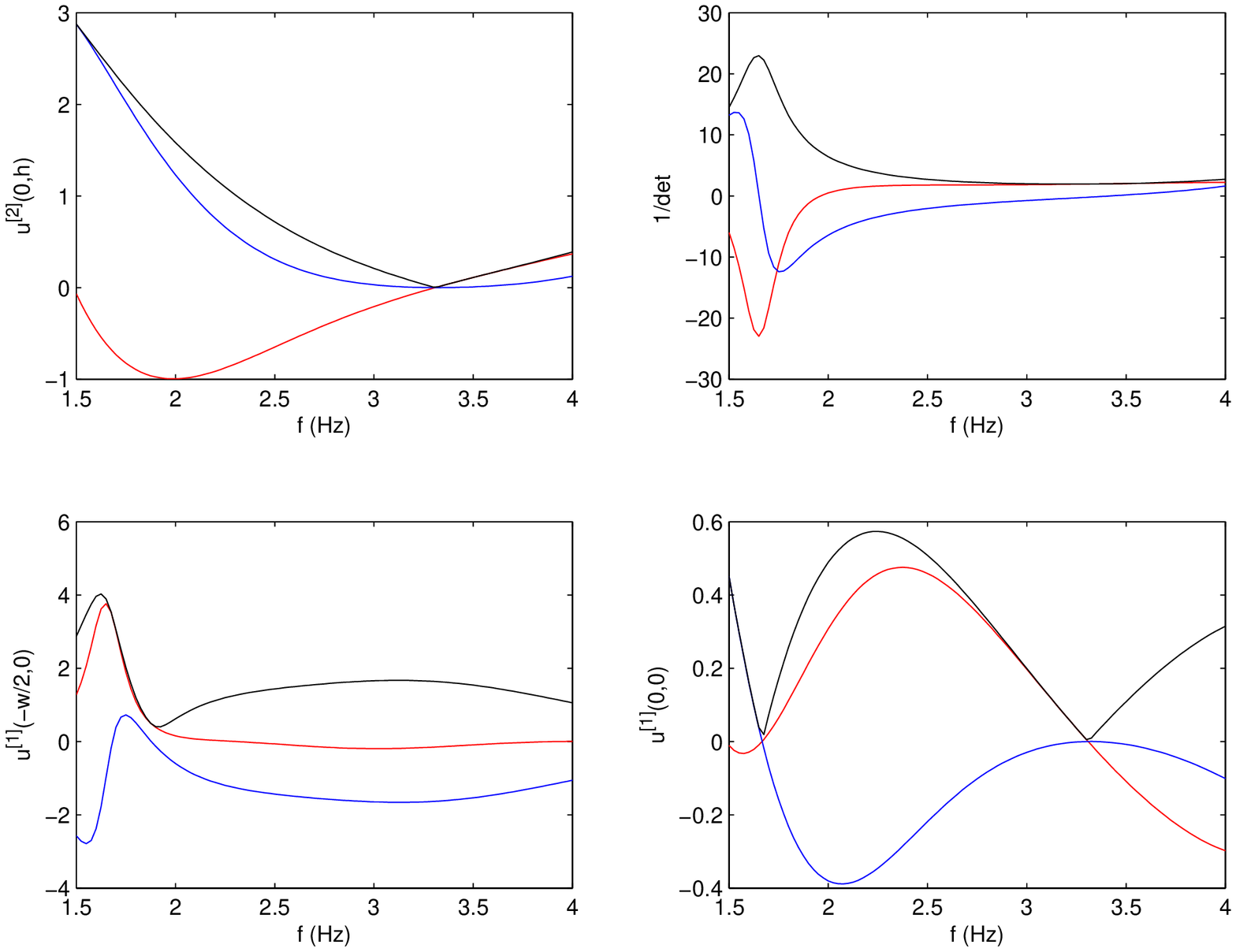}
\caption{Same as fig. \ref{vsc-010} except that  $\mathbf{\mathbf{M=1}}$.}
\label{vsc-020}
\end{center}
\end{figure}

\begin{figure}[ptb]
\begin{center}
\includegraphics[width=0.65\textwidth]{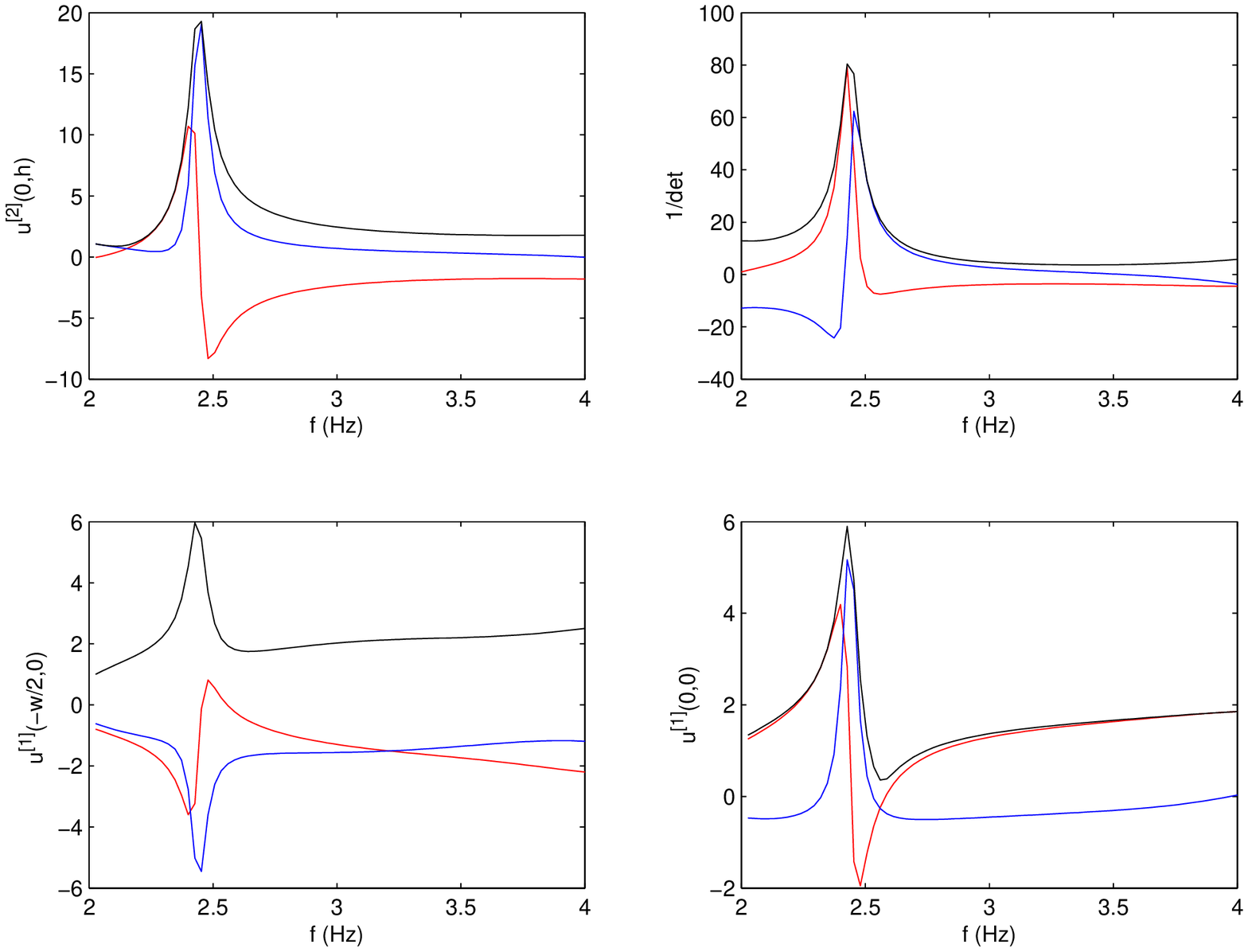}
\caption{Same as fig. \ref{vsc-010} except that  $\mathbf{\mathbf{M=2}}$.}
\label{vsc-030}
\end{center}
\end{figure}

\begin{figure}[ptb]
\begin{center}
\includegraphics[width=0.65\textwidth]{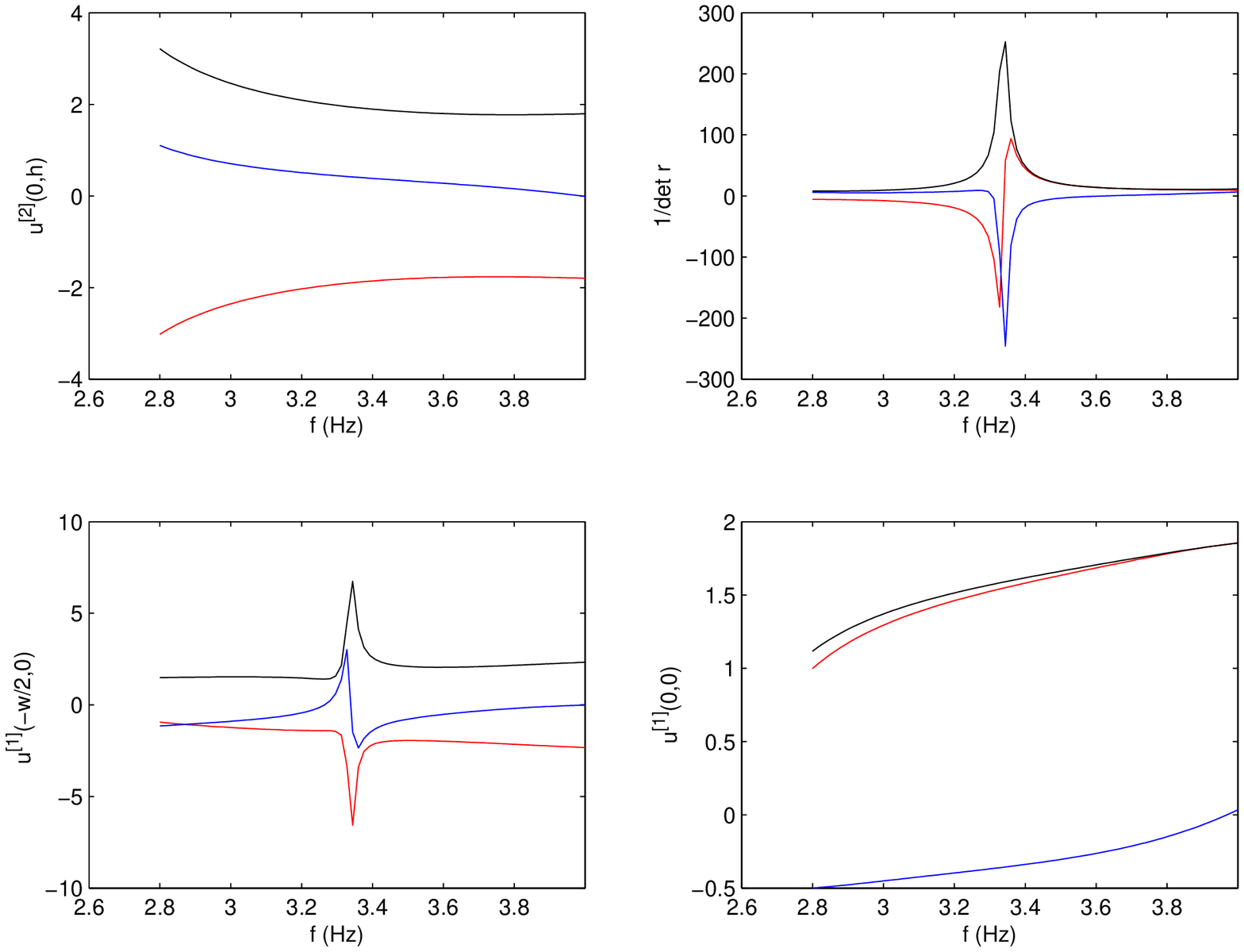}
\caption{Same as fig. \ref{vsc-010} except that  $\mathbf{\mathbf{M=3}}$.}
\label{vsc-040}
\end{center}
\end{figure}

Figs. \ref{vsc-010}-\ref{vsc-040} depict the transfer functions for $M=0,1,2,3$ whereby we spot the resonant frequencies.
\clearpage
\newpage
\noindent The  first four resonant frequencies are thus found to be: $1.28,~1.64,~2.44,~3.337~Hz$. Note that in fig. \ref{vsc-040} there appears no trace of resonant coupling to the response at the specific point $(0,h)$. As we shall see further on, this fact contrasts sharply with what fig. \ref{fm09} reveals as to the appearance of very hot spots on the hilltop at this same frequency and incident angle. {\it This finding underlines the danger of guessing the response at arbitrary points on and within the hill from its  response at a single point on the top its (stress-free) surface}.
%%%%%%%%%%%%%%%%%%%%%%%%%%%%%%%%%%%%%%
\subsubsection{Resonant coupling at $f=2.44~Hz$}
\begin{figure}[ht]
\begin{center}
\includegraphics[width=0.55\textwidth]{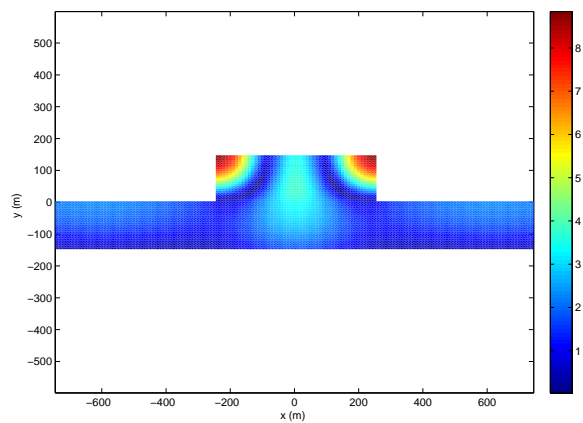}
\caption{Map of $T(x,y;f=2.44~Hz)$. $\mathbf{\theta^{i}=0^{\circ}}$, $h_{1}=0~m$, $h_{2}=150~m$, $w=500~m$, $\mu^{[0]}=6.85~GPa$, $\beta^{[0]}=1629.4~ms^{-1}$,  $\mu^{[2]}=2~GPa$, $\beta^{[2]}=1000-i0~ms^{-1}$.  $M=5$.}
\label{vsc-050}
\end{center}
\end{figure}

\begin{figure}[ht]
\begin{center}
\includegraphics[width=0.55\textwidth]{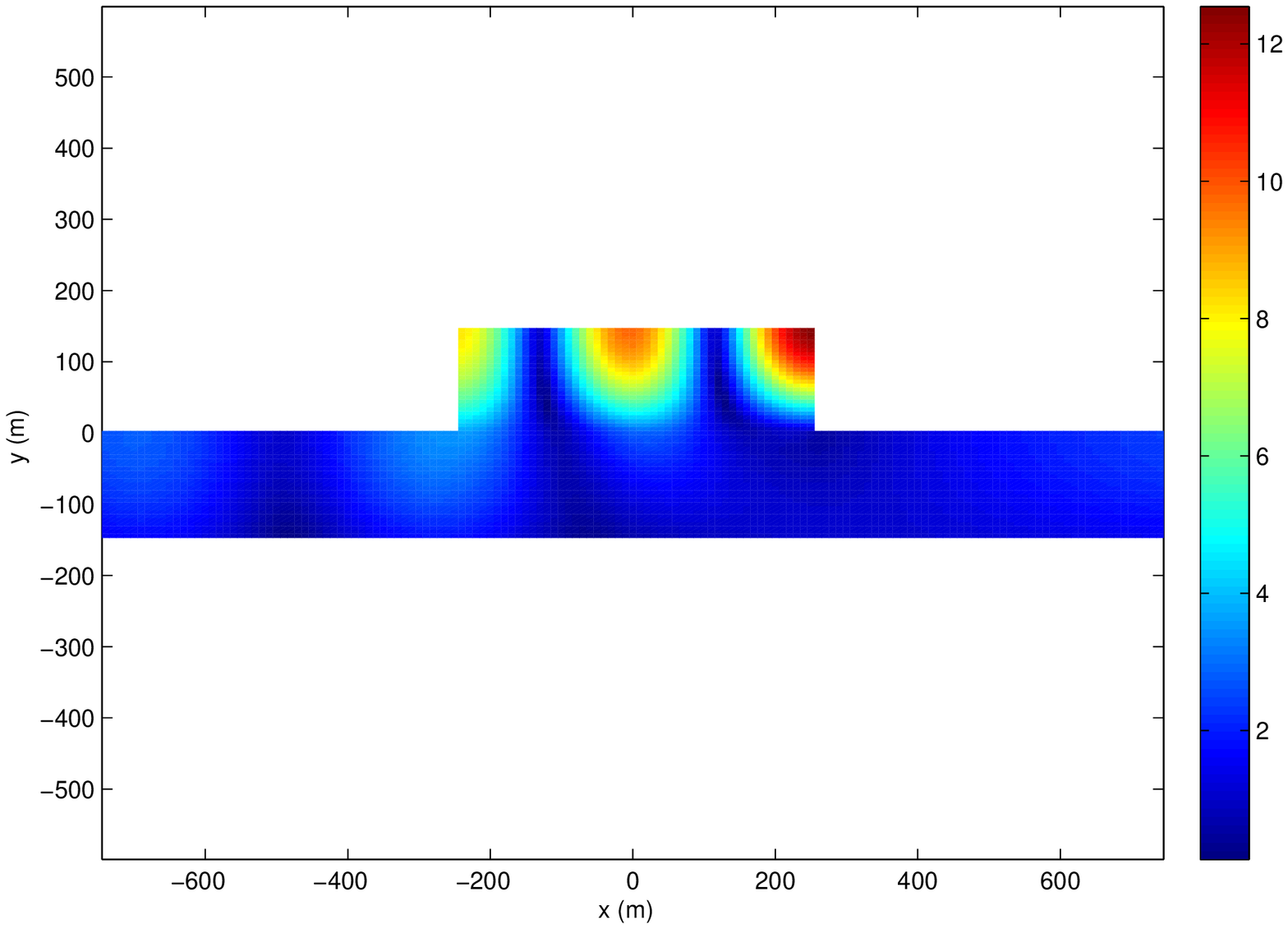}
\caption{Same as fig. \ref{vsc-050} except that $\mathbf{\theta^{i}=40^{\circ}}$.}
\label{vsc-060}
\end{center}
\end{figure}

Figs. \ref{vsc-050}-\ref{vsc-050} are field maps for three incident angles at $f=2.44~Hz$.
\begin{figure}[ht]
\begin{center}
\includegraphics[width=0.55\textwidth]{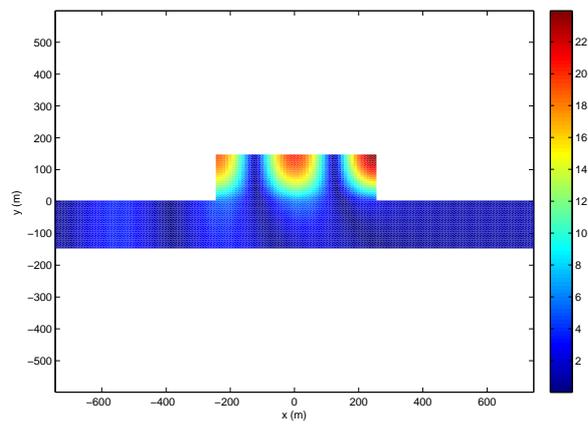}
\caption{Same as fig. \ref{vsc-050} except that $\mathbf{\theta^{i}=80^{\circ}}$.}
\label{vsc-070}
\end{center}
\end{figure}
\clearpage
\newpage
\noindent Notice the strong coupling at this resonant frequency for all incident angles. However the coupling increases with $\theta^{i}$ to attain a very large amount at three hot spots for $80^{\circ}$ incidence (in fig. \ref{vsc-070}).
%%%%%%%%%%%%%%%%%%%%%%%%%%%%%%%%%%%%%%
\subsubsection{Resonant coupling at $f=3.3374~Hz$}
\begin{figure}[ht]
\begin{center}
\includegraphics[width=0.55\textwidth]{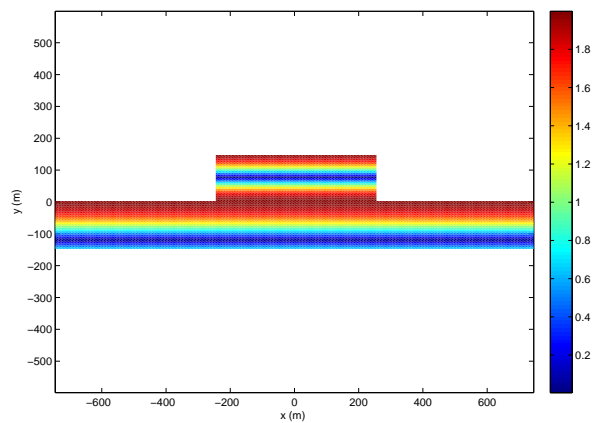}
\caption{Map of $T(x,y;f=3.3374~Hz)$. $\mathbf{\theta^{i}=0^{\circ}}$, $h_{1}=0~m$, $h_{2}=150~m$,  $w=500~m$, $\mu^{[0]}=6.85~GPa$, $\beta^{[0]}=1629.4~ms^{-1}$,  $\mu^{[2]}=2~GPa$, $\beta^{[2]}=1000-i0~ms^{-1}$.  $M=5$.}
\label{fm07}
\end{center}
\end{figure}

\begin{figure}[ht]
\begin{center}
\includegraphics[width=0.55\textwidth]{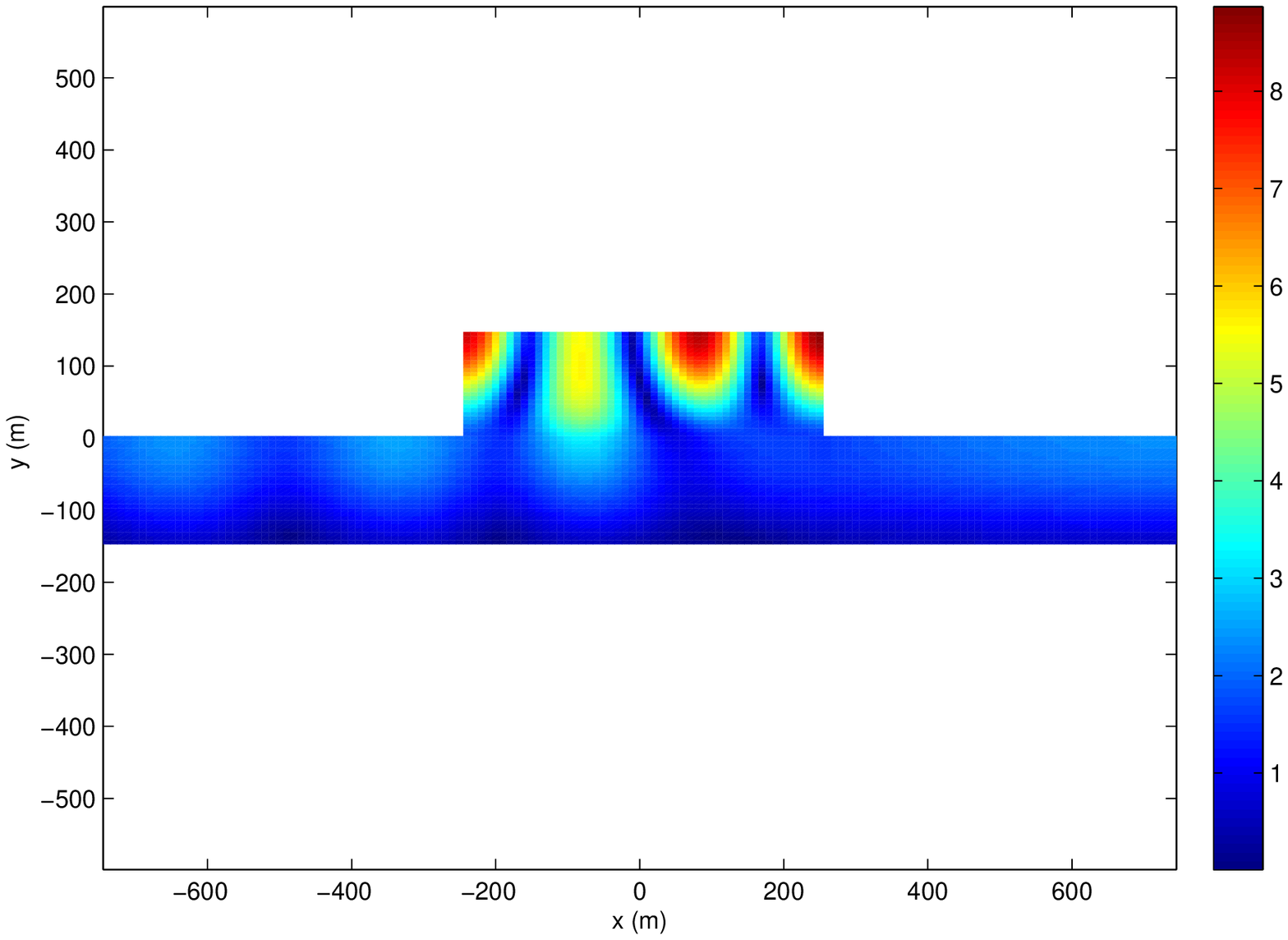}
\caption{Same as fig. \ref{fm07} except that $\mathbf{\theta^{i}=40^{\circ}}$.}
\label{fm08}
\end{center}
\end{figure}

\begin{figure}[ht]
\begin{center}
\includegraphics[width=0.55\textwidth]{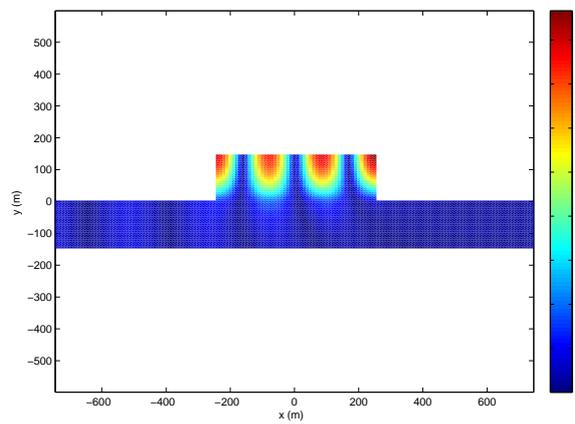}
\caption{Same as fig. \ref{fm07} except that $\mathbf{\theta^{i}=80^{\circ}}$.}
\label{fm09}
\end{center}
\end{figure}

Figs. \ref{fm07}-\ref{fm09} are field maps for three incident angles at $f=3.3374~Hz$.
\clearpage
\newpage
\noindent Notice the very weak coupling at this resonant frequency for normal incidence. However the coupling increases with $\theta^{i}$ to attain a huge amount at four hot spots for $80^{\circ}$ incidence (fig. \ref{fm09}). Again, it is important to underline the fact, illustrated in figs. fig. \ref{vsc-040} and \ref{fm09},  that {\it the transfer function at the midpoint of the top surface of the hill is far from being an adequate indicator of the seismic response within the hill and at other points on its boundary}. The same comments apply to the transfer function at the midpoint of the base of the hill, both of these midpoint transfer functions being often employed (see e.g., \cite{cc02}) to guess what the motion is at arbitrary points within, and on, a convex surface feature such as a mountain, hill or building.
%%%%%%%%%%%%%%%%%%%%%%%%%%%%%%%%%%%%%%%%%%%%%%%
\subsection{With attenuation }
Next, we examine the case in which there are material losses within the  hill. The parameters of the configuration are: $h_{1}=0~m$, $h_{2}=150~m$, $w=500~m$, $\mu^{[0]}=6.85~GPa$, $\beta^{[0]}=1629.4~ms^{-1}$, $\mu^{[2]}=2~GPa$, $\beta^{[2]}=1000-i20~ms^{-1}$.
%%%%%%%%%%%%%%%%%%%%%%%%%%%%%%%%%%%%%%
\subsubsection{Search for the third and fourth resonant frequencies}
\begin{figure}[ht]
\begin{center}
\includegraphics[width=0.65\textwidth]{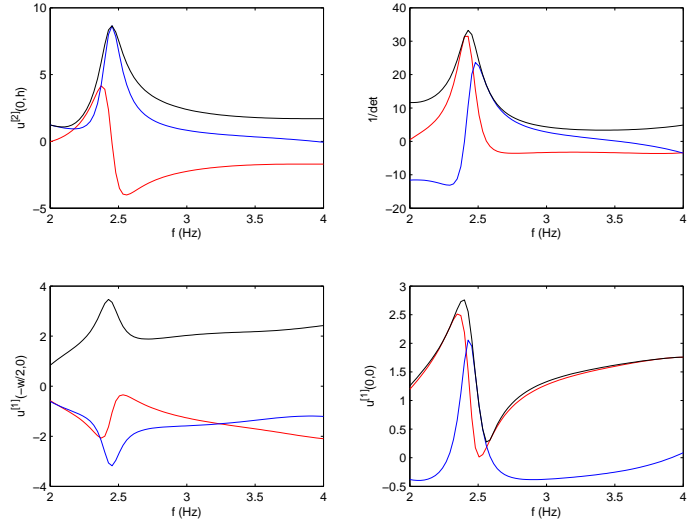}
\caption{The lower right panel is for $T^{(M)}(0,0;f)$, the lower left panel for $T^{(M)}(-w/2,0;f)$ and the upper left panel for $T^{(M)}0,h=h_{2};f)$ whereas the upper right panel depicts $1/D^{(M)}(\omega)$, with $D^{(M)}$ the determinant of the $(M+1)-$by$-(M+1)$ matrix equation involved in the computation of the modal coefficient vector $\mathbf{d}^{(M)}$. The red curves are relative to the real part, the blue curves to the imaginary part and the black curves to the absolute value. $\theta^{i}=80^{\circ}$, $h_{1}=0~m$, $h_{2}=150~m$, $w=500~m$, $\mu^{[0]}=6.85~GPa$, $\beta^{[0]}=1629.4~ms^{-1}$, $\mu^{[2]}=2~GPa$, $\beta^{[2]}=1000-i20~ms^{-1}$. $\mathbf{M=2}$.}
\label{fm10}
\end{center}
\end{figure}

\begin{figure}[ptb]
\begin{center}
\includegraphics[width=0.65\textwidth]{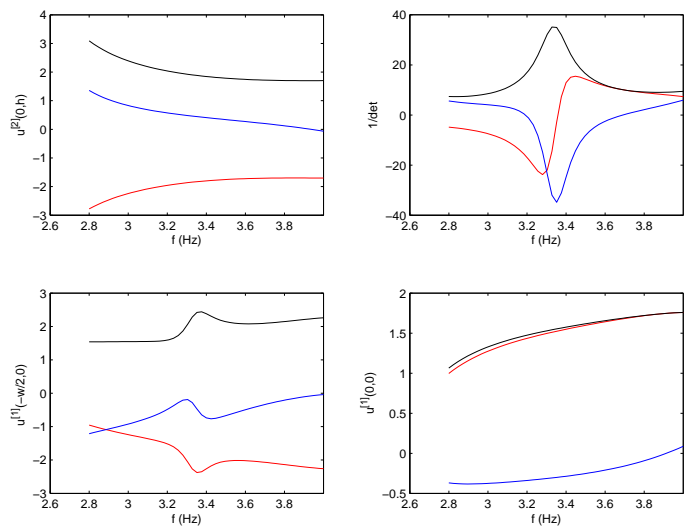}
\caption{Same as fig. \ref{fm10} except that $\mathbf{M=3}$.}
\label{fm11}
\end{center}
\end{figure}

Figs. \ref{fm10}-\ref{fm11} depict the manner of searching for the resonance frequencies.
\clearpage
\newpage
\noindent The third and fourth resonance frequencies are seen to be: $2.44~Hz,~3.3350~Hz$.
%%%%%%%%%%%%%%%%%%%%%%%%%%%%%%%%%%%%%%%%%%%%%%%%%%%%%%%%%%%%%%%%%%%
\subsubsection{Coupling at the third resonance frequency}
\begin{figure}[ht]
\begin{center}
\includegraphics[width=0.55\textwidth]{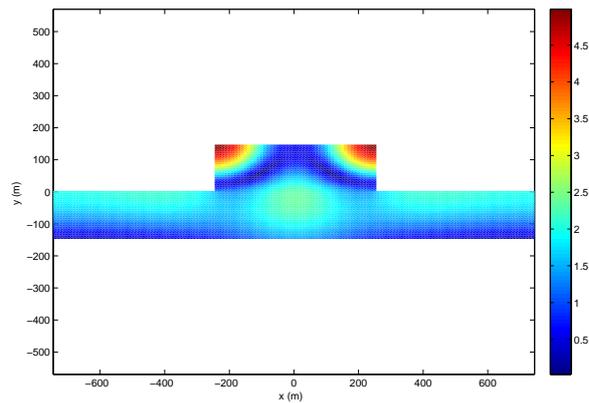}
\caption{Map of $T(x,y;f=2.44~Hz)$. $\mathbf{\theta^{i}=0^{\circ}}$, $h_{1}=0~m$, $h_{2}=150~m$,  $w=500~m$, $\mu^{[0]}=6.85~GPa$, $\beta^{[0]}=1629.4~ms^{-1}$,  $\mu^{[2]}=2~GPa$, $\beta^{[2]}=1000-i20~ms^{-1}$.  $M=5$.}
\label{fm12}
\end{center}
\end{figure}

\begin{figure}[ht]
\begin{center}
\includegraphics[width=0.55\textwidth]{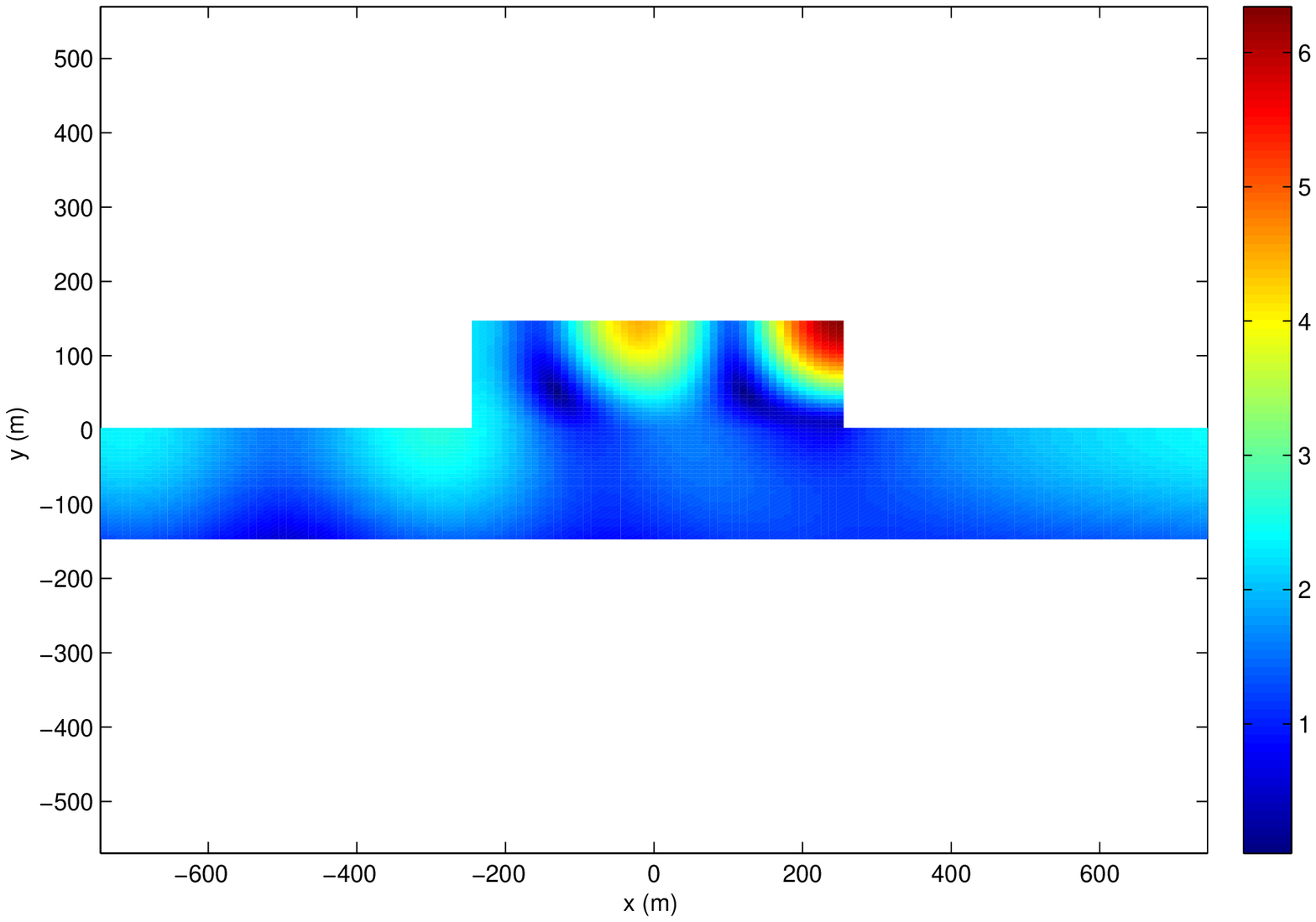}
\caption{Same as fig. \ref{fm12} except that $\mathbf{\theta^{i}=40^{\circ}}$.}
\label{fm13}
\end{center}
\end{figure}

\begin{figure}[ht]
\begin{center}
\includegraphics[width=0.55\textwidth]{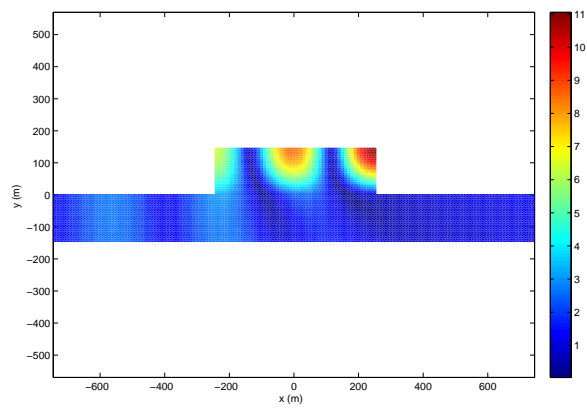}
\caption{Same as fig. \ref{fm12} except that $\mathbf{\theta^{i}=80^{\circ}}$.}
\label{fm14}
\end{center}
\end{figure}

Figs. \ref{fm12}-\ref{fm14} are field maps for three incident angles.
\clearpage
\newpage
\noindent The coupling efficiency again increases  with incident angle and still attains considerable proportions (in fig. \ref{fm14}) for $\theta^{i}=80^{\circ}$.
%%%%%%%%%%%%%%%%%%%%%%%%%%%%%%%%%%%%%%
\subsubsection{Coupling at the fourth resonant frequency}
\begin{figure}[ht]
\begin{center}
\includegraphics[width=0.55\textwidth]{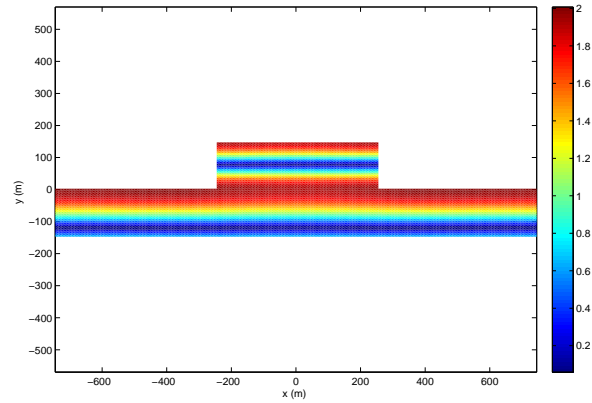}
\caption{Map of $T(x,y;f=3.3350~Hz)$. $\mathbf{\theta^{i}=0^{\circ}}$, $h_{1}=0~m$, $h_{2}=150~m$,  $w=500~m$, $\mu^{[0]}=6.85~GPa$, $\beta^{[0]}=1629.4~ms^{-1}$,  $\mu^{[2]}=2~GPa$, $\beta^{[2]}=1000-i20~ms^{-1}$.  $M=5$. $N=200$.}
\label{fm15}
\end{center}
\end{figure}

\begin{figure}[ptb]
\begin{center}
\includegraphics[width=0.55\textwidth]{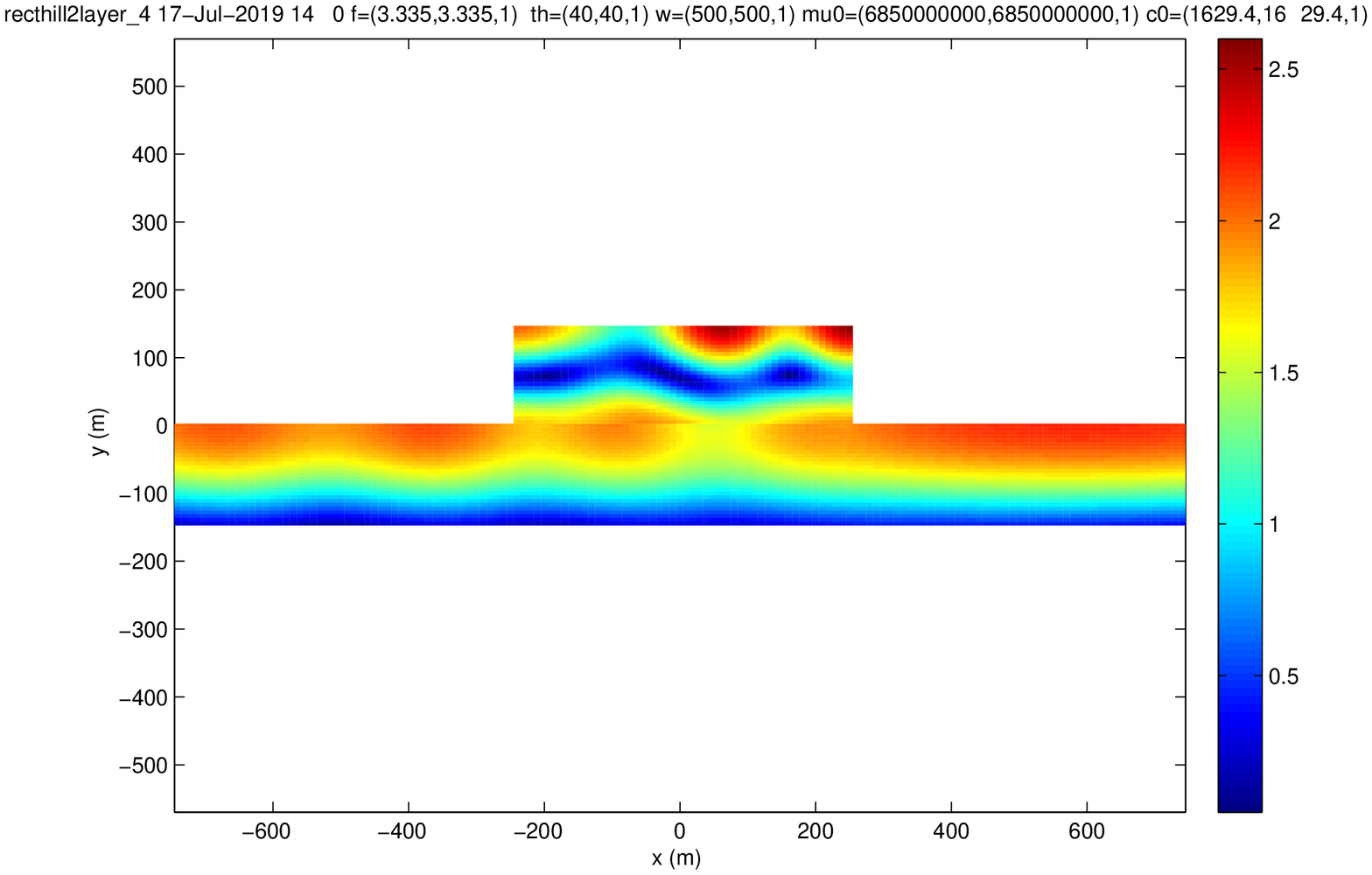}
\caption{Same as fig. \ref{fm15} except that $\mathbf{\theta^{i}=40^{\circ}}$.}
\label{fm16}
\end{center}
\end{figure}

\begin{figure}[ptb]
\begin{center}
\includegraphics[width=0.55\textwidth]{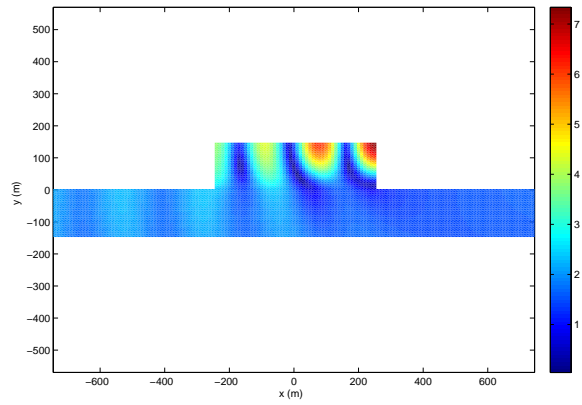}
\caption{Same as fig. \ref{fm15} except that $\mathbf{\theta^{i}=80^{\circ}}$.}
\label{fm17}
\end{center}
\end{figure}

Figs. \ref{fm15}-\ref{fm17} are field maps for three incident angles.
\clearpage
\newpage
\noindent {\it Notice that the lower frequency resonance is much less affected by material loss than the higher frequency resonance at all incident angles}. This will be shown in our second contribution to be a general rule.

Another general rule, already observable in figs. \ref{vsc-020}-\ref{vsc-040} and \ref{fm10}-\ref{fm11}, is that due to the broadening of the resonance peaks in the transfer functions by the introduction of losses, the strong motion  can occur over a wider frequency range than in the absence of losses in the protuberance.

Fig. \ref{fm15} is an illustration of the VS case alluded to in sect. \ref{VS} (which applies, as at present, to normal incidence) since the left-hand side of the first relation in (\ref{4-270}) is equal to $0.1499+i.003$ and the right hand side of the same relation equals $0.1499$, this meaning, together with the fact that the second relation in (\ref{4-270}) is satisfied exactly, that we are are satisfying the conditions (for $N=1$) that define the VS case and are generating by numerical means (with a very small error)  the exact VS solution of the scattering problem alluded-to in sects. \ref{VS}-\ref{VSC}. This is another illustration (in addition to the satisfaction of the conservation of flux relation \cite{wi20}) of the fact that our numerical scheme is sound.
%\clearpage
%\newpage
%%%%%%%%%%%%%%%%%%%%%%%%%%%%%%%%%%%%%%%%%%%%%%%%%%%%%%%%%%%%%%%%%%%%%%%
\section{Seismic response in a small, hard-rock mountain}\label{sgm}
Empirical evidence of wavefield amplification in hard-rock \cite{ac16} mountains has been published in \cite{tk84}. The hard-rock mountain that we now study is higher ($h=1000~m$) and wider ($w=1000~m$) than the previously-studied hills, and the lossless solid (granite) of which it is composed, as well as of the basement, is much stiffer than  those of the hills: $\mu^{[0]}=25~GPa$, $\beta^{[0]}=2650~ms^{-1}$,  $\mu^{[2]}=25~GPa$, $\beta^{[2]}=2650~ms^{-1}$.
%%%%%%%%%%%%%%%%%%%%%%%%%%%%%%%%%%%%%%%%%%%%%%%%
\subsection{Search for the resonances}
\begin{figure}[ht]
\begin{center}
\includegraphics[width=0.65\textwidth]{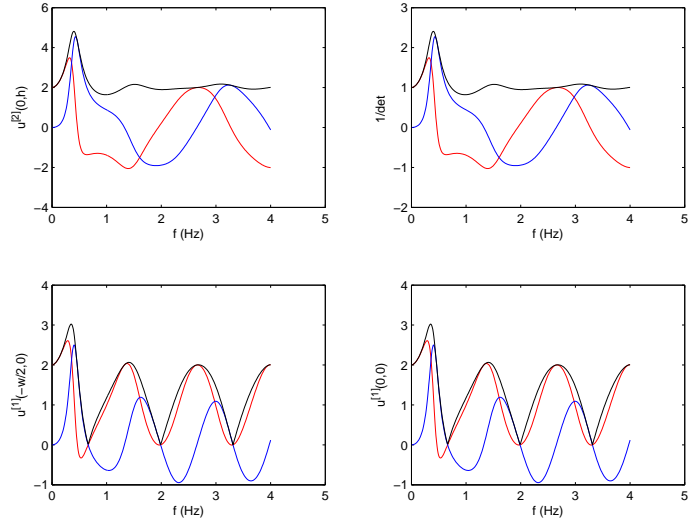}
\caption{$\theta^{i}=80^{\circ}$, $h_{1}=0~m$, $h_{2}=1000~m$, $w=1000~m$, $\mu^{[0]}=25~GPa$, $\beta^{[0]}=2650~ms^{-1}$, $\mu^{[2]}=25~GPa$, $\beta^{[2]}=2650-i0~ms^{-1}$, $\mathbf{M=0}$. The resonance  is at: $f=0.399~Hz$.}
\label{fig-sgm-01}
\end{center}
\end{figure}

\begin{figure}[ptb]
\begin{center}
\includegraphics[width=0.65\textwidth]{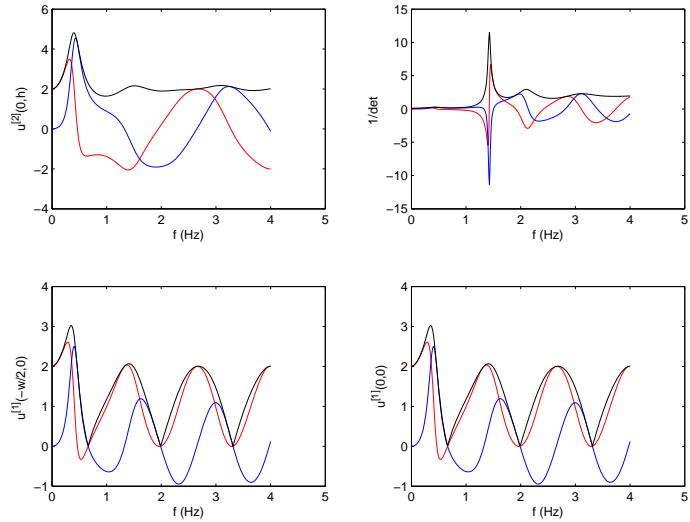}
\caption{Same as fig. \ref{fig-sgm-01} except that $\mathbf{M=1}$. The  resonance is at: $f=1.426~Hz$.}
\label{fig-sgm-02}
\end{center}
\end{figure}

\begin{figure}[ptb]
\begin{center}
\includegraphics[width=0.65\textwidth]{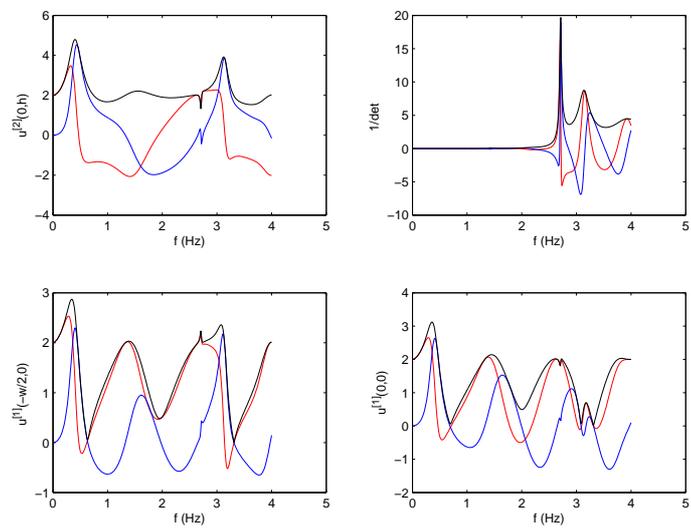}
\caption{Same as fig. \ref{fig-sgm-01} except that $\mathbf{M=2}$. The resonance  is at: $f=2.713~Hz$.}
\label{fig-sgm-03}
\end{center}
\end{figure}

Figs. \ref{fig-sgm-01}-\ref{fig-sgm-03} enable the determination of the first three resonance frequencies of the mountain by location of the maximum of $\|1/D(f)\|$ (the function depicted in the upper right-hand panels of these figures).
\clearpage
\newpage
%%%%%%%%%%%%%%%%%%%%%%%%%%%%%%%%%%%%%%%%%%%%%%%%
\subsection{Response within the mountain at the resonant frequencies and two angles of incidence}
\begin{figure}[ht]
\begin{center}
\includegraphics[width=0.5\textwidth]{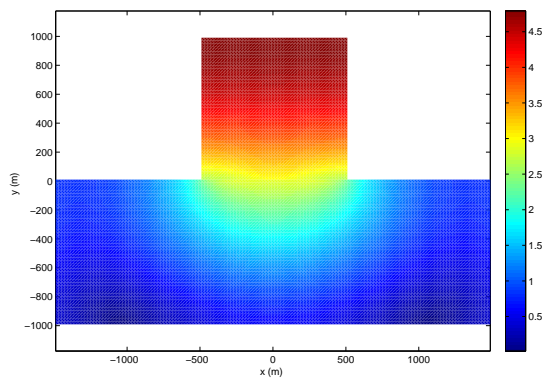}
\caption{$f=0.399~Hz$, $\mathbf{\theta^{i}=0^{\circ}}$, $h_{1}=0~m$, $h_{2}=1000~m$, $w=1000~m$, $\mu^{[0]}=25~GPa$, $\beta^{[0]}=2650~ms^{-1}$,  $\mu^{[2]}=25~GPa$, $\beta^{[2]}=2650-i0~ms^{-1}$, $M=3$.}
\label{fig-sgm-04}
\end{center}
\end{figure}

\begin{figure}[ht]
\begin{center}
\includegraphics[width=0.5\textwidth]{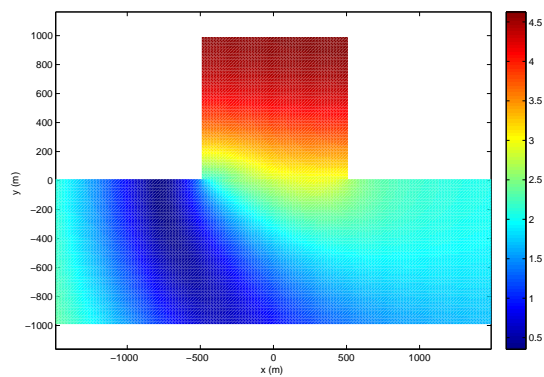}
\caption{Same as fig. \ref{fig-sgm-04} except that $\mathbf{\theta^{i}=80^{\circ}}$.}
\label{fig-sgm-05}
\end{center}
\end{figure}

\begin{figure}[ht]
\begin{center}
\includegraphics[width=0.5\textwidth]{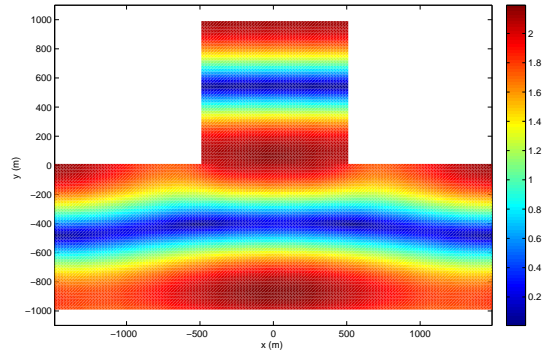}
\caption{$f=1.426~Hz$, $\mathbf{\theta^{i}=0^{\circ}}$, $h_{1}=0~m$, $h_{2}=1000~m$, $w=1000~m$, $\mu^{[0]}=25~GPa$, $\beta^{[0]}=2650~ms^{-1}$,  $\mu^{[2]}=25~GPa$, $\beta^{[2]}=2650-i0~ms^{-1}$, $M=3$}
\label{fig-sgm-06}
\end{center}
\end{figure}

\begin{figure}[ht]
\begin{center}
\includegraphics[width=0.5\textwidth]{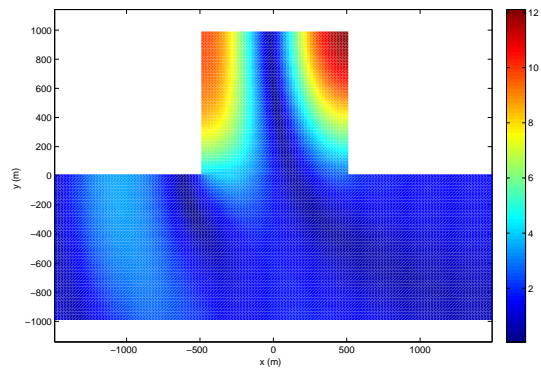}
\caption{Same as fig. \ref{fig-sgm-06} except that $\mathbf{\theta^{i}=80^{\circ}}$.}
\label{fig-sgm-07}
\end{center}
\end{figure}

\begin{figure}[ht]
\begin{center}
\includegraphics[width=0.5\textwidth]{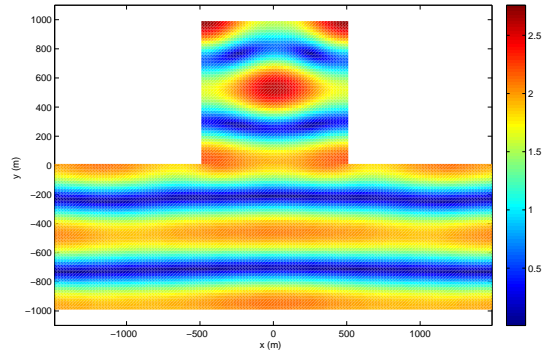}
\caption{$f=2.713~Hz$, $\mathbf{\theta^{i}=0^{\circ}}$, $h_{1}=0~m$, $h_{2}=1000~m$, $w=1000~m$, $\mu^{[0]}=25~GPa$, $\beta^{[0]}=2650~ms^{-1}$,  $\mu^{[2]}=25~GPa$, $\beta^{[2]}=2650-i0~ms^{-1}$, $M=3$.}
\label{fig-sgm-08}
\end{center}
\end{figure}

\begin{figure}[ht]
\begin{center}
\includegraphics[width=0.5\textwidth]{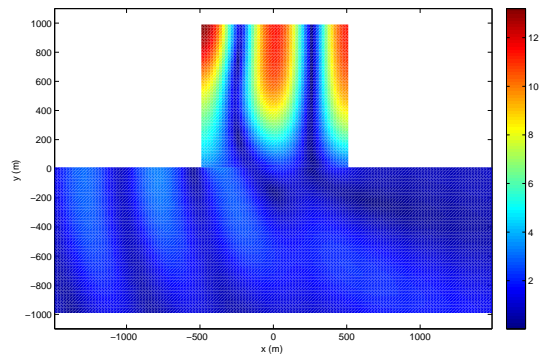}
\caption{Same as fig. \ref{fig-sgm-08} except that $\mathbf{\theta^{i}=80^{\circ}}$.}
\label{fig-sgm-09}
\end{center}
\end{figure}

Figs. \ref{fig-sgm-04}-\ref{fig-sgm-09} constitute field maps of the mountain at several resonant frequencies.

%%%%%%%%%%%%%%%%%%%%%%%%%%%%%%%%%%%%%
\subsubsection{Response at the first resonant frequency}
\clearpage
\newpage
%%%%%%%%%%%%%%%%%%%%%%%%%%%%%%%%%%%%%
\subsubsection{Response at the second resonant frequency}
\clearpage
\newpage
%%%%%%%%%%%%%%%%%%%%%%%%%%%%%%%%%%%%%
\subsubsection{Response at the third resonant frequency}
\clearpage
\newpage
These figures show that mountains respond to seismic waves in much the same manner as hills, i.e., they enable coupling to shape resonances at which frequencies the displacement field can attain significantly-large values within the mountain, especially for large incident angles of the plane-wave solicitation.
%%%%%%%%%%%%%%%%%%%%%%%%%%%%%%%%%%%%%%%%%%%%%%%%
\subsection{Response within the mountain at an off-resonant frequency}
\begin{figure}[ht]
\begin{center}
\includegraphics[width=0.5\textwidth]{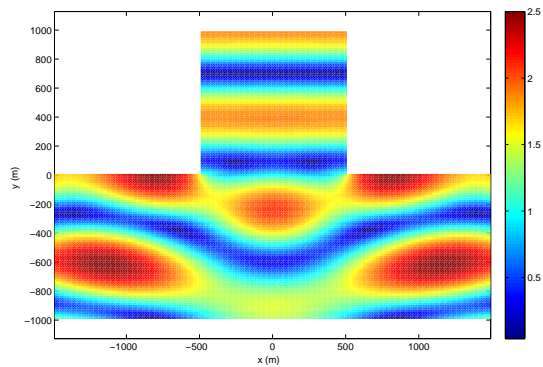}
\caption{$f=2.171~Hz$, $\mathbf{\theta^{i}=0^{\circ}}$, $h_{1}=0~m$, $h_{2}=1000~m$, $w=1000~m$, $\mu^{[0]}=25~GPa$, $\beta^{[0]}=2650~ms^{-1}$,  $\mu^{[2]}=25~GPa$, $\beta^{[2]}=2650-i0~ms^{-1}$, $M=4$.}
\label{fig-sgm-10}
\end{center}
\end{figure}

\begin{figure}[ht]
\begin{center}
\includegraphics[width=0.5\textwidth]{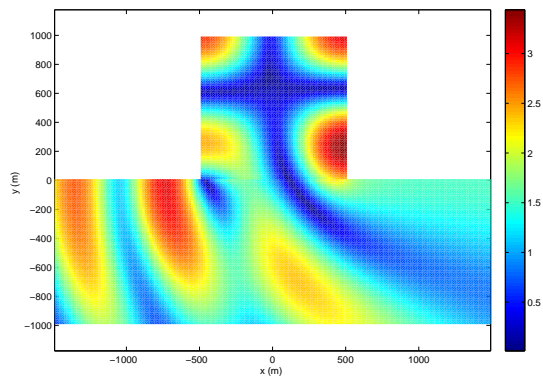}
\caption{Same as fig. \ref{fig-sgm-10} except that $\mathbf{\theta^{i}=80^{\circ}}$.}
\label{fig-sgm-11}
\end{center}
\end{figure}

To appreciate the difference between resonant and non-resonant response, we chose, in figs. \ref{fig-sgm-10}-\ref{fig-sgm-11} (for two incident angles) to offer field maps of the mountain at a frequency rather far away from the previous resonant frequencies..
\clearpage
\newpage
These figures show that large-scale seismic response in a mountain (like that in a hill) is possible only at, or in the neighborhood of, its (shape-) resonant frequencies. Furthermore, at the higher-frequency resonant frequencies, the amplified field is highly-concentrated within the protuberance. At off-resonant frequencies, the field leaks out into the the lower half space, and the modulus of the field within and outside of the protuberance is of the same (modest, i.e., $\sim 2$) order.
%\clearpage
%\newpage
%%%%%%%%%%%%%%%%%%%%%%%%%%%%%%%%%%%%%%%%%%%%%%%%%%%%%%%%%%%%%%%%%%%%%%%%%%%%%%%%%%%%%%%%%%%%%%%%%%%%%%%%%%%%%%%%%%%
\section{Conclusion}
This investigation began with a rather broad evocation of the scientific literature on the problem of the response of a large variety of objects (mostly of geophysical interest) to a wavefield solicitation. The reason for having done this was to show that even though many studies treat this general problem, the ones in one field (specially the one dealing with seismics) usually ignore the results (notably of theoretical nature) obtained in another field. Moreover, the great majority of articles, particularly the recent ones, are of numerical and/or parametric nature wherein it is difficult to discern the underlying physical mecanisms giving rise to the numerical results.

In many of the recent investigations, the effort is directed to configurations that are as close to reality as possible, this being increasingly achievable due to the availability of modern numerical  (boundary-integral, finite difference, finite element, spectral element,...) methods and powerful computers, but although this can have the effect of making possible  the  discovery of  the universal nature of such phenomena as amplified response, it does not usually afford an understanding (which should be of theoretical nature) of this universality.

For this reason, our study was devoted to a quite simple, perhaps unrealistic, geophysical object: a monolayer or bilayer cylindrical hill or mountain of rectangular shape. Moreover, rather than take into account the complex reality of earthquake sources, we assumed a seismic solicitation in the form of a SH plane body wave. For those who think that real hills and mountains are of triangular rather than rectangular shape, we gave references to the many articles, including our own, that treat this problem in a manner (relying on separation-of-variables (SOV) representations of the field) quite similar to the one adopted herein for a rectangular  above-ground structure (AGS). Moreover, we treated the medium in the half-infinite region below the AGS as being homogeneous, although it is well-known that realistic undergrounds are quite inhomogeneous, and often characterized by a layer-like structure, the uppermost layer being softer than the other components of the underground. It is well-known that this layering, especially the one concerning the uppermost layer, has the effect of aggravating the amplifications in the AGS, but this makes it more difficult to define, theoretically, the condition, which we qualify as 'resonant', and related conditions for efficient coupling at the resonance frequencies, that underly  these amplifications (however, the interested reader can refer to Groby's thesis and publications to appreciate more completely the complexity of this problem for the seismic response of an AGS residing on a layered underground). Other sources of complexity that we have avoided are the P or SV polarizations of the 2D problems, and the 3D treatments that are now practically-routinely undertaken in numerical studies. A less common type of problem is one in which the media are nonlinear and anisotropic, but even in the more-recent studies it is rare to encounter ones that deal with these elements of reality.

Having evoked the shortcomings of our study, we now mention some of its achievements. First and foremost, we established the general theoretical framework for (surface shape) resonances and  applied it to the  SOV representation and solution for the chosen AGS. We were thus able to show that the said solution requires the resolution of an infinite-order  linear matrix equation (similar ones occur in a  variety of domain decomposition techniques for the description of  scattering by other types of objects so that the subsequent remarks are actually of quite general nature) for the wavefield in one of the subregions (i.e., its interior) of the rectangular protuberance and that the resonances occur for those shape parameters and frequency for which the determinant of the matrix equation vanishes (for lossless media and neglect of radiation damping) or is very small (for lossy media and/or account being taken of radiation damping), thus resulting in one or a few of the elements of the SOV coefficients becoming large.  We also showed that this condition (for the existence of a resonance) is independent of the characteristics (other than the frequency) of the solicitation, a finding that corroborates and explains such empirically-observed (by Sills and others) independence.

We defined coupling to a resonance as the moment when not only the SOV coefficients (of what we called  'modes'), i.e., the unknowns of the matrix equation, become large, but the characteristics of the solicitation are such as to enable the field within the AGS to become large at certain points within the protuberance (which does not exclude the fact that coupling ti resonances also can have a noticeable effect on the wavefield exterior to the AGS). Thus, a vanishing or very small determinant of the matrix equation is a necessary, but not sufficient condition for the field to become amplified (this amplification is what causes destruction at certain points within AGS's and BGS's during earthquakes) at resonance.

Another achievement of this study is the demonstration that it is possible to obtain an exact solution of the chosen scattering problem for normal incidence and when the frequency and geometic/constitutive parameters satisfy a certain so-called VS condition. This solution, which provides a useful testing device for evaluating subsequent numerical results, amounts to the expression of the scattered field everywhere, i.e., in the protuberance as well as in the underground, as a specularly-reflected wave, so that the total field is everywhere in the form of a standing wave.

Furthermore, under certain low frequency, narrow protuberance, normal-incidence conditions, it is possible to obtain an approximate analytical solution for the wavefield that we called the Homogeneous Shear Wall Resonance (HSWR) which is familiar to all those (starting with Trifunac's shear wall paper) who make use of the so-obtained simple formula (that depends only on the height and constitution of the protuberance) for the resonance frequencies to define what they call 'the resonance frequency of the structure', whatever be the width and shape of the structure. By refining this approximation, we were able to show that when the drastic conditions for its existence are relaxed (as is usually the case in realistic situations) the HSWR frequencies can be rather distant from their actual values and the field at resonance is also different from the HSWR prediction.

Having treated the theoretical aspects, we then undertook the numerical resolution (actually a finite-order version) of the matrix equation, tested the so-obtained solution in the VS case, and by seeing if the conservation of flux relation (refer to our previous paper this year) is,  as it should be, satisfied.

Further confidence in our numerical scheme was provided by the comparison of our response results with those, supposedly-correct, of Sills. Although this author treated the case of a semi-circular cylindrical hill, his transfer functions were found to resemble ours for a rectangular hill that is closest to Sills' semi-circular hill, this being particularly evident at low frequencies. Above all, the said transfer functions were found to possess characteristic resonant behavior (amplifications at certain frequencies), not very different from those of Sills (who was unaware of their resonant nature), which fact, plus  that of the independence of the frequencies of resonance (but not the coupling efficiency at resonance) with respect to the incident angle of the plane wave solicitation, illustrates the universal nature (i.e., for all types of AGS's) of the resonant response of topographical features.

Moreover, as Sills' results are for the transfer functions at selected points on the scattering boundary of his AGS, and since the locations of interest in seismic engineering are rather in the interior of the AGS, we directed our attention to the  internal field of the AGS (this did not exclude the determination of the field in the underground), notably at resonance. This enabled us to show that this field is essentially concentrated (all the more so the resonance frequency and angle of incidence are larger) within (notably near the corners and edges of) the protuberance at resonance, and is much less intense and more dispersed (into the underground) at off-resonance frequencies. All this suggests that the prediction of the seismic response of an AGS by the sole examination of the transfer function at one location (usually taken to be the midpoint of the highest portion of the protuberance) is not foolproof, since the field can, at resonance, be much larger at other points of the boundary, and within, the protuberance. Furthermore, the fact that the efficiency of coupling  to the resonances appears to increase with incident angle, shows that it is dangerous to predict the seismic response of an AGS from the sole response at normal incidence. A final result was that when, as is assumed by Sills, the homogeneous medium within this particular hill is identical to that in the underground, the displacement field can attain values at certain points of the the hill that are nearly four times the value of the field on flat ground (for the same underground) and larger than the amplifications on the boundary of Sills' hill.

Another, although more-hazardous, test of our numerical scheme was provided by a confrontation with the supposely-correct (spectral element) numerical and experimental results of Paolucci for a specific hill in Italy. We encountered a difficulty with Palolucci's paper in  that the constitutive parameters of the hill and the underground are not well-defined therein and the shape of the hill is only approximately rectangular (which makes it difficult to determine its most-appropriate aspect ratio). Contrary to Sills, Paolucci speaks of resonant response, notably near $1~Hz$, but we were unable to obtain transfer functions similar to his for what we considered our rectangular hill to be closest to his. As with Sills, Paolucci assumes the medium in the hill to be homogeneous and the same as the medium in the underground, and, from what we gathered, to be lossless, but to obtain a response similar to that of Paolucci, we had to introduce substantial losses into the medium.

Thus, the confrontation with Paolucci's results was not very satisfactory, but certain remarks, of historical nature, made by the author of this work, concerning the absence of damage to structures on the ground and damage initiated at  the edge(s) of the hill (i.e., landslides), led us to associate the causes of these damages, or absence of damages, with resonances. Again, the field maps within and below the structure proved to be a useful tool for examining this issue. In this way, we showed that the field, rather far away from the $1~Hz$ resonance, is concentrated at the two upper edges of the hill when (as is assumed by Paolucci) the incidence is normal, and at only one of these edges when the incidence is oblique. This led us to the speculation that the damage (i.e., landslides, initiated  predominantly at one edge of the hill) or absence of damage (to towns on the ground outside of the hill) is due to a resonance initiated by a seismic wave  of  higher frequency (i.e., $4.2~Hz$) and larger angle of incidence than what was assumed by Paolucci.

This speculation was based on the perhaps-dubious assumption that an amplification of approximately 2 at an edge of the hill is sufficient to initiate a landslide, so that we were led to inquire as to whether other hills might give rise to much larger amplifications (recall that our Sills-compatible hill gave rise to amplifications about twice this amount). To do this, we  concentrated our attention on a monolayer hill, whose aspect ratio was 3.33, filled with a medium that is softer than that of the underground. We first treated the case in which the filler medium is lossless, and subsequently when it is lossy. In the lossless case we found, locations within the hill at which the resonant response attains a value of more than 40 (i.e. an amplification with respect to the flat ground field of 20) which is not very different from the empirically-obtained values of Davis and West, among others, but for mountains. This large amplification was obtained at rather high resonance frequencies and for large incident angles, and often not at the midpoint of the top portion of the boundary, which again shows that it is dangerous to predict the overall  seismic response of an AGS from its response at such a point.

The next question was whether large amplifications are maintained for the case of a lossy hill. The answer turned out to be that the amplifications at resonance are attenuated all the more so the higher the frequency (they are still largest at 'hotspots' for oblique incidence).

Finally, we turned our attention to the seismic response of a mountain. Our working hypothesis was that a typical mountain: 1) is characterized by a shape which  is such that its aspect ratio (i.e., the ratio of the width $w$ to the height $h$, which we chose to be equal to 1) is  smaller than that of a hill, and 2) by a composition that is such that the solid in the mountain is the same as in the underground and is hard rock, non lossy-like (we chose granite). We then proceeded as previously by searching for the resonance frequencies via the minima of the determinant of the matrix equation for the coefficient vector $\mathbf{d}$ (or $\boldsymbol{\mathcal{F}}$) and then plotted the displacement field maps, inside and underneath the mountain, at these resonance frequencies as well as at a non-resonance frequency. The successive resonance frequencies were found to be lower than their counterparts of the hill, and the fields at these frequencies to be less-intense than for a hill, except at the large angle of incidence. Moreover, as for the hills, the field was found to be non-uniformly distributed within and underneath the protuberance, with a clear tendency for it (i.e., its modulus) to be largest near or on  the top segment of the boundary, this occurring  at all resonance frequencies,  and at the top edges for the higher resonance frequencies and incident angle. This again means that there exist regions within the protuberance at which the field is significantly amplified at resonance and our last two figures show that such is not the case at an off-resonance frequency (i.e., at these frequencies, the field within and underneath the mountain is  close to what it would be (=2) in the absence of the mountain.

All this shows that the principal (i.e., qualitative) characteristics of the seismic response of a mountain are quite similar to those of a hill, and that the occurrence of significative amplification of the displacement field in certain internal regions of both of these structures is due to the coupling of the incident wave to (surface shape) resonances.

Although we have provided a positive answer to Celebi's question (Topographical amplifications-a reality?), there still exists a host of yet unanswered questions, such as those raised by Assimaki \& Jeong on the one hand and Burjanek, F\"ah et al. on the other hand,  concerning the relative importance of the shape and composition (notably 'weathering') of the protuberance in the amplification phenomena. Other, related, questions have to do with how the radiation damping varies with the protuberance parameters, the role of the aspect ratio and material losses of the protuberance in the level of shaking, as well as how to deal with multiple protuberances spread out on the ground. We shall address these questions in subsequent contributions, mostly in the numerical, parametric manner afforded by our domain decomposition separation of variables formulation.
\clearpage
\newpage
%%%%%%%%%%%%%%%%%%%%%%%%%%%%%%%%%%%%%%%%%%%%%%%%%%%%%%%%%%%%%%%%%%%%%%%%%%%%%%%%%%%%%%%%%%%%%%%%%%%%%%%%%%%%

\end{document}